\documentclass[aps,prb,twocolumn,nofootinbib,superscriptaddress,reprint,floatfix]{revtex4-2}
\usepackage{amsmath,amssymb,braket}
\usepackage{bbm}
\usepackage{ifpdf}
\usepackage{color}
\usepackage{xcolor}
\usepackage{graphicx,epsfig}
\usepackage{hyperref}
\usepackage{enumitem}
\usepackage{mathtools}
\usepackage{mathrsfs}

\hypersetup{
	colorlinks = True,
	linkcolor = {teal},
	linkbordercolor = {teal},
	citecolor = {blue},
	citebordercolor = {blue},
	urlcolor = {blue},
}

\usepackage[utf8]{inputenc}
\usepackage{tikz}
\usepackage{pgfplots} 
\usepackage{graphicx}
\usepackage{soul}

\newcommand{\bigket}[1]{\scalebox{1.5}{$|$}\,#1\,\scalebox{1.5}{$\rangle$}}

\usepackage[normalem]{ulem}

\pgfplotsset{compat=1.16} 
\begin{document}
	
	\title{Rydberg Atoms in a Ladder Geometry: Quench Dynamics and Floquet Engineering}
	
	\author{Mainak Pal}
	\email[]{intdydx@gmail.com}
	\affiliation{School of Physical Sciences, Indian Association for the Cultivation of Science, Kolkata 700032, India}
	\affiliation{International Centre for Theoretical Sciences, Tata Institute of Fundamental Research, Bangalore 560089, India}
	
	\author{Tista Banerjee}
	\email[]{banerjeemou09@gmail.com}
	\affiliation{School of Physical Sciences, Indian Association for the Cultivation of Science, Kolkata 700032, India}
	\affiliation{International Centre for Theoretical Sciences, Tata Institute of Fundamental Research, Bangalore 560089, India}
	
	\begin{abstract}
		Rydberg atom quantum simulator platforms are novel quantum simulators for physical systems ranging from condensed matter to particle physics. In this paper, we study out-of-equilibrium quantum dynamics in a model of Rydberg atoms arranged in ladder geometries, with a semi-staggered detuning profile. As the staggering strength ($\Delta) $ is varied from $0\rightarrow\infty$, the model exhibits a wide class of dynamical phenomena, ranging from  quantum many-body scars (QMBS) ($\Delta \sim 0,1$) to integrability induced slow dynamics and approximate Krylov fractures ($\Delta \ge 2$).  We study the robustness of these dynamical features against inevitable influences from the environment in the form of pure dephasing and the finite lifetime of the Rydberg excited state. Additionally, by leveraging an underlying spectral reflection symmetry, we design Floquet protocols having dynamical signatures reminiscent of discrete-time-crystalline (DTC) order and exact Floquet flat bands, and study their stability under protocol imperfections. Finally we consider long-range van der Waals interactions and investigate the validity of the kinetic constraints in an out-of-equilibrium scenario.
	\end{abstract}
	
	\maketitle
	
	%\tableofcontents
	
	\section{\label{sec:intro}Introduction}
	
	Over the past decade, Rydberg atom quantum simulators \cite{Weimer2010,Bernien2017,Nguen_PhysRevX.8.011032,Browaeys2020,Scholl2021,Ebadi2021,Bluvstein2022} have become a versatile testbed for studying fundamental aspects of quantum matter. This includes study of quantum phase transitions \cite{Bernien2017,Keesling2019,Ebadi2021,Kalinowski_PhysRevB.105.174417,Browaeys2020}, topological phases of matter \cite{Kanungo2022_topological_phases,doi:10.1126/science.aav9105,doi:10.1126/science.abi8794},  out-of-equilibrium quantum dynamics \cite{Bernien2017,Bluvstein2022,Maskara_PhysRevLett.127.090602}, quantum information science \cite{Xu_PhysRevLett.127.050501,Xu2024}, quantum state preparation \cite{Carr_PhysRevLett.111.033607,Quantum_control_PhysRevApplied.20.034019,Bell_state_prep_Scholl2023,Roghani_2018,doi:10.1126/science.adh9932} as well as lattice gauge theories (LGTs) \cite{Lattice_gauge_theory_Dalmonte_PhysRevX.10.021041,Gonzalez-Cuadra2025_stringbreaking_2D,borla2025stringbreaking21dmathbbz2,Meth2025_2DLGT,halimeh2022stabilizinggaugetheoriesquantum_stability_quantum_matter}. Such platforms have been particularly helpful in direct demonstration of the equilibration processes of an  out-of-equilibrium initial state under unitary Hamiltonian dynamics and preparation of novel out-of-equilibrium phases of matter \cite{Bernien2017,Bluvstein2022}.\\
	
	Our focus will be on understanding the nature of the non-equilibrium dynamics of such many-body quantum systems, in a variety of settings. In generic isolated interacting many-body quantum systems, where the total energy is the only conserved quantity, it is expected that any out-of-equilibrium initial state (having sub-extensive energy fluctuations) relaxes quickly to a late-time thermal equilibrium that is entirely controlled by the initial energy density of the state with respect to the underlying Hamiltonian \cite{Rigol2008}. Such a thermal equilibrium achieved via a completely unitary quantum evolution, can be understood within the paradigm of the eigenstate thermalization hypothesis (ETH) \cite{Deutsch_PhysRevA.43.2046,MarkSrednicki_1999,Rigol2008,Deutsch_2018}. However, it was realized in Ref.~\cite{Bernien2017} that under certain circumstances, even simple initial states evolving via a generic Hamiltonian, may fail to thermalize in the above sense: certain initial states (for e.g., charge density wave states of various periods) undergo persistent many-body oscillations, evading thermalization within the time scales up to which coherent unitary evolution could be maintained. Consequently, these many-body oscillations were attributed to the presence of quantum many-body scars (QMBS) \cite{Turner2018,Turner_PhysRevB.98.155134}. QMBSs are special eigenstates of the Hamiltonian which have a anomalously large overlap with these special initial states and form an almost equidistant tower in the many-body spectrum, giving rise to near perfect single frequency many-body oscillations. The appearance of QMBSs in the Rydberg atom quantum simulator platform in the so-called Rydberg blockade regime \cite{Rydberg_blockade_ZOLLER_PhysRevLett.87.037901,Jaksch_PhysRevLett.85.2208} (i.e., no two atoms within a certain radius can be simultaneously in the Rydberg excited state due to high energy penalty) can be most easily understood by studying the paradigmatic PXP model \cite{Fendley2004,Sachdev_PhysRevB.66.075128}. Furthermore, it was shown in Ref.~\cite{Maskara_PhysRevLett.127.090602} that by leveraging an underlying chirality operator, which manifests itself as a many-body $\pi$-pulse in this system, one can design Floquet protocols that lead to the realization of a period-two discrete time crystalline order, stabilized by the presence of QMBS. It is worth emphasizing here that the presence of a chirality operator that anti-commutes with the Hamiltonian and commutes with the spatial reflection symmetry operations associated with the Hamiltonian implies the existence of an exponentially large number of exact zero modes, which are protected by an index theorem \cite{Schecter_PhysRevB.98.035139}. \\
	
	The Rydberg atom platforms offer a remarkable degree of tunability in the geometric arrangement of the atoms and control parameters, such as Rabi frequency and detuning strength, via light-matter interactions \cite{Browaeys2020,labuhn:tel-01346557}. Owing to this tunability, it becomes relevant and also interesting to ask how the above features, such as the presence of QMBS and DTC order, manifest themselves when the atomic arrangements and control parameters are tweaked. To explore this, we focus on an arrangement of Rydberg atoms on a 2-leg square ladder geometry (see Fig.~\ref{fig:model}), which aims to understand the properties of two closely spaced atomic chains, with a mutual separation smaller than the Rydberg blockade radius. A PXP model on any bipartite lattice preserves the structure of the chirality operator \cite{Maskara_PhysRevLett.127.090602}, and it carries over to the 2-leg square ladder geometry with zero detuning. Hence, the spectral reflection symmetry and the exponentially large number of exact mid-spectrum zero modes follow immediately. Furthermore, in experiments, it is also possible to implement site-dependent detuning profiles \cite{Bernien2017,Bluvstein_doi:10.1126/science.abg2530,Browaeys2020,labuhn:tel-01346557}; however, in this case, it is no longer guaranteed that the above features, such as the existence of a chirality operator and an exponentially large number of exact mid-spectrum zero modes, will be preserved under an arbitrary detuning profile. In Ref.~\cite{Pal_PhysRevB.111.L161101} a simple semi-staggered detuning profile (see Fig.~\ref{fig:model}) was utilized to preserve these desirable features. This choice led to the appearance of QMBSs, which are qualitatively different in nature compared to those of the paradigmatic PXP chain.\\
	
	In this paper, we have extensively studied the nature of non-equilibrium dynamics of the model considered in Ref.~\cite{Pal_PhysRevB.111.L161101} for an extended parameter regime. We find that tuning the strength of staggering ($\Delta$) leads to a broad range of dynamical phenomena. As already pointed out in Ref.~\cite{Pal_PhysRevB.111.L161101}, the model hosts QMBS for $\Delta \sim 0,1$. We shall provide evidence that, in addition to hosting QMBS, the model also exhibits slow dynamics and approximate Krylov fractures induced by approximate emergent integrability arising for large staggering strengths $\Delta \ge 2.5$. This is a consequence of the fact that the second order low-energy effective Hamiltonian becomes exactly integrable, and higher order non-integrable terms only contribute very weakly in this parameter regime. The particular form of the second-order effective Hamiltonian allows us to explicitly identify the extensive number of quasi-conserved charges. We briefly illustrate the occurrence of similar anomalous dynamical phenomena in a 3-leg generalization of the model.\\
	
	Additionally, by leveraging the spectral reflection symmetry of the Hamiltonian, one can design Floquet protocols leading to dynamical signatures reminiscent of DTC order \cite{Else_PhysRevLett.117.090402,Yao_PhysRevLett.118.030401,Khemani_PhysRevLett.116.250401,Keyserlingk_PhysRevB.94.085112,Zhang2017,Choi2017,Sacha_2018,Else_annurev:/content/journals/10.1146/annurev-conmatphys-031119-050658,khemani2019briefhistorytimecrystals} and exact Floquet flat bands, which are important in quantum control and quantum information processing applications \cite{Bomantara_PhysRevB.104.L180304,danieli2025quantumstorageflatbands}. We  study the robustness of these distinct classes of dynamical phenomena against possible imperfections in the implementation of the Floquet protocols. \\
	
	The physical Rydberg atom quantum simulator platforms are inevitably influenced by environmental loss channels. Hence it is important to study the robustness of some of the dynamical features exhibited in these models under purely unitary evolution, against environmental effects such as pure dephasing and the finite lifetime of the Rydberg excited state. We do this by assuming that the dynamics is governed by the Gorini-Kossakowski-Sudarshan-Lindblad (GKSL) master equation \cite{Lindblad1976,GKS_10.1063/1.522979}, under dissipative processes such as depasing and spontaneous emission from the Rydberg excited state. \\
	
	Finally, we take into account how the presence of long-range van der Waals (vdW) repulsive interactions (which are present in the actual experimental setup) influences the nature of quantum dynamics in such models. Our results suggest that, for the 2-leg model, the vdW repulsive interactions between the diagonally placed atoms are not negligible (compared to immediate neighbors). This automatically brings into question the validity of simple kinetic constraints used to analyze the non-equilibrium dynamics, which have been helpful in analyzing low-energy features both theoretically \cite{Samajdar_PhysRevLett.124.103601,Ebadi2021,Sarkar_10.21468/SciPostPhys.14.1.004,Liao_PhysRevB.111.165154,Soto-Garcia_PhysRevResearch.7.013215,Yuzhou2025,Pierre_PhysRevB.106.155411} and experimentally \cite{Labuhn201_Long_range_Ising_model_expt,Schaub2012}. We present a quantitative comparison between fully connected long-range interacting systems and kinetically constrained systems to illustrate this clearly. Motivated by our results on the ladder systems, we briefly revisit the persistent many-body oscillations stabilized by QMBS observed in two-dimensional PXP models \cite{Lin_PhysRevB.101.220304} and study their stability against long-range interactions. \\
	
	The rest of this paper is organized as follows: in Sec.~\ref{sec:model} we introduce the model Hamiltonian and discuss some of its special properties. In Sec.~\ref{subsec:Delta_0} we briefly recapitulate the QMBS phenomenology of the model as presented in Ref.~\cite{Pal_PhysRevB.111.L161101}. Next, we focus on the emergent integrable behavior (Sec.~\ref{subsec:emergent_integrability}), the nature of short-range spectral correlations (Sec.~\ref{subsec:short_range_spectral_correlations}), slow dynamics (Sec.~\ref{subsec:slow_dynamics}) and stronger forms of ergodicity breaking (Sec.~\ref{subsec:stronger_erogidicity_violations}). In Sec.~\ref{main-sec:floquet-engineering}, we design Floquet protocols that (i) lead to sub-harmonic responses with a state-dependent revival period (Sec.~\ref{sec:DTC}) and (ii) give rise to exact Floquet flat bands (Sec.~\ref{sec:floquet-flat-bands}). In Sec.~\ref{open-quantum-system} we analyze the stability of QMBS (Sec.~\ref{stability-of-QMBS_environment}) and emergent conservation laws (Sec.~\ref{emergent_conservation_open}) in the presence of environmental loss mechanisms, such as pure-dephasing and spontaneous decay of the Rydberg excited states. Finally, in Sec.\ref{sec:long-range} we relax the strict kinetic constraint and consider long-range van der Waals interactions with $nS$-type excited orbitals. We comment on the validity of the constraints assumed in the rest of the article for the 2-leg model (Sec.~\ref{subsec:2leg-blockade}) and the 2D square lattice generalizations (Sec.~\ref{subsec:2D_pxp_staggered_detuning}).
	
	\section{Model\label{sec:model}}
	
	In this paper, we focus on the nature of out-of-equilibrium quantum dynamics in the model governed by the Hamiltonian \eqref{main:eq:hamiltonian_ladder}. This model is a non-perturbative generalization of the paradigmatic PXP chain \cite{Fendley2004,Sachdev_PhysRevB.66.075128,Turner2018,Turner_PhysRevB.98.155134} to the case of a 2-leg square ladder geometry with a semi-staggered detuning profile along the longer direction. The Hamiltonian for this system is given as,
	
	\begin{equation}
		\frac{\hat{H}}{\hbar}= \sum_{j=1}^L\sum_{a=1}^2 \Omega\;\hat{P}^\downarrow_{j-1,a}\hat{P}^\downarrow_{j+1,a}\hat{P}^\downarrow_{j,\overline{a}} \hat{\sigma}^{x}_{j,a} -\Delta (-1)^j \;\hat{\sigma}_{ja}^z
		\label{main:eq:hamiltonian_ladder}
	\end{equation}
	
	Here $\hat{\sigma}^{\alpha}_{j,a}$ are spin-$\frac{1}{2}$ operators at site $(j,a)$, with $j=1,2,...,L$, $a=1,2$, and $\alpha=x,y,z$. The operators $\hat{P}^\downarrow_{j,a}=(1-\hat{\sigma}^z_{j,a})/2$ are local projection operators onto the spin-down state ($\ket{\downarrow}_{j,a})$ i.e. the electronic ground-state ($\ket{\circ}_{j,a}$) at site $(j,a)$. For brevity, we use the notation $\hat{\tilde{\sigma}}^{\alpha}_{j,a}=\hat{P}^\downarrow_{j-1,a}\hat{P}^\downarrow_{j+1,a}\hat{P}^\downarrow_{j,\overline{a}} \hat{\sigma}^{\alpha}_{j,a}$ where $\overline{a}$ denotes the leg opposite to $a$, i.e. $\overline{a}=1$ when $a=2$ and vice-versa, meaning the site $(j,\overline{a})$ is the neighbor of site $(j,a)$ in the vertical direction. The total number of atoms is $N=2L$. We also denote the diagonal and off-diagonal parts of the Hamiltonian (in the computational basis) as $\hat{H}_z=-\Delta\sum_{j,a} (-1)^j \hat{\sigma}^z_{j,a}$ and $\hat{H}_x=\Omega\sum_{j,a} \hat{\tilde{\sigma}}^x_{j,a}$, so that $\hat{H}=\hat{H}_x+\hat{H}_z$. In the language of Rydberg atom quantum simulator platforms \cite{Bernien2017,Bluvstein_doi:10.1126/science.abg2530,Browaeys2020}, the Rabi frequency in Eq.~\eqref{main:eq:hamiltonian_ladder} is $2\Omega$, and the detuning at the site labeled by $(j,a)$ is $(-1)^j2\Delta$. Henceforth, we set $\hbar=1$, $\Omega=1$, and vary $\Delta$ to access different dynamical regimes of the model. \newline

	\begin{figure}[!htpb]
		\centering
		\includegraphics[width=0.5\textwidth]{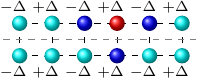}
		\caption{Schematic representation of the 2-leg Rydberg atom square ladder setup with a semi-staggered detuning profile $\Delta_{j,a}=(-1)^j\Delta$. The dashed gray line shows spatial reflection symmetry axes. The kinetic constraint i.e., the strong Rydberg blockaded regime is illustrated as follows: if an atom is in the $\ket{\bullet} \equiv \ket{\uparrow}$ state (Rydberg excited state, denoted by the red sphere), then all its neighboring atoms must be in the $\ket{\circ} \equiv \ket{\downarrow}$ state (electronic ground-state, denoted by blue spheres) and cannot be flipped to the $\ket{\bullet}$ state under the action of the Hamiltonian \eqref{main:eq:hamiltonian_ladder}.}
		\label{fig:model}
	\end{figure}
	
	The generalization of the PXP chain ($\hat{H}_{\text{PXP}}=\Omega\sum_{j=1}^L \hat{\tilde{\sigma}}^x_j$) to the model \eqref{main:eq:hamiltonian_ladder} preserves some important features found in the PXP chain, such as (i) the spectral reflection symmetry of $\hat{H}$ and (ii) an exponentially large number of zero modes protected by an index theorem \cite{Schecter_PhysRevB.98.035139}. These properties can be understood by considering the chirality operators $\hat{C}_{1,2}$ 
	
	\begin{subequations}
		\begin{align}
			\hat{C}_1 & = \hat{T}_x \hat{C} \label{eq:C1-defn} \\
			\hat{C}_2 & = \hat{T}_x\hat{T}_y \hat{C} \label{eq:C2-defn}
		\end{align}
	\end{subequations}
	
	which anti-commutes with the Hamiltonian, i.e. $\{\hat{C}_{1,2},\hat{H}\} = 0$. Here $\hat{C} = \prod_{j,a} \hat{\sigma}^z_{j,a}$ and $\hat{T}_{x,y}$ denote the lattice-translation operators by one unit along the longer and the shorter directions, respectively. The anti-commutation implies that eigenstates of $\hat{H}$ appear in pairs $\ket{\pm E}$, where $\ket{-E}=\hat{C}_{1,2}\ket{E}$, and thus we refer to $\hat{C}_{1,2}$ as chirality operators. Moreover, as the operators $\hat{H}$, $\hat{C}_{1,2}$ all commute with the spatial reflection operation about the horizontal line (dashed gray line in Fig.~\ref{fig:model}), it follows that the Hamiltonian \eqref{main:eq:hamiltonian_ladder} hosts an exponentially large number (in system size $N$) of exact mid-spectrum zero modes, which are protected by an index theorem \cite{Schecter_PhysRevB.98.035139} for any value of $\Delta$.\newline 
	
	As the model \eqref{main:eq:hamiltonian_ladder} shares a number of key features with the paradigmatic PXP chain, it is interesting to ask whether it hosts anomalous non-equilibrium dynamics, similar to QMBS \cite{Bernien2017,Turner2018,Turner_PhysRevB.98.155134} or DTC order observed in the periodically kicked PXP chain \cite{Maskara_PhysRevLett.127.090602}. A part of this question was answered in Ref.~\cite{Pal_PhysRevB.111.L161101}, where the QMBS phenomenology of this model was studied for $\Delta\in[0,1]$. Our goal in this paper is to present evidence of further anomalous dynamical behavior, both in the quench dynamics for a range of parameters and to consider the possibility of engineering Floquet protocols by leveraging the underlying spectral reflection symmetry.
	
	\section{Quench Dynamics\label{main-sec:quench_dynamics}}
	
\begin{figure}[!htpb]
		\centering
		\includegraphics[width=0.485\textwidth]{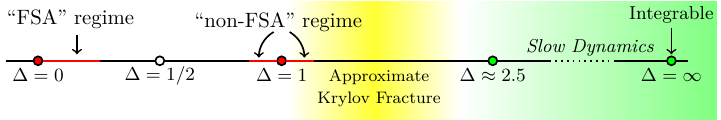}
		\caption{Schematic representation of the phenomenology of quench dynamics of model \eqref{main:eq:hamiltonian_ladder} when a single parameter $\Delta$ is varied.}
		\label{fig:quench_dynamics_schematics_of_dynamical_phases}
	\end{figure}

	Recently, it has been argued that the model \eqref{main:eq:hamiltonian_ladder} hosts a variety of QMBS for parameter choices $\Delta \sim 0,1$ \cite{Pal_PhysRevB.111.L161101}. In this paper, we present substantial evidence for the fact that, in addition to QMBS, the model \eqref{main:eq:hamiltonian_ladder} hosts other classes of anomalous quantum dynamics for a broad range of parameters. As per our current understanding, the schematic Fig.~\ref{fig:quench_dynamics_schematics_of_dynamical_phases} illustrates the broad range of quench dynamics that the model \eqref{main:eq:hamiltonian_ladder} exhibits.
	
	As $\Delta$ is varied from $0\rightarrow\infty$, one encounters (i) weak ergodicity violation due to the presence of quantum many-body scars (QMBS) \cite{Pal_PhysRevB.111.L161101}, (ii) emergent approximate integrability-induced slow dynamics, and approximate Krylov fractures. In what follows, we present a detailed description of (ii) and only provide a brief recapitulation of (i).
	
	\subsection{Weak violation of Ergodicity \label{subsec:Delta_0}}
	
	At $\Delta=0$, the model \eqref{main:eq:hamiltonian_ladder} becomes a ``PXP" type model on the 2-leg square ladder. Since this lattice is bipartite, it is expected to host QMBS \cite{Omiya_PhysRevA.107.023318}. Explicit numerical evidence for persistent oscillations in the 2-leg square ladder ``PXP" model was presented in Ref.~\cite{Pal_PhysRevB.111.L161101}, where it was shown that the quantum quench dynamics under the Hamiltonian \eqref{main:eq:hamiltonian_ladder} starting from the N\'eel or $\ket{\mathbb{Z}_2}$ state (i.e. $\ket{\substack{\circ\bullet\circ\bullet\circ\bullet...\\\bullet\circ\bullet\circ\bullet\circ...}}$), or equivalently its translated partner $\ket{\overline{\mathbb{Z}}_2}$ state (i.e., $\ket{\substack{\bullet\circ\bullet\circ\bullet\circ...\\\circ\bullet\circ\bullet\circ\bullet...}}$), exhibit persistent many-body oscillations that last several cycles. This behavior was attributed to the presence of QMBS. The QMBS observed in Ref.~\cite{Pal_PhysRevB.111.L161101} for $\Delta\sim 0$ were found to be qualitatively similar to those in the 1D PXP chain in the sense that, they could be understood from a semi-analytical forward scattering approximation (FSA) picture \cite{Turner2018,Turner_PhysRevB.98.155134}. This has been labeled as ``FSA"-like QMBS in the schematic Fig.~\ref{fig:quench_dynamics_schematics_of_dynamical_phases}. In addition to this ``FSA" regime, Ref.~\cite{Pal_PhysRevB.111.L161101} also presented numerical evidence of many-body persistent revivals starting from the $\ket{\mathbb{Z}_2}$ (or equivalently $\ket{\overline{\mathbb{Z}}_2}$) state and the electronic ground-state at all sites, $\ket{\text{vac.}}=\ket{\substack{\circ\circ...\circ\\\circ\circ...\circ}}$, at $\Delta\sim 1$. However, these revivals cannot be understood through a straightforward adaptation of the FSA scheme, due to the presence of finite longitudinal fields, which makes the energy-landscape of the Fock space non-trivial. A qualitative picture leading to this second class of revivals remains to be understood. We label these revivals as ``non-FSA"-like QMBS in Fig.~\ref{fig:quench_dynamics_schematics_of_dynamical_phases}.\\
	
	\subsection{Emergent Integrability\label{subsec:emergent_integrability}} 
	
	In the limit $\Delta\rightarrow\infty$, only the staggered detuning term in the Hamiltonian is relevant, rendering the Hamiltonian completely integrable: all Fock states in the $\sigma^z$ basis allowed by the Hilbert space constraint are eigenstates of the Hamiltonian. Moreover, constructing a perturbative low-energy effective Hamiltonian for large finite values of $\Delta$ via a Schrieffer-Wolff (SW) transformation reveals that, up to second-order in $\Omega/\Delta$, the low-energy effective Hamiltonian is exactly integrable. The effective Hamiltonian at second-order is $\hat{H}_{\text{eff}}^{[2]}=H_z + \hat{H}_{\text{eff}}^{\left(2\right)}$, where $\hat{H}_{\text{eff}}^{\left(2\right)}$ takes the following form (see Appendix-\ref{app:Heff_SW} for details of the derivation).

	\begin{widetext}
		\begin{equation}
			\begin{split}
				\hat{H}_{\text{eff}}^{\left(2\right)} & = -\frac{\Omega^2}{2\Delta} \sum_{j=1}^{L} \sum_{a=1}^2 (-1)^j  \hat{\tilde{\sigma}}^z_{j,a} -\frac{\Omega^2}{4\Delta} \sum_{j=1}^L \sum_{a=1}^2 (-1)^j \hat{P}^\downarrow_{j-1,a} \hat{P}^\downarrow_{j-1,\overline{a}} \hat{P}^\downarrow_{j+1,a} \hat{P}^\downarrow_{j+1,\overline{a}} \hat{P}^\downarrow_{j,a} \left( \hat{\sigma}^x_{j,a}\hat{\sigma}^x_{j,\overline{a}}+\hat{\sigma}^y_{j,a}\hat{\sigma}^y_{j,\overline{a}}\right) 
				\label{eq:Heff_SW}
			\end{split}
		\end{equation}     
	\end{widetext}    
	
	At this order in perturbation theory, there are spin-flip processes (see Eq.~\eqref{eq:Heff_SW}) with an extensive number of conservation laws, namely, 
	
	\begin{subequations}
		\begin{align}
			\hat{Z}_\pi=\sum_{j,a}(-1)^j\hat{\sigma}^z_{j,a} \label{eq:main:conservation_laws_Zpi} \\
			\hat{Q}_j=\hat{\sigma}^z_{j,1}\hat{\sigma}^z_{j,2}, \:\forall j=1,2,...,L
			\label{eq:main:conservation_laws_Qj}
		\end{align}
	\end{subequations}
	
	The presence of these additional conservation laws can be understood by considering the action of $\hat{H}_{\text{eff}}^{(2)}$ on the Fock states: the terms in the second summation of Eq.~\eqref{eq:Heff_SW} essentially perform ``blockaded spin-flips", i.e., $\ket{\substack{...\circ_{j-1}\bullet_{j}\circ_{j+1}...\\...\circ_{j-1}\circ_{j}\circ_{j-1}...}} \leftrightarrow \ket{\substack{...\circ_{j-1}\circ_{j}\circ_{j+1}...\\...\circ_{j-1}\bullet_{j}\circ_{j+1}...}}$ where the ``blockade" is due to the $\hat{P}^\downarrow$ projectors at sites $j\pm1$ on both legs of the ladder. This leads to the realization of an effective two-level oscillation of a Rydberg excitation between the two legs at the $j^{\text{th}}$ rung, provided that this rung is ``blockaded" appropriately by $\ket{\circ}$ states at surrounding sites. As a consequence of this, all eigenstates of $\hat{H}_{\text{eff}}^{(2)}$ can be labeled by the quantum numbers corresponding to these conserved quantities, and a recipe for such an enumeration procedure is given in Appendix-\ref{subsec:Heff2spectrum}. As special cases, we note that the states $\ket{\mathbb{Z}_2},\ket{\text{vac.}}$ remain frozen under the action of $\hat{H}_{\text{eff}}^{(2)}$.
	
	Going beyond $\hat{H}_{\text{eff}}^{\left(2\right)}$, we find that the third-order effective Hamiltonian vanishes exactly, and the next non-trivial contribution comes from the fourth-order terms in perturbation theory, which break the emergent integrability (see Appendix-\ref{app:Heff4_integrability_breaking} for details). However, the strengths of these integrability breaking terms are $\sim \Omega^3/\Delta^4$, and they contribute very weakly for the parameter range $\Delta \ge 2$ when spectral correlations and time-evolution from simple initial states are considered.

	\subsection{Short-range spectral correlations} \label{subsec:short_range_spectral_correlations}
	
\begin{figure}[!htpb]
		\centering
		\rotatebox{0}{\includegraphics[width=0.435\linewidth]{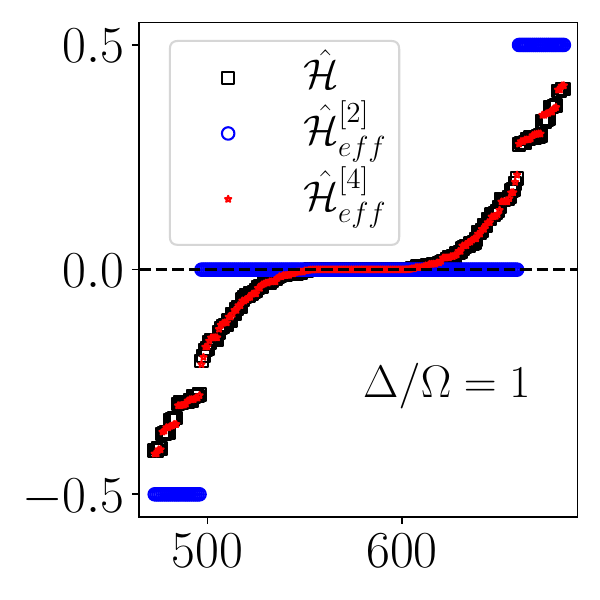}}
		\rotatebox{0}{\includegraphics[width=0.435\linewidth]{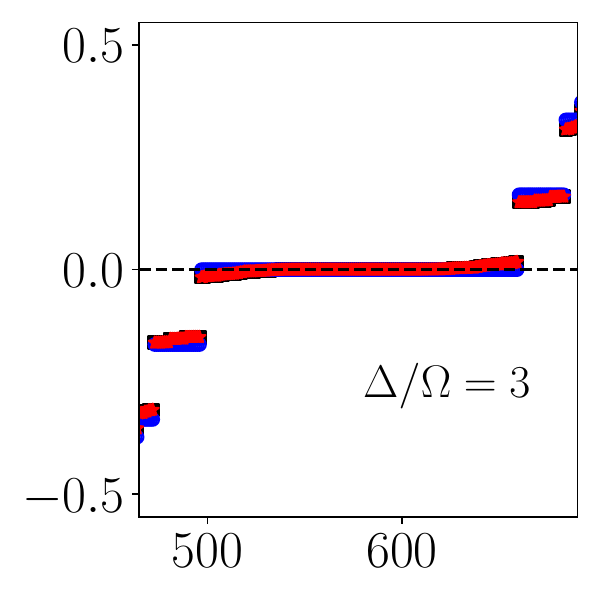}}
		\caption{Comparison of the many-body spectrum of $\hat{H}$, $\hat{H}^{[2]}_{\text{eff}}$ and $\hat{H}^{[4]}_{\text{eff}}$ for $\Delta=1,3$ with $N=16$ atoms obtained via ED. (only the middle part of the spectrum is shown).}
		\label{fig:compare_exact_vs_SW2nd_SW4th_spectrum}
	\end{figure}

\begin{figure}[!hbtp]
		\centering
		\rotatebox{0}{\includegraphics[width=0.5\textwidth]{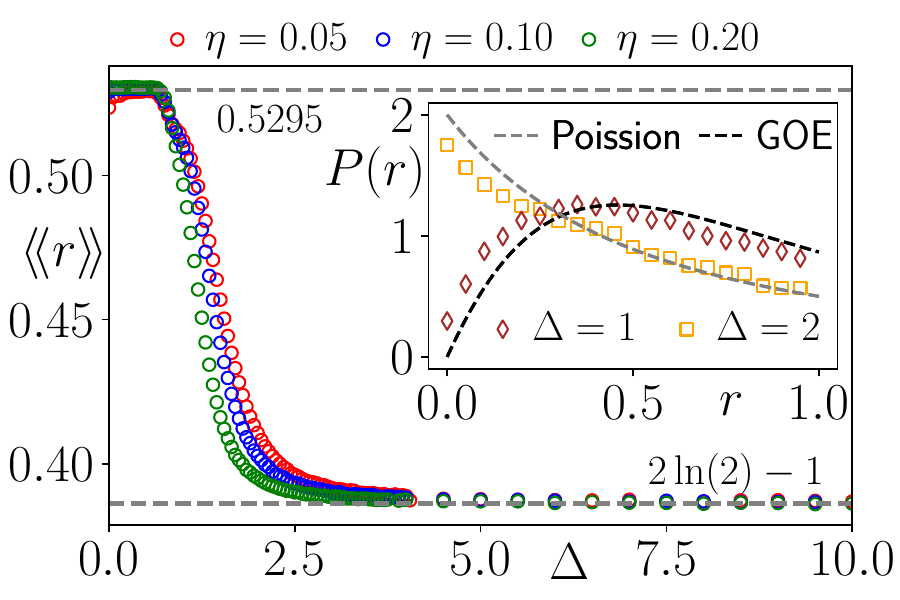}}
		\caption{Inset: Distribution of consecutive level spacing ratios ($P(r)$) for $\Delta=1,2$ with translation quantum numbers $k_{x,y}=\pi$ (see Appendix-\ref{app-hsd}) obtained via ED. Main panel: Disorder averaged mean consecutive level spacing ratio $\langle\!\langle r \rangle\!\rangle$ as a function of staggered detuning strength $\Delta$ for $N=24$ sites (averaged over 100 disorder realizations).}
		\label{fig:spectral_statistics}        
	\end{figure}

	The fact that the second-order perturbative Hamiltonian $\hat{H}^{[2]}_{\text{eff}}$ remains exactly integrable has some implications for the short range spectral correlations of the system, as well as for the relaxation dynamics from certain non-equilibrium initial states. As a measure of short-range spectral correlations, we have studied the distribution of the ratio ($r_n$) of consecutive level spacings ($s_n$), defined as $r_n=\text{min}(s_n,s_{n-1})/\text{max}(s_n,s_{n-1})$, where $s_n=E_{n+1}-E_n$ denotes the gap between the $n$-th and the $(n+1)^\text{th}$ eigenstates of $\hat{H}^{[2]}_{\text{eff}}$ \cite{Oganesyan_PhysRevB.75.155111}. The eigenvalues $E_n$ are obtained by performing numerical exact diagonalization (ED) of the Hamiltonian $\hat{H}$ (Eq.~\eqref{main:eq:hamiltonian_ladder}) in the basis of constrained Fock states which satisfy the strong Rydberg-blockade condition on the 2-leg ladder \cite{Pal_PhysRevB.111.L161101,Turner_PhysRevB.98.155134,Turner2018}. At values of $\Delta\sim 2$, due to the integrability breaking term being weak, the consecutive level-spacing ratio distribution ($P(r)$) resembles a Poissonian nature (see inset of Fig.~\ref{fig:spectral_statistics} ). For $\Delta\sim 1$ (see inset of Fig.~\ref{fig:spectral_statistics}), $P(r)$ exhibits level repulsion, which is a consequence of the fact that the fourth-order integrability breaking terms are not weak in this regime, and the effective Hamiltonian is non-integrable. As Fig.~\ref{fig:compare_exact_vs_SW2nd_SW4th_spectrum} (right panel) demonstrates, the eigenvalues of $\hat{H}^{[2]}_{\text{eff}}$ and that of the full Hamiltonian $\hat{H}$, obtained by numerical exact diagonalization (ED) agree very well for $\Delta\sim 3$, whereas for $\Delta\sim 1$ (Fig.~\ref{fig:compare_exact_vs_SW2nd_SW4th_spectrum}, left panel) one needs to consider higher order processes such as $\hat{H}^{[4]}_{\text{eff}}$, to capture the spectral features of the full Hamiltonian $\hat{H}$. We have obtained the matrix elements of $\hat{H}^{[4]}_{\text{eff}}$ in the constrained (strong Rydberg-blockaded) Fock basis through a numerical implementation of Eq.~\eqref{app:eq:Heff_formulas}.\newline
	
	The emergent integrability is not related to any lattice symmetries of the Hamiltonian \eqref{main:eq:hamiltonian_ladder}. This can be justified from the spectral analysis of the many-body energy levels of a modified Hamiltonian (see Eq.~\eqref{main:eq:hamiltonian_ladder_disorder}) having additional weak onsite disorder on top of the existing detuning profile.
	
	\begin{equation}
		\hat{H}_{\text{dis}}= \sum_{j=1}^L\sum_{a=1}^2 \Omega\hat{\tilde{\sigma}}^{x}_{j,a} - (-1)^j \left(\Delta+\eta \mathcal{R}_j\right)\;\hat{\sigma}_{ja}^z
		\label{main:eq:hamiltonian_ladder_disorder}
	\end{equation}    
	
	Here, $\eta$ is the disorder strength and $\mathcal{R}_j \sim \mathcal{U}[0,1)$, i.e. for $j=1,2,...,L$, $\mathcal{R}_j$'s are random numbers which are uniformly distributed between $0$ and $1$. The Hamiltonian given by Eq.~\eqref{main:eq:hamiltonian_ladder_disorder} does not possess any lattice symmetries; nevertheless, the disorder-averaged mean consecutive level spacing ratio $\langle\!\langle r \rangle\!\rangle$ becomes close to $2\;\text{ln}2-1\sim0.386$ for $\Delta \ge 2$, which indicates Poissonian level statistics at these couplings (see Fig.~\ref{fig:spectral_statistics}). Thus, the observed Poissonian nature of the short-range spectral correlations can be attributed to the existence of an extensive number of approximate emergent conservation laws. \newline

	\subsection{Slow Dynamics and Ensemble of relaxation} \label{subsec:slow_dynamics}
	
\begin{figure}[!htpb]
		\centering
		\rotatebox{0}{\includegraphics*[width=0.48\textwidth]{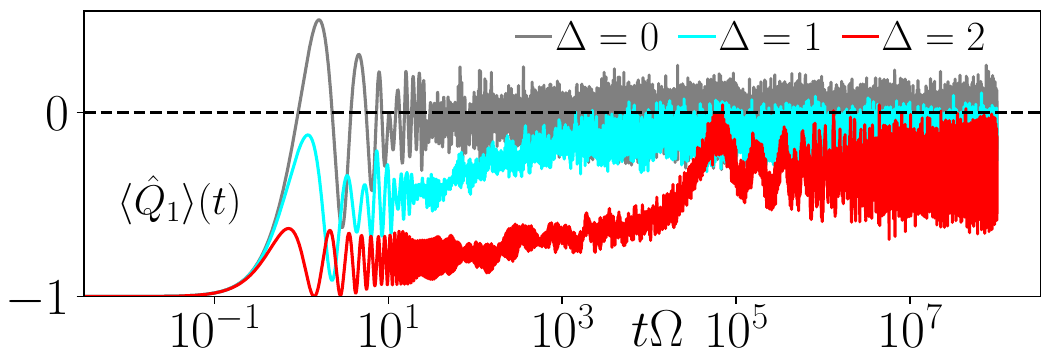}}
		\rotatebox{0}{\includegraphics*[width=0.48\textwidth]{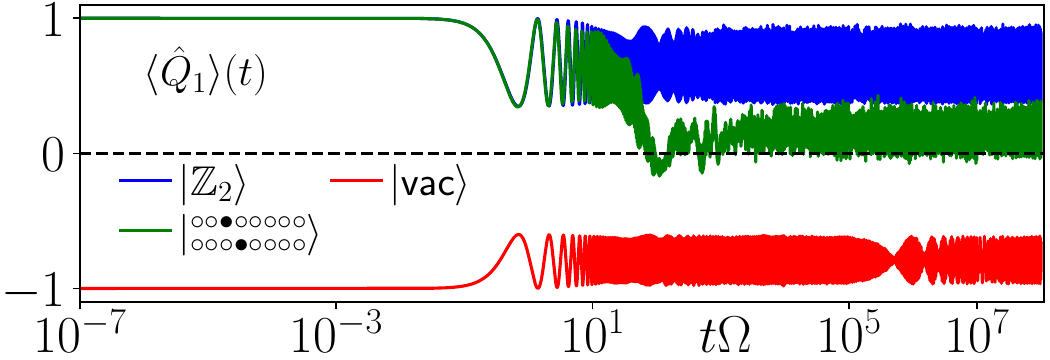}}
		\caption[]{Top panel: Evolution of quasi-conserved charge $\langle\hat{Q}_1\rangle(t)$ with time under the full Hamiltonian \texorpdfstring{\eqref{main:eq:hamiltonian_ladder}}{} at various values of $\Delta$ starting from the initial state $\ket{\substack{\bullet\circ\circ\bullet\circ\bullet\circ\circ\\\circ\circ\circ\circ\bullet\circ\circ\circ}}$ for $N=16$ sites. Bottom panel: Slow dynamics from other initial states at $\Delta=2$. Time-evolution of the states have been performed via ED method.}
		\label{fig:slow_dynamics}
	\end{figure}

	For $\Delta\sim 2$ and beyond, the strength of $\hat{H}^{(4)}_{\text{eff}}$ becomes negligible, and the evolution of the approximately conserved charges $\hat{Q}_j, \forall j=1,2,...,L$ under the full Hamiltonian becomes more constrained near their initial values. As a consequence, certain initial states (which can be labeled by the eigenvalues of $\{\hat{Q}_j\}$, say $\{q_{j}\}$), exhibit exceptionally slow dynamics, as can be seen in Fig.~\ref{fig:slow_dynamics}, where the instantaneous expectation values of these operators under the full Hamiltonian, i.e., $\{ \langle Q_j(t) \rangle \}$, remain close to their initial values for very long times. The time-evolution in these cases are performed using the ED method. For quench dynamics from such initial states, observables should relax to a generalized Gibbs ensemble (GGE) when an extensive number of exact conservation laws are present. In contrast, in the absence of any exact conservation laws (apart from the energy itself), the system should eventually relax to the Gibbs ensemble (GE).\\
	
\begin{figure}[!htpb]
		\centering
		\rotatebox{0}{\includegraphics*[width=0.48\textwidth]{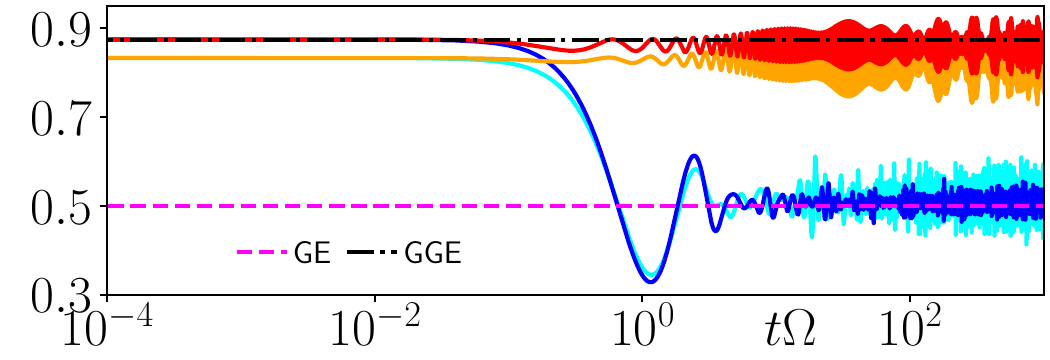}}
		\caption[]{Ensemble of Relaxation for different values of $\Delta$ and system sizes starting from the $\ket{\text{1P}} \equiv \ket{\substack{\bullet\circ\circ...\circ\\\circ\circ\circ...\circ}}$ state. The dashed lines represent values of $\langle \hat{h}_{1,1}(t) \rangle$. The solid lines represent the results of the evolution under the full Hamiltonian \eqref{main:eq:hamiltonian_ladder} for (i) $\Delta=0$ (blue solid line represents $N=16$ sites, cyan solid line represents $N=12$ atoms) and (ii) $\Delta=5$ (red solid line represents $N=16$ sites, orange solid line represents $N=12$ atoms). Time-evolution of the states have been performed via ED method.}
		\label{fig:GE_vs_GGE}
	\end{figure}

	The exact quantum dynamics of the model \eqref{main:eq:hamiltonian_ladder} illustrates that, in the regime where the emergent conservation laws are approximate and not exact, the exact quantum dynamics of observables still do not relax to a GE, and remain very well describable at late times by a GGE (until time scales set by the higher order effective Hamiltonians). This is illustrated in Fig.~\ref{fig:GE_vs_GGE}, where we show the late-time relaxation of a one-body operator ($\hat{h}_{j,a}\equiv \hat{\sigma}^z_{j,a}-\hat{\tilde{\sigma}}^x_{j,a}$) for $N=12$ and $N=16$ sites at $\Delta=0,5$ starting from a one-particle product state, i.e., $\ket{\text{1P}}=\ket{\substack{\bullet\circ\circ\circ...\circ\\\circ\circ\circ\circ...\circ}}$. For $\Delta\sim0$, only the total energy, i.e., $\langle \psi(t) | \hat{H} | \psi(t) \rangle$, is an exactly conserved quantity; hence, the system relaxes to the corresponding Gibbs ensemble with the appropriate temperature. In contrast, for $\Delta\gg1$, in this case $\Delta=5$, as there are an extensive number of approximate conservation laws, the system fails to relax to the Gibbs ensemble even at extremely long times. Instead, the system relaxes to a GGE defined by the Lagrange multipliers corresponding to all the conserved charges (exact or approximate) . These Lagrange multipliers are determined by requiring that the initial values of all the conserved charges be equal to the expectation value of the operators with respect to the GGE (see Appendix \ref{app:GGE} for details).

	\begin{figure}[!htpb]
		\centering
		\rotatebox{0}{\includegraphics*[width=0.444\textwidth]{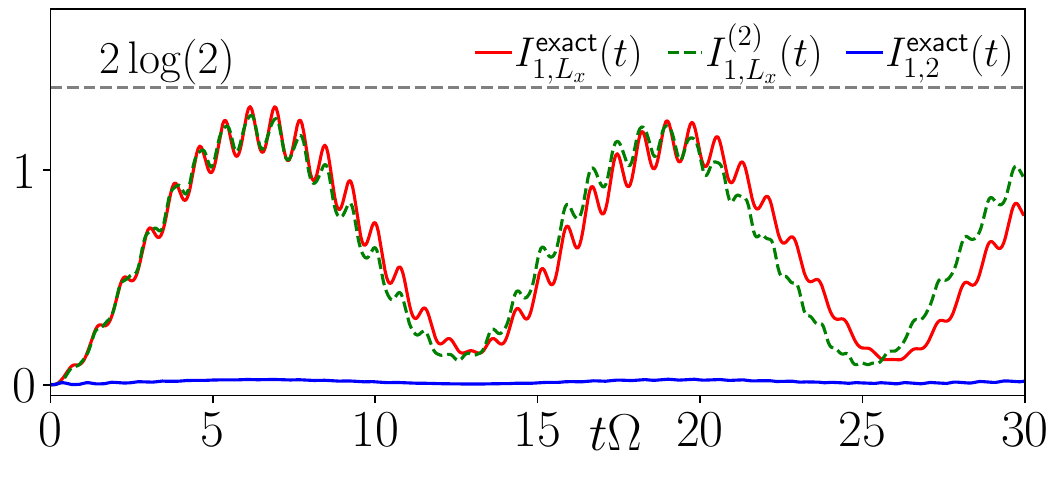}}
		\caption{Time-evolution of mutual information along horizontal (blue solid line) and vertical (red solid line) nearest neighbor bonds computed via ED with the full Hamiltonian $\hat{H}$ for $\Delta=4$. The green dashed curve is obtained from $\hat{H}_{\text{eff}}^{[2]}$.}
		\label{fig:mutual_info_dynamics}
	\end{figure}

	Moreover, in some special initial states, the growth of the correlations is extremely slow along the horizontal bonds. This is evident from Fig.~\ref{fig:mutual_info_dynamics}, where the mutual information content (see Appendix-\ref{app:mutual_info}) of the horizontal ($I_{1,2}$) and vertical bond ($I_{1,L_x}$) is shown as a function of time for the $\ket{\text{1P}}$ state (i.e. $\ket{\substack{\bullet\circ...\circ\\\circ\circ...\circ}}$) with $\Delta=4$ for $N=16$ sites. While $I_{1,2}$ remains close to zero, $I_{1,L_x}$ shows large amplitude oscillations. Such contrasting behaviors along the two directions can be traced back to the fact that for large enough $\Delta$, an almost exact effective Rabi flip-flop $\ket{\substack{\bullet\circ...\circ\\\circ\circ...\circ}} \leftrightarrow \ket{\substack{\circ\circ...\circ\\\bullet\circ...\circ}}$ takes place, and $I_{1,L_x}$ almost reaches the theoretical maximum value of $2\log(2)$ at the middle of the Rabi cycle. The second order effective Hamiltonian also describes the mutual information dynamics qualitatively. The off-diagonal terms in Eq.~\eqref{eq:Heff_SW} act as a spin-flip term which introduces a Rabi oscillation between states $\ket{\substack{\bullet\circ...\circ\\\circ\circ...\circ}}$ and $\ket{\substack{\circ\circ...\circ\\\bullet\circ...\circ}}$ on the same rung, with frequency $\Omega^2/\Delta$. Furthermore, taking into account the perturbative rotation of the initial state due to the SW transformation, one is also able to qualitatively understand the high-frequency oscillations in $I_{1,L_x}(t)$. The perturbative rotation weakly hybridizes different $\hat{Z}_\pi$ sectors and the interference from these different sectors gives rise to the high-frequency ($\sim\Delta$), small-amplitude oscillations on top of the aforementioned low-frequency ($\sim 1/\Delta$), fixed amplitude oscillations. See Appendix-\ref{app:mutual_information_dynamics} for details. \\
	
	The aforementioned features of slow dynamics, along with the existence of simple local quasi-conserved charges $\{q_j\}$, could, in principle, be utilized to store $L/2$ ``classical" bits of information on a ladder of $N=2L$ atoms (see Appendix~\ref{app:classical_bit_storage}). \newline 
	
	\subsection{Stronger violations of ergodicity} \label{subsec:stronger_erogidicity_violations}
	
	\subsubsection{ETH violation for other initial states}
	
\begin{figure}[!htpb]
		\centering
		\includegraphics[width=0.485\textwidth]{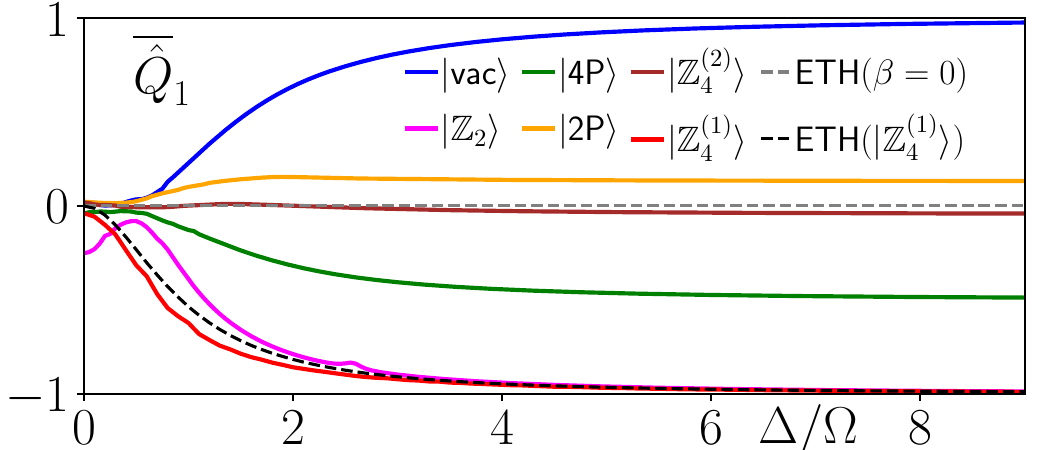}
		\caption{Variation of the infinite-time average value $\overline{\langle \hat{Q}_1 \rangle}$ and the corresponding Gibbs ensembles values predicted by ETH for the quasi-conserved charges with $\Delta$. See text for details. }
		\label{fig:dea}
	\end{figure}

	In Fig.~\ref{fig:dea}, we show the variation of the infinite-time average and the Gibbs ensemble ETH values of the quasi-conserved charges with $\Delta$. Both quantities can be estimated from the knowledge of the exact eigenstates of the Hamiltonian (see Appendix-\ref{app:sec:ETH_infty_time} for details). In what follows, we shall consider the following states: $\ket{\text{vac.}}\equiv \ket{\substack{\circ\circ\circ\circ\circ\circ\circ\circ\\\circ\circ\circ\circ\circ\circ\circ\circ}}$, $\ket{\mathbb{Z}_2}\equiv \ket{\substack{\circ\bullet\circ\bullet\circ\bullet\circ\bullet\\\bullet\circ\bullet\circ\bullet\circ\bullet\circ}}$, $\ket{\text{4P}} \equiv \ket{\substack{\bullet\circ\circ\bullet\circ\bullet\circ\circ\\\circ\circ\circ\circ\bullet\circ\circ\circ}}$, $\ket{\text{2P}}\equiv\ket{\substack{\circ\circ\bullet\circ\circ\circ\circ\circ\\\circ\circ\circ\bullet\circ\circ\circ\circ}}$, $\ket{\mathbb{Z}_4^{(2)}} \equiv \ket{\substack{\circ\circ\circ\bullet\circ\circ\circ\bullet\\\bullet\circ\circ\circ\bullet\circ\circ\circ}}$, and $\ket{\mathbb{Z}_4^{(1)}} \equiv \ket{\substack{\circ\bullet\circ\circ\circ\bullet\circ\circ\\\circ\circ\circ\bullet\circ\circ\circ\bullet}}$. 
	
	For the state $\ket{\mathbb{Z}_4^{(1)}}$, the energy expectation value is non-zero; $\langle \mathbb{Z}_4^{(1)} | \hat{H}| \mathbb{Z}_4^{(1)} \rangle = -L\Delta$ and hence, $\beta\ne0$. This, in turn, implies that the ETH value of the quasi-conserved charges corresponding to this state depends on the value of $\Delta$, shown as a black dashed line, while the infinite-time average values is shown as a solid red curve. For all other states shown, the energy expectation value is zero, which, combined with the fact that the many-body spectrum has a reflection symmetry, implies that for these states $\beta=0$ and the ETH value is independent of $\Delta$. This figure illustrates the fact that for certain initial states with zero energy density, corresponding to the infinite temperature Gibbs ensemble, the infinite-time average value drifts from the ETH value as $\Delta$ is increased from $\Delta=0$, signaling a form of ETH violation. For other states, the ETH and the infinite-time average value coincide, implying that these states do not violate the ETH. \newline

	\subsubsection{Approximate Krylov Fractures}
	
\begin{figure}[!htpb]
		\centering
		\includegraphics[width=0.485\textwidth]{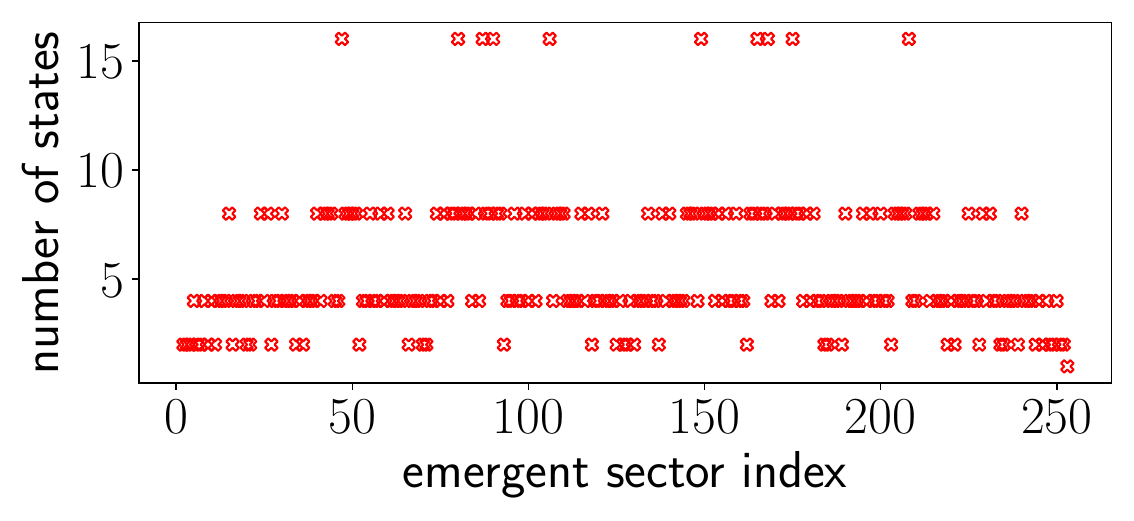}
		\caption{Emergent sector sizes within the $Z_\pi=0$ Hilbert space for $N=20$ site system. Each index (horizontal axis) corresponds to a specific charge sector labeled by $\{\hat{Q}_1,\hat{Q}_2,...\hat{Q}_L\}$.}
		\label{main:fig:emergent_sector_sizes}
	\end{figure}

	We have seen in earlier sections that for values of $\Delta \ge 2.5$, the Hilbert space of the model breaks down into several approximately disconnected sectors which are labeled by the emergent conserved charges $\{\hat{Q}_1,\hat{Q}_2,...\hat{Q}_L\}$, even within a fixed $\hat{Z}_\pi$ sector. In such a scenario, typically one has sectors of various sizes and there is a very broad distribution of the sector sizes --- there are many small sectors and very few large sectors \cite{Khemani_PhysRevB.101.174204}. This is illustrated in Fig.~\ref{main:fig:emergent_sector_sizes}, where the horizontal axis enumerates the distinct charge sectors by an index constructed from the charges $\{\hat{Q}_1,\hat{Q}_2,...\hat{Q}_L\}$ and the vertical axis shows the corresponding number of states.
	
\begin{figure*}[!htpb]
		\centering
		\rotatebox{0}{\includegraphics*[width=0.224\textwidth]{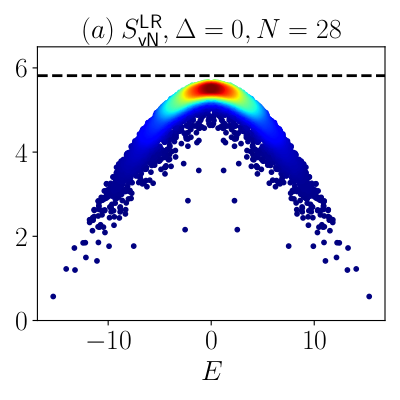}}
		\rotatebox{0}{\includegraphics*[width=0.224\textwidth]{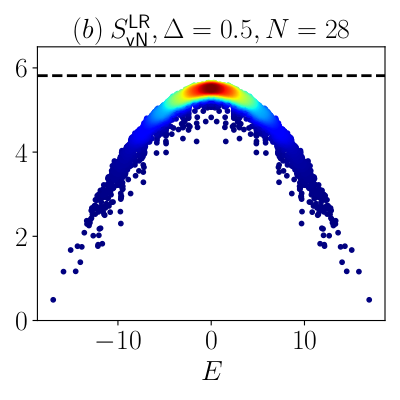}}
		\rotatebox{0}{\includegraphics*[width=0.224\textwidth]{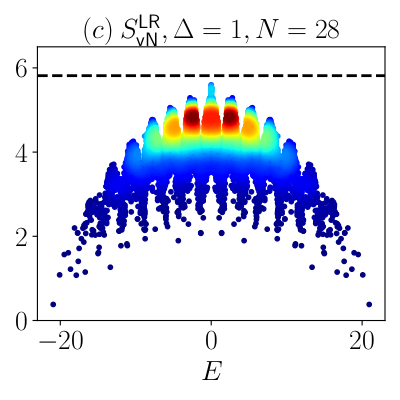}}
		\rotatebox{0}{\includegraphics*[width=0.224\textwidth]{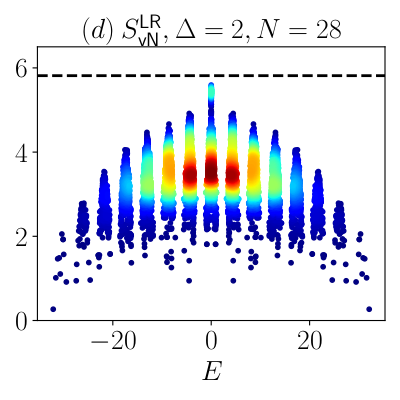}}
		\rotatebox{0}{\includegraphics*[width=0.224\textwidth]{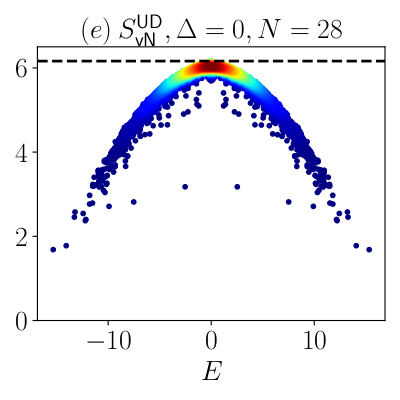}}
		\rotatebox{0}{\includegraphics*[width=0.224\textwidth]{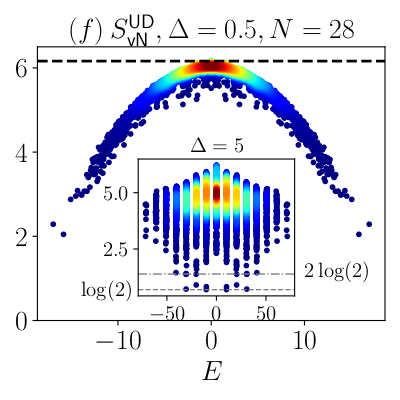}}        
		\rotatebox{0}{\includegraphics*[width=0.224\textwidth]{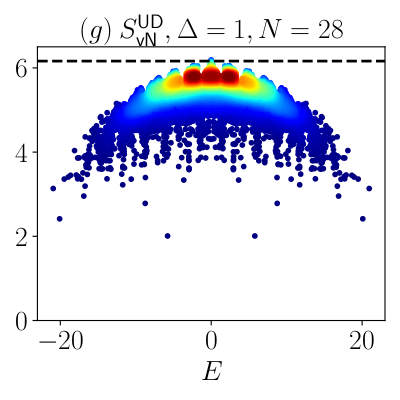}}
		\rotatebox{0}{\includegraphics*[width=0.224\textwidth]{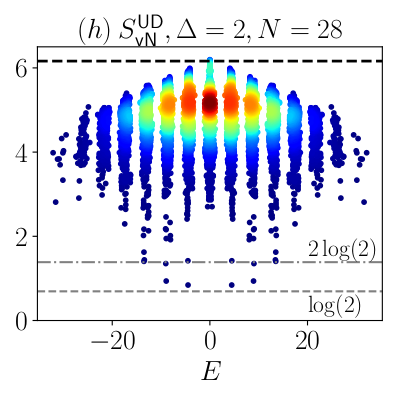}}
		\caption{Bipartite von Neumann entanglement entropy (see Appendix \ref{app:entanglement-entropy}) of all many-body eigenstates of $\hat{H}$ with respect to the LR partition (a)-(d) and UD partition (e)-(h) respectively for $N=28$ atoms in the $k_x=k_y=0$ sector for $\Delta=0,0.5,1,2$. The solid dashed black line denotes the Page value (i.e. the average entanglement entropy of Haar-random states) in this specific symmetry sector. The broad distribution of values of entanglement entropies for $\Delta=1,2$ suggest that there is a Krylov fracture of the Hilbert space, albeit approximate. In the inset of the panel (f), the entanglement entropies of analytically constructed eigenstates of $H_{\text{eff}}^{[2]}$ with one and two Bell pairs shared between the legs are shown as dashed gray lines at $\log(2)$ and $2\log(2)$}
		\label{fig:EE}
		% /mnt/ntfs1/asus-Vivobook-backup/Documents/check_EE_LR --> location of plotcodes
	\end{figure*}

	Due to this approximate disconnectedness, the many-body eigenstates of the Hamiltonian are very weakly mixed between these different $\{\hat{Q}_1,...,\hat{Q}_L\}$ sectors. As a result, we expect the entanglement entropy of the eigenstates $\ket{E}$ of the Hamiltonian Eq.~\eqref{main:eq:hamiltonian_ladder}, which are related to the size of these approximately disconnected sectors also to have a very broad distribution of values, even in the middle of the spectrum (i.e. the $\hat{Z}_\pi=0$ sector for $\Delta \gg \Omega$). In Fig.~\ref{fig:EE} we indeed observe such a broad distribution of the values of the entanglement entropies for larger values of $\Delta$ within a specific exact symmetry sector. The different mutually separated vertical stripes for larger values of $\Delta$ in Fig.~\ref{fig:EE} correspond to different $\hat{Z}_\pi$ sectors.
	
	In Appendix-\ref{app:Heff_SW}, we have analytically constructed all eigenstates of the second-order effective Hamiltonian $\hat{H}^{[2]}_{\text{eff}}$. These eigenstates have simple ``Bell-pair"-like states embedded in some of the rungs, while the other rungs remain completely unentangled when the entanglement entropy is measured across the vertical cut (henceforth denoted as UD cut), designated as dashed gray line in Fig.~\ref{fig:model}. Such a structure can help us understand, to some degree, the nature of the bi-partite von Neumann entanglement entropy of the eigenstates of the full Hamiltonian at $\Delta=2$. We note here that this only explains the entanglement from the UD cut. The entanglement entropies across the left-right bi-partition (henceforth denoted as the LR cut) are exactly zero for all the analytically constructed eigenstates (labeled by a collection of conserved charges $\{z_\pi,q_1,q_2,...,q_L\}$) of $\hat{H}^{[2]}_{\text{eff}}$. However, as these eigenstates are highly degenerate, we suspect that a full numerical diagonalization scheme that is unaware of such approximate symmetries outputs arbitrary hybridizations of these analytically constructed eigenstates, leading to non-zero values of entanglement entropies for the eigenstates across the LR cut. The finite bandwidth of the tower of states in Fig.~\ref{fig:EE}(d),(h) is due to the influence of higher order processes and can be captured quantitatively via $\hat{H}^{[4]}_{\text{eff}}$ (see Fig.~\ref{fig:compare_exact_vs_SW2nd_SW4th_spectrum}(b)). That being said, we do not have any understanding of the eigenstates of $\hat{H}^{[4]}_{\text{eff}}$ from an analytical point of view, as the construction and diagonalization of $\hat{H}^{[4]}_{\text{eff}}$ were performed using numerical schemes. \newline
	
	We end this section by noting that although the system governed by the Hamiltonian Eq.~\eqref{main:eq:hamiltonian_ladder} was shown to have anomalous quantum dynamics, in the thermodynamic limit, the system will ultimately relax to the GE, with the effective inverse temperature $\beta$ set by the initial energy density. This can be inferred from the scaling behavior of the regularized norm of the adiabatic gauge potential, which implies that the system becomes chaotic at very late times \cite{Pandey_PhysRevX.10.041017}. See Appendix-\ref{app:AGP_norm_scaling} for details.
	
	\section{Floquet Engineering\label{main-sec:floquet-engineering}}
	
	In Ref.~\cite{Maskara_PhysRevLett.127.090602}, it was shown that by leveraging the underlying chirality operator associated with the PXP model (on any bipartite lattice), it is possible to design Floquet protocols, such as periodic kicks in the form of a many-body $\pi$-pulse, leading to dynamical signatures reminiscent of DTC order which are stabilized by QMBS. This naturally raises the question of whether it is possible to design interesting Floquet protocols with the help of the two chirality operators $\hat{C}_{1,2}$ introduced earlier in Sec.~\ref{sec:model} (see Eqs.~\eqref{eq:C1-defn},\eqref{eq:C2-defn}). As expected, the protocol proposed in \cite{Maskara_PhysRevLett.127.090602} (see Eq.~\eqref{eq:protocol-0}), henceforth referred to as protocol-0, gives rise to a subharmonic response (in the form of period-2 exact revivals) in the model \eqref{main:eq:hamiltonian_ladder} for $\Delta=0$, starting from \textit{any} initial state
	
	\begin{equation}
		\hat{U}_F(\tau) = \hat{C} e^{-i\tau\hat{H}_{\Delta=0}}
		\label{eq:protocol-0}
	\end{equation}
	
	This happens as $\{\hat{H}_{\Delta=0},\hat{C}\}=0$ and $\hat{C}^2=\hat{\mathbbm{1}}$ which implies that $\hat{U}_F^2(\tau) = \hat{\mathbbm{1}}$. However, protocol-0 does not give rise to exact revivals for $\Delta\ne0$, since $\{\hat{H}_{\Delta\ne0},\hat{C}\} \ne 0$. In what follows, we shall show that by leveraging the appropriate chirality operators valid for $\Delta \ne 0$, namely $\hat{C}_{1,2}$ (Eqs.~\eqref{eq:C1-defn},\eqref{eq:C2-defn}), we can generate two new classes of Floquet protocols, allowing us to realize exact revivals for any $\Delta \ne 0$.
	
	\subsection{Protocol-I: Subharmonic response \label{sec:DTC}}
	
	We now consider a modified version of protocol-0, which is obtained by substituting $\hat{C}_1$ for $\hat{C}$ in Eq.~\eqref{eq:protocol-0}. This gives rise to a new protocol (henceforth referred to as protocol-I) defined by the following unitary evolution operator
	
	\begin{equation}
		\hat{U}^{\text{I}}_F(\tau) = \hat{C}_1\;e^{-i\hat{H}\tau}
		\label{eq:protocol-1}
	\end{equation}

\begin{figure}[!htpb]
		\centering
		\rotatebox{0}{\includegraphics[width=0.495\textwidth]{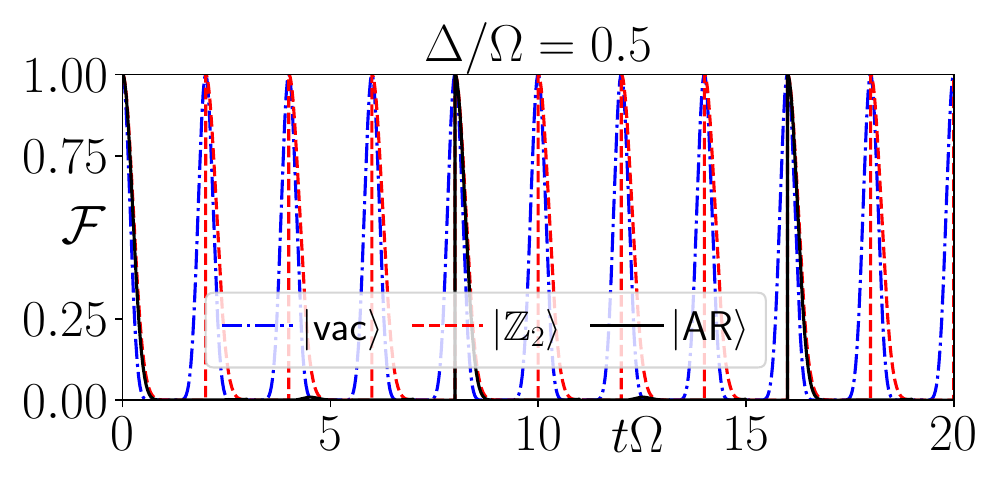}}
		\caption[]{Exact revivals under protocol-I (Eq.~\eqref{eq:protocol-1}). Return probability $\mathcal{F}(t) = |\langle \psi(0)|\psi(t) \rangle|^2$ is plotted as a function of time starting from different initial states: (i) $\ket{\text{vac.}}$ (blue dashed), (ii) $\ket{\mathbb{Z}_2}$ (red dashed) and (iii) a generic Fock state $\ket{\text{AR}}\equiv\ket{\substack{\bullet\circ\circ\bullet\circ\bullet\circ\circ\\\circ\circ\circ\circ\bullet\circ\circ\circ}}$ (black solid) at $\Delta=0.5$ for $N=16$ sites.}
		\label{fig:protocol-I-DTC}
	\end{figure}

	In the context of Rydberg atom quantum simulators, such a protocol can be thought of as unitary evolution via the Hamiltonian $\hat
	{H}$ for time $\tau$, followed by action of the chirality operator $\hat{C}_1=\hat{T}_x\hat{C}$, which itself is composed of a many-body $\pi$-pulse and spatial translation of the atoms by one site along the longer direction, respecting periodic boundaries. Repeated action of this unitary on any initial state of the system, defines for us a Floquet protocol, with $\hat{U}^{\text{I}}_F(\tau)$ as the Floquet unitary evolution operator with the time-period $\tau$. Since $\{\hat{C}_{1,2},\hat{H}\}=0$, one can see that $\hat{U}_F^{\text{I}}(2m\tau) = (\hat{U}_F^{\text{I}}(\tau))^{2m} = \hat{T}_x^{2m}$ ($m\ge1$). This relation implies that if an initial state $\ket{\psi(0)}$, satisfies $\hat{T}_x^{2m}\ket{\psi(0)}=\ket{\psi(0)}$, then it revives exactly after $2m$ cycles under the Floquet protocol \eqref{eq:protocol-1}. In other words, \textit{every} Fock state returns to itself exactly after a specific number of cycles, which is set by the lattice-translation properties of that state with respect to $\hat{T}_x$. As a result of this delayed exact revivals, the interacting quantum many-body system described above, exhibits a state-dependent subharmonic response which is suggestive of DTC order \cite{Zalatel_RevModPhys.95.031001,khemani2019briefhistorytimecrystals}. This state-dependent exact revivals are illustrated in Fig.~\ref{fig:protocol-I-DTC} by studying (numerically) the micromotion associated with protocol-I. \newline

\begin{figure}[!htpb]
		\centering
		\includegraphics[width=0.42\textwidth]{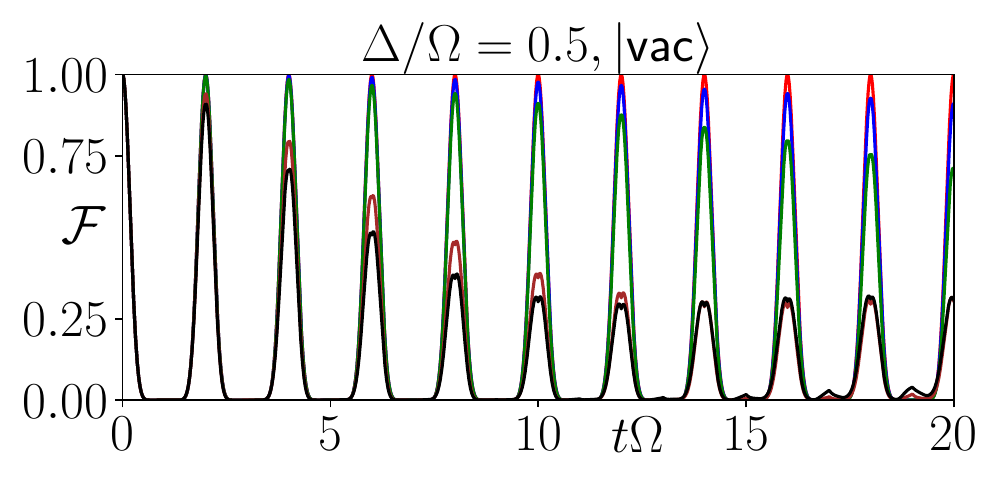}
		\includegraphics[width=0.42\textwidth]{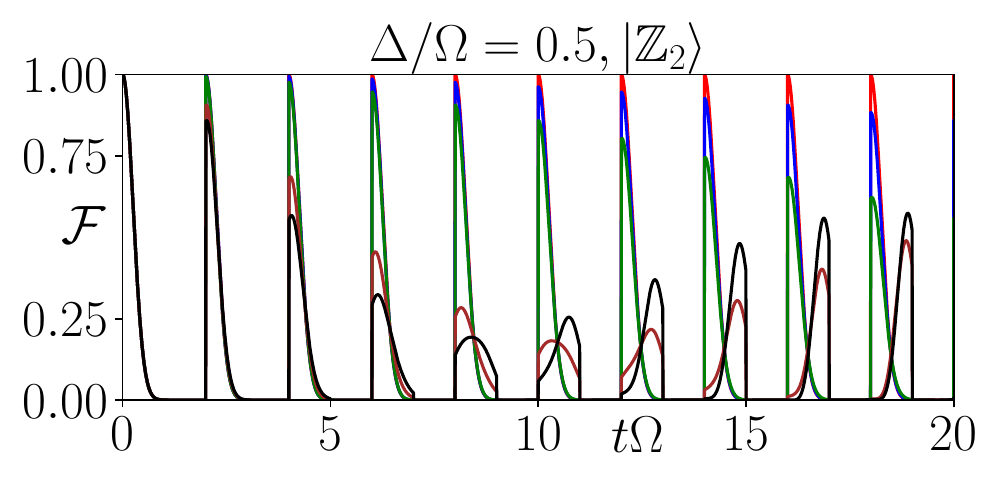}
		\includegraphics[width=0.42\textwidth]{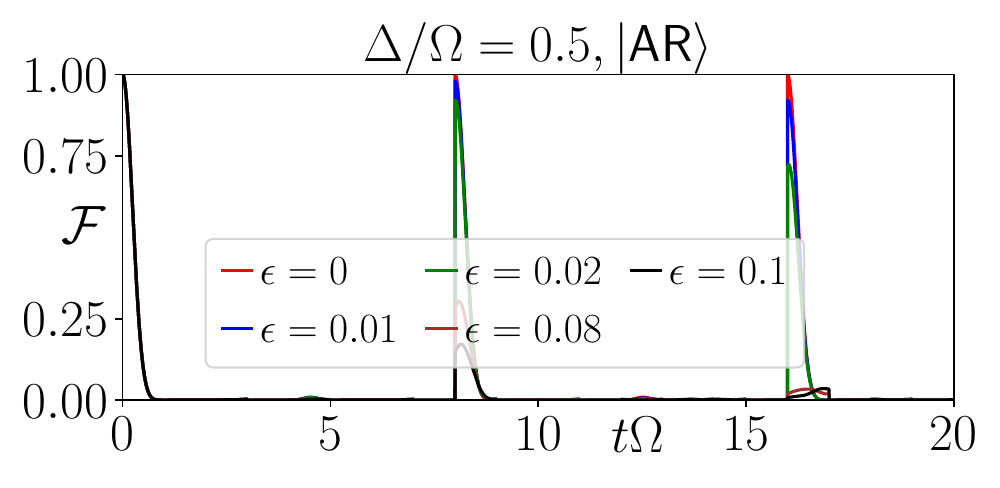}
		\caption[]{This figure represents the degree of robustness of exact revivals under protocol-I due to imperfections in the many-body $\pi$-pulse. Return probability $\mathcal{F}(t) = |\langle \psi(0)|\psi(t) \rangle|^2$ is plotted as a function of time for $\Delta=0.5,\tau=2$ with $N=16$ sites starting from the states $\ket{\text{vac.}}$ (top panel) and $\ket{\mathbb{Z}_2}$ (middle panel) and $\ket{\text{AR}}\equiv\ket{\substack{\bullet\circ\circ\bullet\circ\bullet\circ\circ\\\circ\circ\circ\circ\bullet\circ\circ\circ}}$ (bottom panel) at different values of imperfection $\epsilon$ in the $\pi$-pulse, denoted as, $\epsilon=0$ (red), $\epsilon=0.01$ (blue), $\epsilon=0.02$ (green), $\epsilon=0.08$ (brown) and $\epsilon=0.1$ (black) respectively. As the figure illustrates, the $\ket{\text{vac.}}$ state is more stable against imperfections in the application of the exact many-body $\pi$-pulse.}
		\label{fig:protocol-I-pipulse-stability}
	\end{figure}

	In our numerical simulations we implement the action of the chirality operator $\hat{C}_1$ and the lattice-translation operator $\hat{T_x}$ as an \textit{instantaneous} operation on the many-body quantum state $\ket{\psi(t)}$ at integer multiples of time $t=\tau$. This choice leads to an artificial sharp jump in the return probability $\mathcal{F}(t)$ in Fig.~\ref{fig:protocol-I-DTC} and also Fig.~\ref{fig:protocol-I-pipulse-stability} later on. However, in reality the action of these individual operators would be themselves implemented as finite time unitary processes, implying these artificial jumps would be replaced by a smooth rise in $\mathcal{F}(t)$. Although not required for our protocol, we note that there have been some recent attempts at making such translation operations fast by employing quantum optimal control strategies \cite{PhysRevResearch.6.033282,7r3w-8m61}. \\
	
	As already mentioned earlier, implementing the protocol-I described above in actual Rydberg atom simulator platforms, requires the technological ability to apply a many-body $\pi$-pulse and to translate the atoms to a new location without acquiring additional kinematical phase. Although this is extremely challenging, in light of recently demonstrated technological progress in \cite{Bluvstein2022,Xu2024}, such operations may be feasible in the near future.\newline
	
	We now focus on how the imperfections in the action of the many-body $\pi$-pulse can influence the subharmonic response shown in Fig.~\ref{fig:protocol-I-DTC}. We consider the influence of an imperfect many-body $\pi$-pulse, where the phase rotation can deviate from $\pi$ to $\pi(1-\epsilon)$. In this case, the single cycle unitary evolution operator for protocol-I reads 
	
	\begin{equation}
		\hat{U}_F^{\text{I}}(\tau;\epsilon) = \hat{T}_x e^{-i\pi\left(1-\epsilon\right)\hat{\mathcal{N}}} e^{-i\tau\hat{H}}
		\label{eq:protocol-I-imperfect-pi-pulse}
	\end{equation}
	
	where $\hat{\mathcal{N}}=\sum_{j,a} \hat{n}_{j,a}$, $\hat{n}_{j,a}=\left(\hat{\sigma}^z_{j,a}+1\right)/2$. In Fig.~\ref{fig:protocol-I-pipulse-stability}, we show the stability of the exact revivals starting from the initial state $\ket{\text{vac.}}$ (top panel), $\ket{\mathbb{Z}_2}$ (middle panel) and $\ket{\text{AR}}\equiv \ket{\substack{\bullet\circ\circ\bullet\circ\bullet\circ\circ\\\circ\circ\circ\circ\bullet\circ\circ\circ}}$ (bottom panel) for $\Delta=0.5,\tau=2$ for $N=16$ sites at different values of $\epsilon$, $\epsilon=0$ (red) , $\epsilon=0.01$ (blue) , $\epsilon=0.02$ (green) , $\epsilon=0.08$ (brown) and $\epsilon=0.1$ (black). \\

	Furthermore, as the chirality operator $\hat{C}_2$ also anti-commutes with the Hamiltonian $\hat{H}$ for any value of $\Delta$, one can substitute $\hat{C}_1$ in Eq.~\eqref{eq:protocol-1}, by $\hat{C}_2$ to generate an equivalent protocol with similar features: i.e., $\hat{U}_F(2m\tau) = \hat{T}_x^{2m}\hat{T}_y^{2m}$ ($m\ge1$). In this case, the revival period would depend on the lattice-translation properties of the initial state along both directions.\\

	\subsection{Protocol-II: Exact Floquet flat bands \label{sec:floquet-flat-bands}}
	
	We now introduce a second class of Floquet protocols (henceforth called protocol-II), which gives rise to exact revivals after every cycle, starting from any initial state for any value of $\Delta\ne0$ and hence constitutes a different class of dynamical phenomena compared to protocol-I. The one-cycle Floquet unitary for protocol-II reads
	
	\begin{equation}
		\hat{U}_F^{\text{II}}(\tau) = \hat{C} e^{-i\hat{H}_{-\Delta_0}\tau/2}\hat{C} e^{-i\hat{H}_{+\Delta_0}\tau/2}
		\label{eq:protocol-II}
	\end{equation}
	
	Where $\hat{H}_{+\Delta_0}$, $\hat{H}_{-\Delta_0}$ are Hamiltonians with staggered detuning profiles $\Delta_{j,a}=(-1)^j \Delta_0$ and $\Delta_{j,a}=-(-1)^j \Delta_0$ respectively. Owing to the identity $\hat{C} e^{-i\hat{H}_{-\Delta_0}\tau/2} \hat{C} = e^{+i\hat{H}_{+\Delta_0}\tau/2}$, we have $\hat{U}_F^{\text{II}}(\tau)=\hat{\mathbbm{1}}$. The identity can be verified by expanding the LHS and using the properties $\hat{C}^{-1} \hat{H}_z \hat{C} = \hat{H}_z $ and $\hat{C}^{-1} \hat{H}_x \hat{C} = -\hat{H}_x$. 
	
	This equivalence for the $m^{\text{th}}$ power is shown below 
	
	\begin{widetext}
		\begin{equation}
			\begin{split}
				& \hat{C} \; \frac{\left(-i\tau\right/2)^m}{m!} \overbrace{\left( \hat{H}_z + \hat{H}_x \right) ... \left( \hat{H}_z + \hat{H}_x \right)}^{m\:\text{terms}} \; \hat{C} = \frac{\left(-i\tau\right)^m}{2^m\;m!}\; \overbrace{\hat{C}^{-1} \left( \hat{H}_z + \hat{H}_x \right) \hat{C} \;...\; \hat{C}^{-1}\left( \hat{H}_z + \hat{H}_x \right) \hat{C} \;...\; \hat{C}^{-1} \left( \hat{H}_z + \hat{H}_x \right)\; \hat{C} }^{m\:\text{terms}}\\
				& = \frac{\left(-i\tau\right)^m}{2^m\;m!}\; \overbrace{\left( \hat{H}_z - \hat{H}_x \right) \;...\; \left( \hat{H}_z - \hat{H}_x \right) \;...\; \left( \hat{H}_z - \hat{H}_x \right)}^{m\:\text{terms}} = \frac{\left(+i\tau\right)^m}{2^m\;m!}\;\hat{H}_{-\Delta_0}
				%& = \frac{\left(-i\tau\right)^m}{2^m\;m!}\;(-1)^m \overbrace{\left( -\hat{H}_z + \hat{H}_x \right) \;...\; \left( -\hat{H}_z + \hat{H}_x \right) \;...\; \left( -\hat{H}_z + \hat{H}_x \right)}^{m\:\text{terms}} = \frac{\left(+i\tau\right)^m}{2^m\;m!}\;\hat{H}_{-\Delta_0}
			\end{split}
			\label{eq:protocol-C1-switch-identity-proof}
		\end{equation}
	\end{widetext}   
	
	As we can see from Eq.~\eqref{eq:protocol-C1-switch-identity-proof}, the term corresponding to the $m$-th power of $\tau$ of $\hat{C} e^{-i\hat{H}_{-\Delta_0}\tau/2} \hat{C}$ is equal to the $m$-th power of $e^{+i\hat{H}_{+\Delta_0}\tau/2}$, which implies that the identity $\hat{C} e^{-i\hat{H}_{-\Delta_0}\tau/2} \hat{C} = e^{+i\hat{H}_{+\Delta_0}\tau/2}$ holds. This in turn completes the proof that for drive protocol \eqref{eq:protocol-II}, $\hat{U}_F^{\text{II}}(\tau)=\hat{\mathbbm{1}}$.\newline
	
\begin{figure}[!htpb]
		\centering
		\rotatebox{0}{\includegraphics[width=0.495\textwidth]{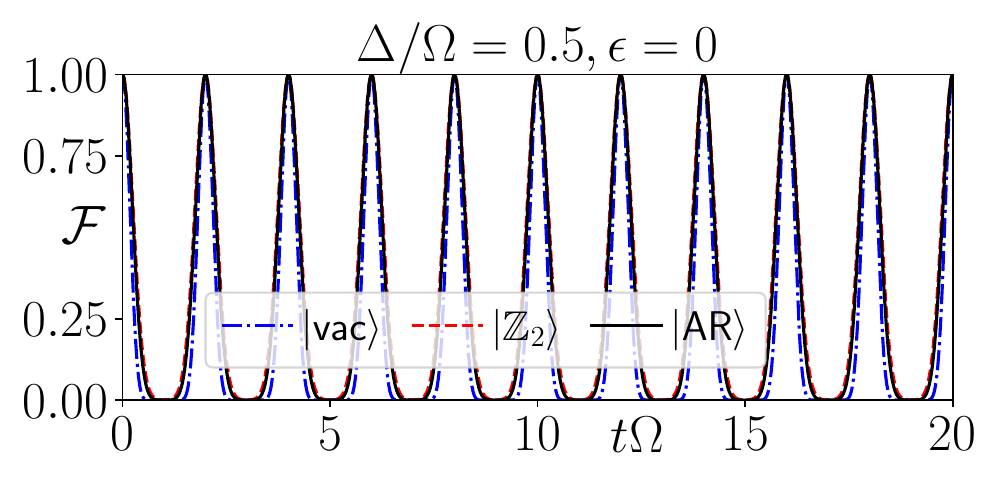}}
		\caption[]{Illustration of realization of exact Floquet flat band via protocol-II starting from different initial states. Return probability $\mathcal{F}(t) = |\langle \psi(0)|\psi(t) \rangle|^2$ is plotted as a function of time starting (i) $\ket{\text{vac.}}$ ( blue dashed line), (ii) $\ket{\mathbb{Z}_2}$ (red dashed line) and (iii) a generic Fock state $\ket{\text{AR}}$=$\ket{\substack{\bullet\circ\circ\bullet\circ\bullet\circ\circ\\\circ\circ\circ\circ\bullet\circ\circ\circ}}$ (black dashed line) at $\Delta_0=0.5$ for $N=16$ sites. As evident from the figure, all states revive after exactly one cycle of the protocol described in Eq.~\eqref{eq:protocol-II}.}
		\label{fig:protocol-II-Floquet-flat-bands}
	\end{figure}

	Since $\hat{U}_F^{\text{II}}(\tau)=\hat{\mathbbm{1}}$, under protocol-II, every initial state exhibits exact revivals with a period equal to the time period of the protocol ($\tau$) and the system hosts exact Floquet flat bands at any value of $\Delta_0\ne0$. This is illustrated in Fig.~\ref{fig:protocol-II-Floquet-flat-bands} by studying (numerically) the micro-motion associated with the protocol-II for a system with $N=20$ atoms, $\Delta_0=0.5,\tau=2$ for three different initial states (same as those in Fig.~\ref{fig:protocol-I-DTC}).\newline
	
\begin{figure}[!htpb]
		\centering
		\includegraphics[width=0.5\textwidth]{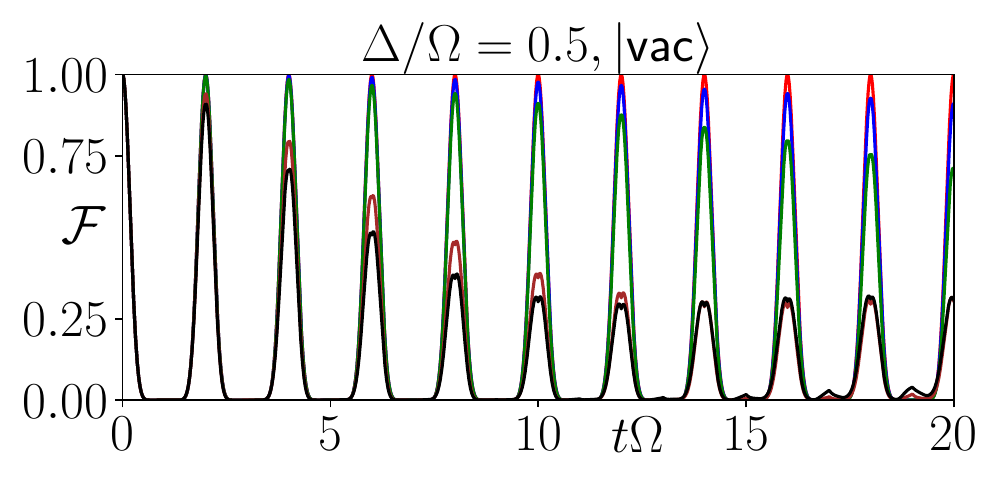}
		\includegraphics[width=0.5\textwidth]{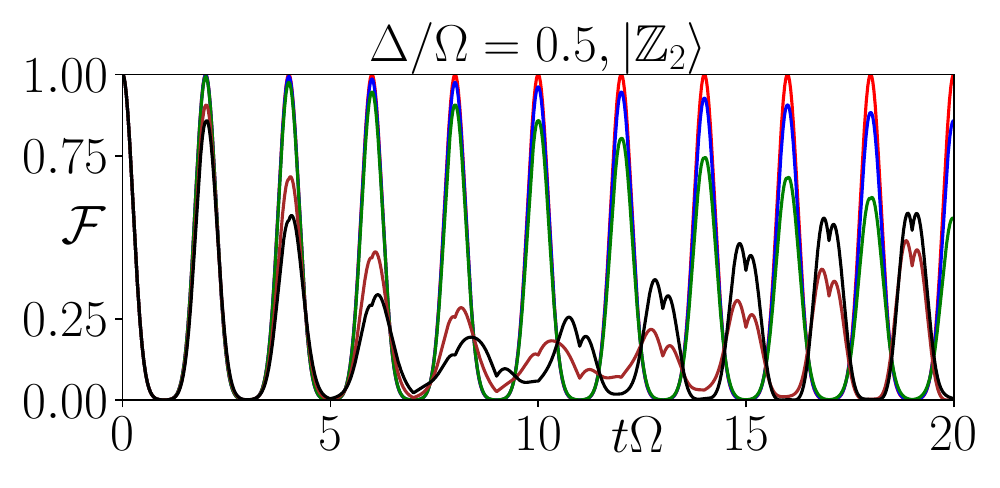}
		\includegraphics[width=0.5\textwidth]{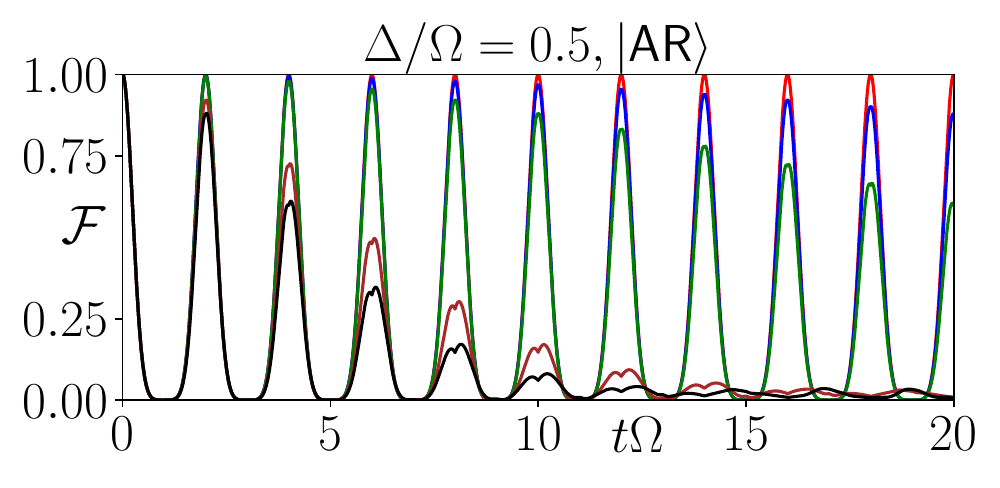}
		\caption[]{Robustness of exact revivals under protocol-II due to imperfections in the many-body $\pi$-pulse: the time-evolution of return probability $\mathcal{F}(t) = |\langle \psi(0)|\psi(t) \rangle|^2$ is shown at $\Delta_0=0.5\Omega,\tau=2$ with $N=16$ atoms for initial the states $\ket{\text{vac.}}$ (top panel), $\ket{\mathbb{Z}_2}$ (middle panel) and $\ket{\text{AR}}\equiv\ket{\substack{\bullet\circ\circ\bullet\circ\bullet\circ\circ\\\circ\circ\circ\circ\bullet\circ\circ\circ}}$ (bottom panel) at different values of imperfection $\epsilon$ in the $\pi$-pulse, denoted as, $\epsilon=0$ (red), $\epsilon=0.01$ (blue), $\epsilon=0.02$ (green), $\epsilon=0.08$ (brown) and $\epsilon=0.1$ (black) respectively.}
		\label{fig:protocol-II-pipulse-stability}
	\end{figure}

	Similar to the case of protocol-I, we will now return to the question of how the imperfections in the action of the many-body $\pi$-pulse influences the appearance of exact Floquet flat band as depicted in Fig.~\ref{fig:protocol-II-Floquet-flat-bands}. As mentioned in the above section, we consider an imperfect many-body $\pi$-pulse, where the deviation in the phase rotation from $\pi$ is symbolically represented by an amount $\epsilon$. In this case, the one-cycle unitary takes the following form 
	
	\begin{equation}
		\hat{U}_F^{\text{II}}(\tau;\epsilon) = e^{-i\pi\left(1-\epsilon\right)\hat{\mathcal{N}}} e^{-i\hat{H}_{-\Delta_0}\tau/2}e^{-i\pi\left(1-\epsilon\right) \hat{\mathcal{N}}} e^{-i\hat{H}_{\Delta_0}\tau/2}
		\label{eq:protocol-II-imperfect-pi-pulse}
	\end{equation}
	where $\hat{\mathcal{N}}=\sum_{j,a} \hat{n}_{j,a}$, $\hat{n}_{j,a}=\left(\hat{\sigma}^z_{j,a}+1\right)/2$. In Fig.~\ref{fig:protocol-II-pipulse-stability}, we show the nature of the revivals starting from the initial state $\ket{\text{vac.}}$ (top panel), $\ket{\mathbb{Z}_2}$ (middle panel) and $\ket{\text{AR}}\equiv \ket{\substack{\bullet\circ\circ\bullet\circ\bullet\circ\circ\\\circ\circ\circ\circ\bullet\circ\circ\circ}}$ (bottom panel) for $\Delta_0=0.5,\tau=2$ for $N=16$ atoms at different values of $\epsilon$, $\epsilon=0$ (red), $\epsilon=0.01$ (blue), $\epsilon=0.02$ (green), $\epsilon=0.08$ (brown) and $\epsilon=0.1$ (black).

	Figs.~\ref{fig:protocol-I-pipulse-stability}, \ref{fig:protocol-II-pipulse-stability} indicate that the state with all atoms in the electronic ground-state (see top panels of Figs.~\ref{fig:protocol-I-pipulse-stability}, \ref{fig:protocol-II-pipulse-stability}) is more stable against imperfections in the application of the many-body $\pi$-pulse for both protocol-I and protocol-II. We note however, that this is not a generic feature and it can happen that in some regimes of the parameter space the $\ket{\mathbb{Z}_2}$ state is indeed more stable (see Appendix-\ref{app:protocol-II-additional-results}).  \\
	
	We note here that while the realization of exact Floquet flat-bands in non-interacting systems are well studied \cite{Leykam01012018,Rhim01012021,Vicencio_Poblete01012021,IvanDutta_2024,Banerjee_PhysRevB.110.085431}, protocols I and II provide a new route towards their realization in interacting many-body quantum systems which are much less explored \cite{Banerjee_10.1088/1367-2630/adfd07,ghosh2025heatingsuppressiontworaterandom}.
	
	\section{Effects of environment\label{open-quantum-system}}
	
	Rydberg atom quantum simulators which host Hamiltonians akin to \eqref{main:eq:hamiltonian_ladder}, have inevitable interactions with the environmental degrees of freedom \cite{Weimer2010,Daley_2014,Tamura_PhysRevA.101.043421}. The Rydberg excited states have a finite lifetime and may undergo spontaneous emission, projecting the system onto the ground-state of that atom stochastically during the real-time evolution stage \cite{Bernien2017}. For a single atom in such a platform, incoherent processes may also appear as a result of fluctuations in the laser intensities, leading to fluctuations in Rabi frequency and detuning parameters from one shot to another in a real experiment. These environmental effects at a single atom level can accumulate in an interacting many-body setup and give rise to evolution of the system which cannot be effectively captured by a pure Hamiltonian evolution. It is thus important to investigate the robustness of QMBS and existence of quasi-conserved charges in presence of such environmental loss channels. Effects of imperfections in detuning/Rabi frequency profile which may arise, can be taken into account by a modified disordered version of the Hamiltonian \eqref{main:eq:hamiltonian_ladder}. However, in this section we only focus on imperfections which lead to non-unitary time evolution.\newline
	
	In this section, we focus on the effects of the inevitable coupling to an external environment on the nature of the anomalous non-equilibrium dynamics discussed in Sec.~\ref{main-sec:quench_dynamics}. We study two such environmental loss channels: (i) pure-dephasing and (ii) spontaneous decay of the Rydberg excited state by evolving the Gorini–Kossakowski–Sudarshan–Lindblad (GKSL) master equation \cite{Lindblad1976,GKS_10.1063/1.522979} with appropriate Lindblad jump operators (denoted by $\hat{J}_{\{\alpha\}}$), which describes both coherent and dissipative parts of the evolution within the paradigm of Born-Markov approximation \cite{Daley_2014}. The GKSL master equation reads
	
	\begin{equation}
		\frac{d\hat{\rho}(t)}{dt} =-i[\hat{H},\hat{\rho}] + \sum_{\alpha} \left( \hat{J}_{\alpha}\hat{\rho}\hat{J}_{\alpha}^{\dagger} -\frac{1}{2}\left\{\hat{J}_{\alpha}^{\dagger}\hat{J}_{\alpha},\hat{\rho}\right\} \right)
		\label{eq:Lindblad-master-equation}
	\end{equation}    
	
	\begin{figure*}[!htpb]
		\centering
		\rotatebox{0}{\includegraphics*[width=0.329\textwidth]{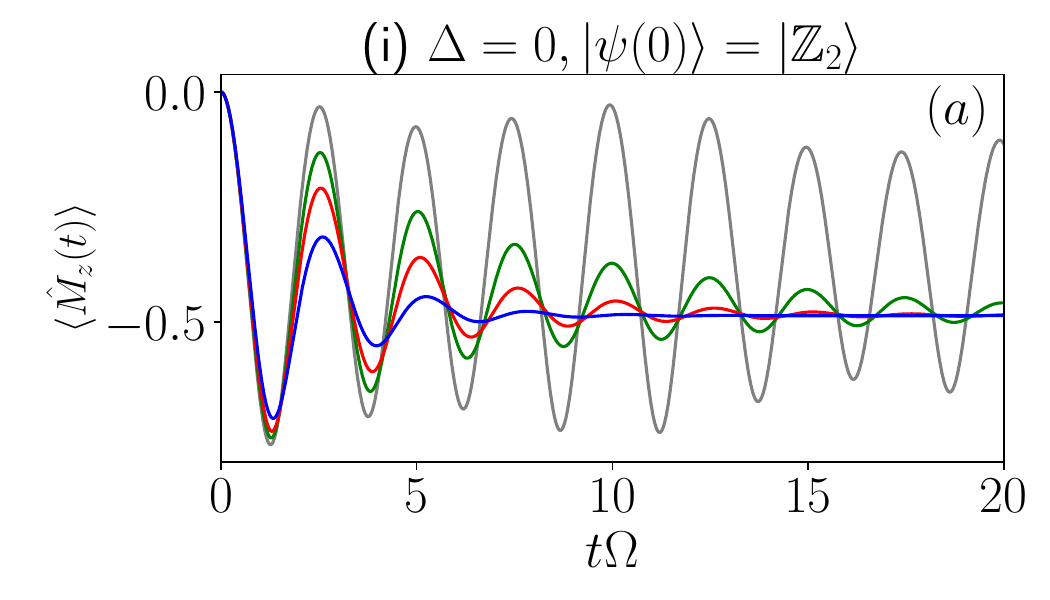}} 
		\rotatebox{0}{\includegraphics*[width=0.329\textwidth]{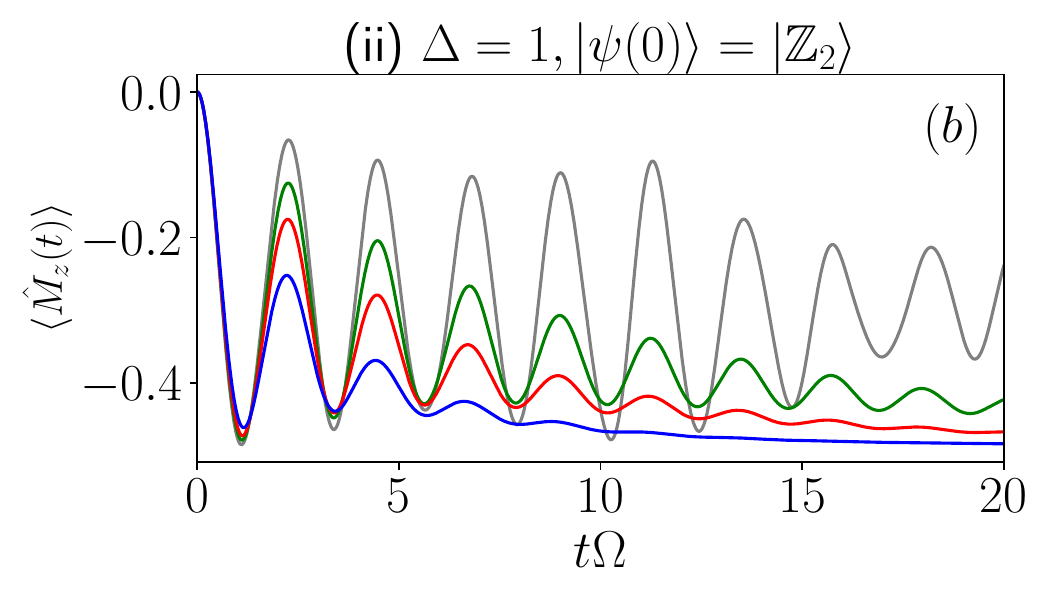}}
		\rotatebox{0}{\includegraphics*[width=0.329\textwidth]{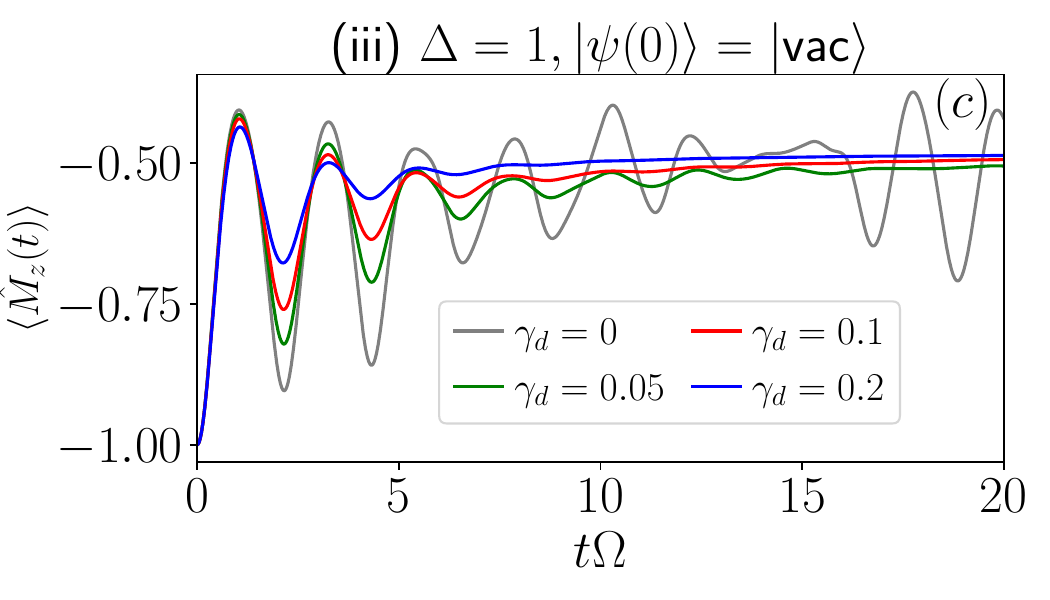}}
		\rotatebox{0}{\includegraphics*[width=0.329\textwidth]{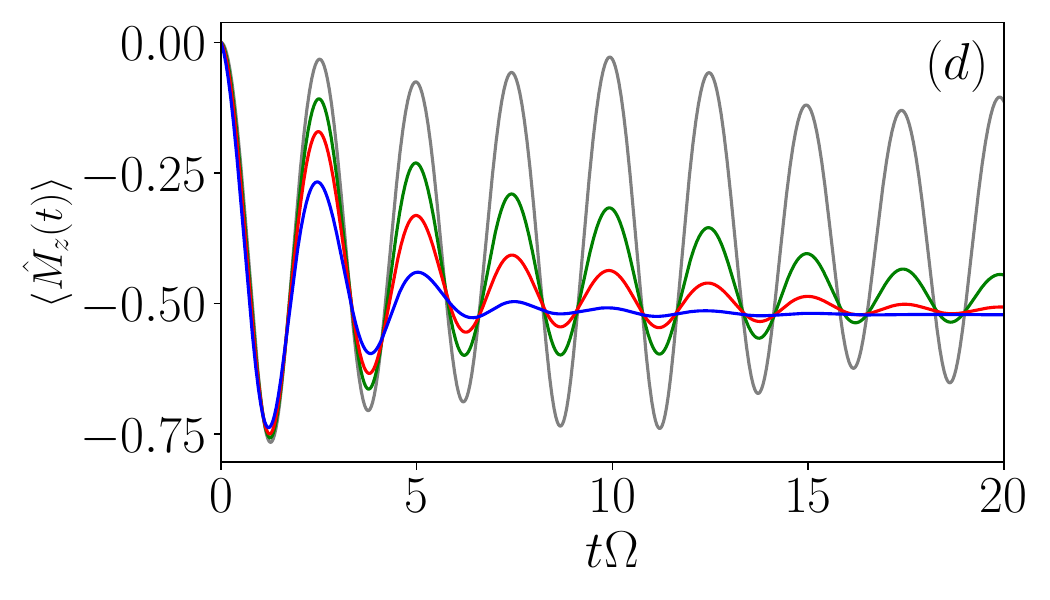}}
		\rotatebox{0}{\includegraphics*[width=0.329\textwidth]{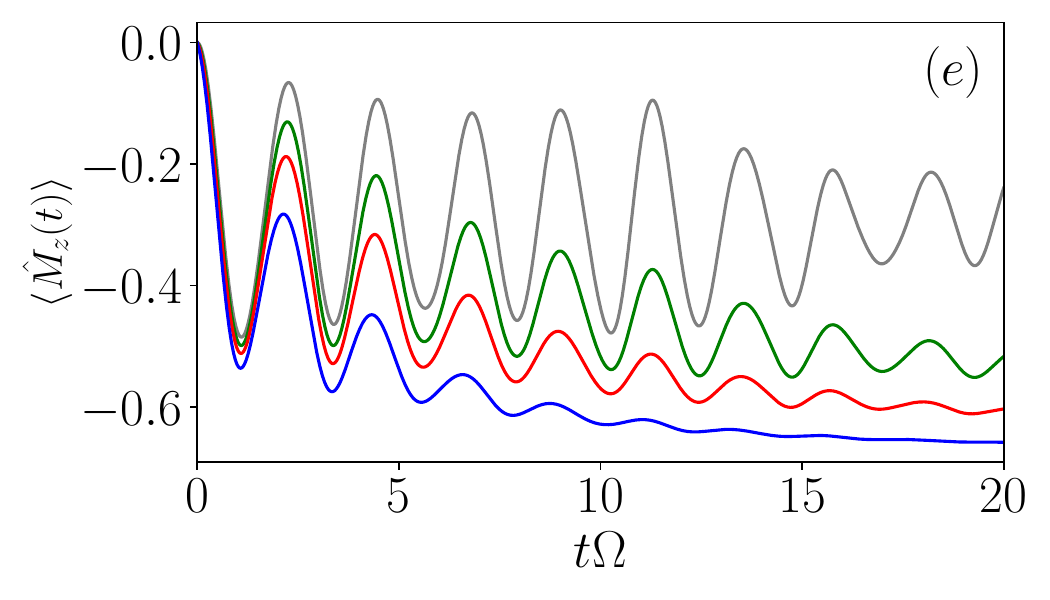}}
		\rotatebox{0}{\includegraphics*[width=0.329\textwidth]{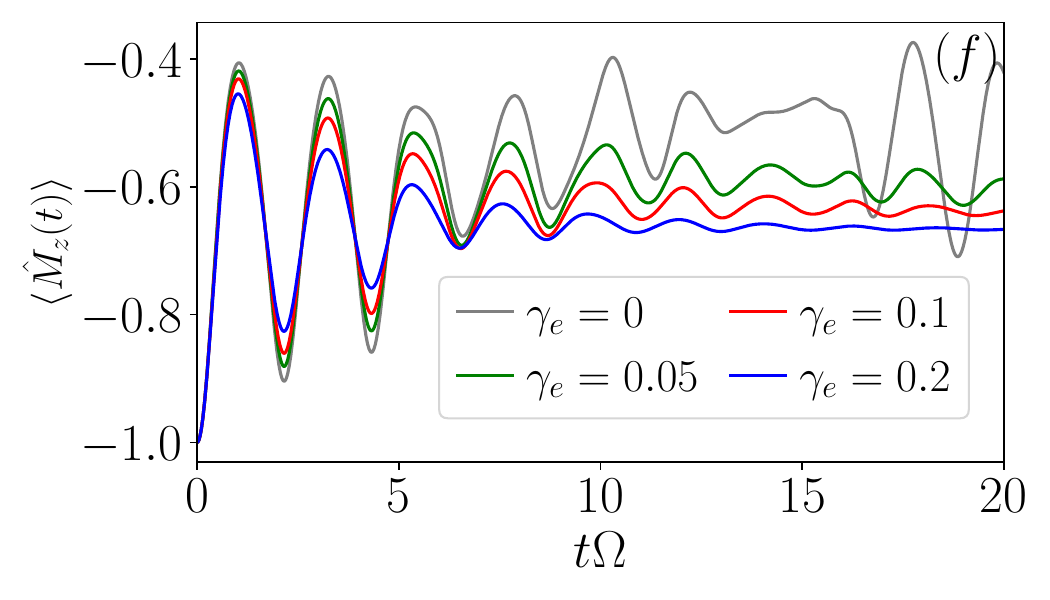}}
		\caption{Evolution of $\langle \hat{M}_z(t) \rangle$ as a function of time $t$ for (i) $\Delta=0,\ket{\psi(0)}=\ket{\mathbb{Z}_2}$, (ii) $\Delta=1,\ket{\psi(0)}=\ket{\mathbb{{Z}}_2}$ and (iii) $\Delta=1,\ket{\psi(0)}=\ket{\text{vac.}}$, considering that the Rydberg atoms experience dephasing [(a)-(c)] and spontaneous emission [(d)-(f)]}
		\label{fig:env_stability_QMBS}
	\end{figure*}     
	
	where $\hat{H}$ is given by Eq.~\eqref{main:eq:hamiltonian_ladder} and $\{\cdot,\cdot\}$ denotes the anti-commutator. The jump-operators take the form $\hat{J}^{(d)}_{j,a}=\sqrt{\gamma_d}\;\hat{\sigma}^z_{j,a}$, $\hat{J}^{(e)}_{j,a}=\sqrt{\gamma_e}\;\hat{\sigma}^-_{j,a}$, for dephasing and spontaneous emission respectively, with rates $\gamma_{d}$ and $\gamma_e$. \\
	
	In what follows, we study the stability of persistent revivals (Sec.~\ref{stability-of-QMBS_environment}) and quasi-conserved charges (Sec.~\ref{emergent_conservation_open}) considering these environmental loss channels.
	
	\subsection{Stability of persistent revivals\label{stability-of-QMBS_environment}}
	
	In Figs.~\ref{fig:env_stability_QMBS}(a)-(c), we show the evolution of average magnetization $\langle \hat{M}_z(t) \rangle \equiv \text{Tr}[\hat{\rho}(t)\hat{M}_z]$, where $\hat{\rho}(t)$ is the instantaneous (normalized) density operator of the system governed by Eq.~\eqref{eq:Lindblad-master-equation} and $\hat{M}_z \equiv \sum_{j,a} \hat{\sigma}^z_{j,a}/N$. The time-evolution of the density matrix $\hat{\rho}(t)$ is obtained by numerically solving the GKSL master equation (Eq.~\eqref{eq:Lindblad-master-equation}) using fourth-order Runge-Kutta integration scheme. We consider three scenarios where the system undergoes persistent revivals under Hamiltonian evolution: (i) $\Delta=0,\ket{\psi(0)}=\ket{\mathbb{Z}_2}$, (ii) $\Delta=1,\ket{\psi(0)}=\ket{\mathbb{Z}_2}$ and (iii) $\Delta=1,\ket{\psi(0)}=\ket{\text{vac.}}$ with finite dephasing rate $\gamma_d$. In all cases, the initial density matrix reads $\rho(0)=\ket{\psi(0)}\bra{\psi(0)}$. GKSL evolution always leads to a steady state, i.e. $d\hat{\rho}/dt=0$ in a finite dimensional Hilbert space \cite{Frigerio1978,Fagnola_10.1063/1.1340870}, and hence the persistent oscillations cannot be robust against dephasing. Pure dephasing diminishes the off-diagonal components of the density matrix (in the Fock basis), resulting in the loss of coherent oscillations. In the steady-state, all the Fock states have the same weight. This fact, when combined with the kinetic constraint implies that the steady-state magnetization value $M_z^{\text{ss}} \sim - 1/2$ in the presence of finite dephasing (see Fig.~\ref{fig:env_stability_QMBS}(a)-(c)). For very weak dephasing strengths e.g., $\gamma_d/\Omega=0.05$, a few cycles of many-body oscillation still appear with the frequency of the unitary dynamics, which is eventually damped at later times due to dephasing.
	
	In Fig.~\ref{fig:env_stability_QMBS}(d)-(f), we show the time-evolution of $\langle \hat{M}_z(t) \rangle$ for the same three scenarios, but now considering spontaneous decay of the Rydberg excited state. If there are no intrinsic spin-flip dynamics, the spontaneous emissions will continuously destroy all the Rydberg excitations of the initial state, if any, resulting in a steady state with no Rydberg excitations and $\hat{M}_z^\text{ss}=-1$. Naturally, a higher $\gamma_e$ leads to faster relaxation to this state. In presence of single spin-flip processes in the Hamiltonian dynamics, there is competition between spontaneous emission and coherent dynamics. As a result, the steady state depends crucially on system parameters. When $\gamma_e \ll 1$, $M_z^{\text{ss}} \sim -1/2$ following the arguments given for pure dephasing, while for $\gamma_e \gg 1$, $M_z^{\text{ss}} \sim -1$. For intermediate decay rates, the steady state will depend strongly on the parameter point and has to be computed by finding the fixed-point of GKSL evolution. In Fig.~\ref{fig:env_stability_QMBS}(d)-(f), $\langle \hat{M}_z(t) \rangle$ is displayed for decay rates $\gamma_e=0$ (gray), $\gamma_e=0.05$ (green),  $\gamma_e=0.1$ (red), and  $\gamma_e=0.2$ (blue) respectively. Our results demonstrate that the persistent oscillations are, in general, not robust against spontaneous emission. However, as observed in Fig.~\ref{fig:env_stability_QMBS}(f), magnetization oscillations from the $\ket{\text{vac.}}$ state are more robust against spontaneous emission, as there are no Rydberg excitations to start with which naturally suppresses decay events.

	\subsection{Stability of Approximate Emergent Conservation Laws\label{emergent_conservation_open}}
	
\begin{figure}[!htpb]
		\centering
		\rotatebox{0}{\includegraphics*[width=0.429\textwidth]{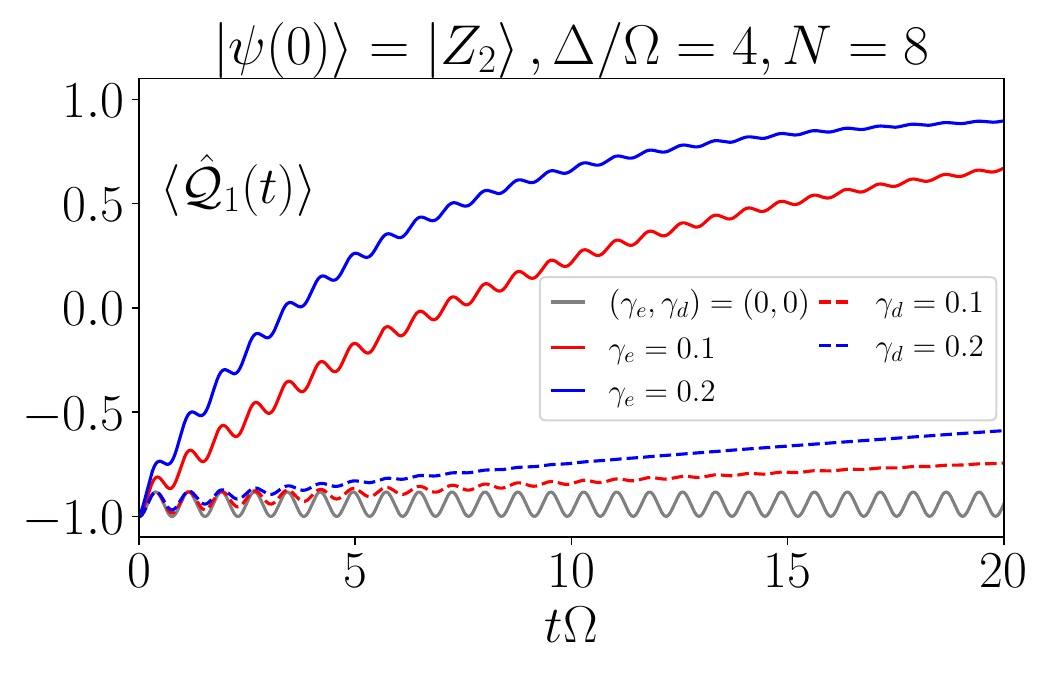}}
		\rotatebox{0}{\includegraphics*[width=0.429\textwidth]{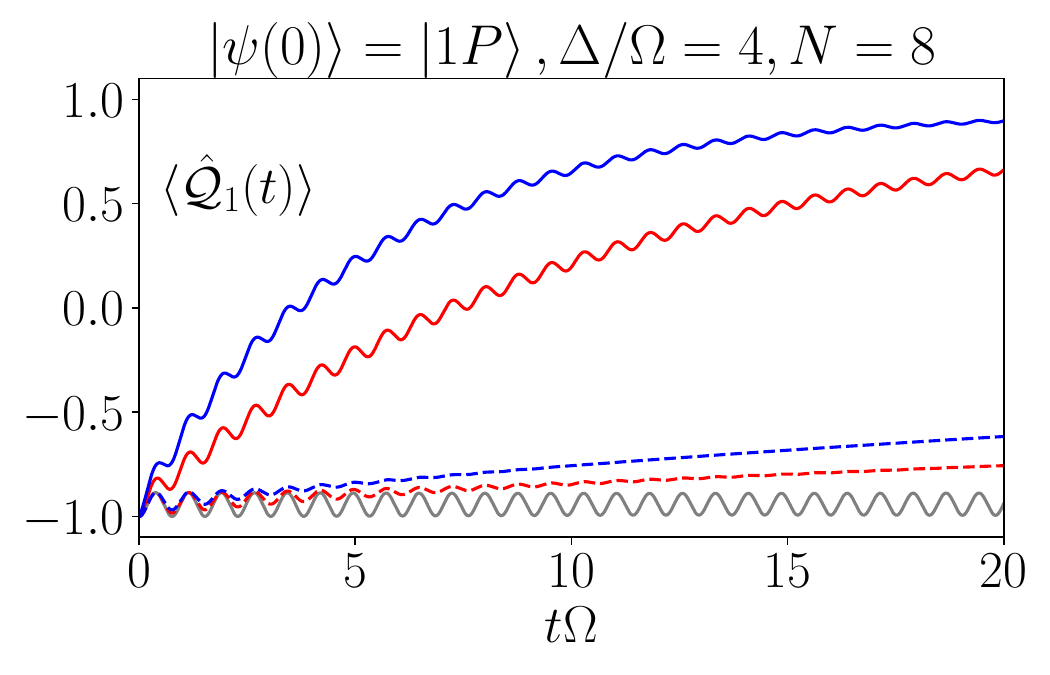}}
		\rotatebox{0}{\includegraphics*[width=0.429\textwidth]{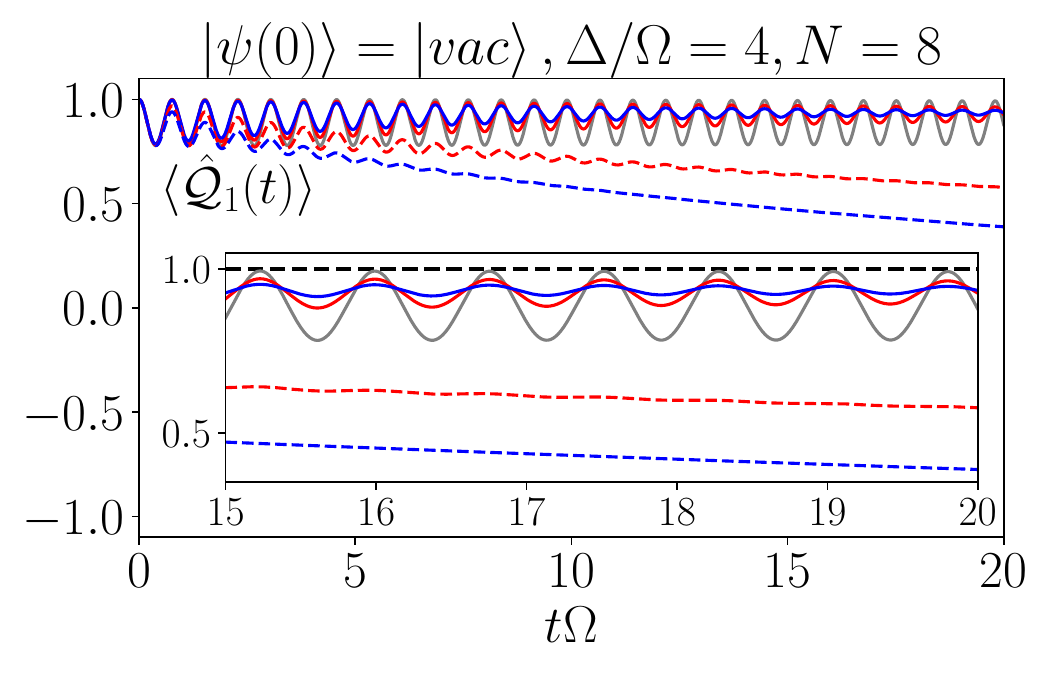}}
		\caption{Illustration of the stability of the emergent quasi-conserved charges $\hat{Q}_1$ at $\Delta=4$ starting from three different initial states ($\ket{\mathbb{Z}_2}$ top panel, $\ket{\text{1P}}$ middle panel and $\ket{\text{vac.}}$ bottom panel) under dephasing ($\gamma_d \ne 0$) and spontaneous emission ($\gamma_e \ne 0$) for a 2-leg ladder with $N=8$ atoms.}
		\label{fig:app-emission-dephasing-conservation-law-stability}
	\end{figure}

	We now consider the behavior of the quasi-conserved charge $\langle \hat{Q}_1(t)\rangle \equiv \text{Tr}[\hat{\rho}(t)\hat{Q}_1]$ under the time-evolution governed by the GKSL master equation (see Eq.~\eqref{eq:Lindblad-master-equation}). In Fig.~\ref{fig:app-emission-dephasing-conservation-law-stability} we consider three initial states, namely (i) $\ket{\psi_0} =\ket{\mathbb{Z}_2}$ (top panel), (ii) $\ket{\psi_0} =\ket{\text{1P}}$ (middle panel) and (iii) $\ket{\psi_0} = \ket{\text{vac.}} $ (bottom panel) for a 2-leg ladder with $N=8$ atoms at $\Delta=4$ and different strengths of environmental loss channels: $\gamma_d/\Omega=0.1$ (red solid curve), $\gamma_d = 0.2$ (blue solid curve), $\gamma_e=0.1$ (red dashed curve), $\gamma_e= 0.2$ (blue dashed curve), $\gamma_{d}=\gamma_e=0$ (gray solid curve). This figure exhibits the robustness of the approximate emergent conservation laws against dephasing and spontaneous emission. \newline
	
	It is expected and also evident from Fig.~\ref{fig:app-emission-dephasing-conservation-law-stability} that for generic initial states, the emergent conservation laws are more stable against dephasing than compared to spontaneous emission, since one event of spontaneous emission of Rydberg-excited atoms on any of the sites $(j,1)$ or $(j,2)$ changes the sign of the quasi-conserved charge $\langle \hat{Q}_j \rangle(t)$. Furthermore, we note that pure dephasing does not alter the sign of the quasi-conserved charge $\langle \hat{Q}_1(t)\rangle$ for all the initial states considered here up to times $t\Omega \sim 20$. Hence, the storage of $L/2$ classical bits of information based on values of $\langle \hat{Q}_1(t) \rangle/|\langle \hat{Q}_1(t) \rangle|$, as discussed in Appendix-\ref{app:classical_bit_storage} is a robust mechanism even in the presence of pure-dephasing. There is an important exception to this generic expectation: for the $\ket{\text{vac.}}$ state, quasi-conserved charges are more robust against spontaneous emission compared to dephasing. From the inset of the bottom panel of Fig.~\ref{fig:app-emission-dephasing-conservation-law-stability} it becomes evident that for the $\ket{\text{vac.}}$ state, larger decay rates lead to enhancement in the degree of conservation of these charges. This happens since after each event of spontaneous emission, one excited Rydberg atom is converted into the electronic ground-state and hence the expectation value of the quasi-conserved charges shift towards the initial time expectation value with respect to the $\ket{\text{vac.}}$ state.\newline

	Alternatively, environmental effects can be understood by the Monte Carlo wave function (MCWF) approach \cite{Zoller_PhysRevA.35.198,Zoller:92,Dum_PhysRevA.45.4879,Molmer:93,Daley_2014}. In this approach, one averages over a large number of stochastically sampled quantum trajectories, where the stochasticity comes from randomness in the occurrence of environmental losses, and allows one to track the evolution of the open system as a function of time in terms of an ensemble of pure quantum states.  Additionally, the MCWF approach also enables the possibility of monitoring quantum entanglement of the system under different continuous measurement protocols \cite{PhysRevLett.128.243601_PICHLER_TAITANA_2022,Tatiana_PhysRevA.110.012207_Pichler}, uncovering important physical processes. In a single realization, an open quantum system under the influence of environmental loss channels, does not follow the GKSL master equation. Instead, a single quantum trajectory is realized in a single shot. This makes the analysis of individual quantum trajectories very important and relevant, when the goal is to understand stability of dynamical phenomena influenced by loss processes. 
	
	\begin{figure}[!htpb]
		\centering
		\includegraphics[width=0.5\textwidth]{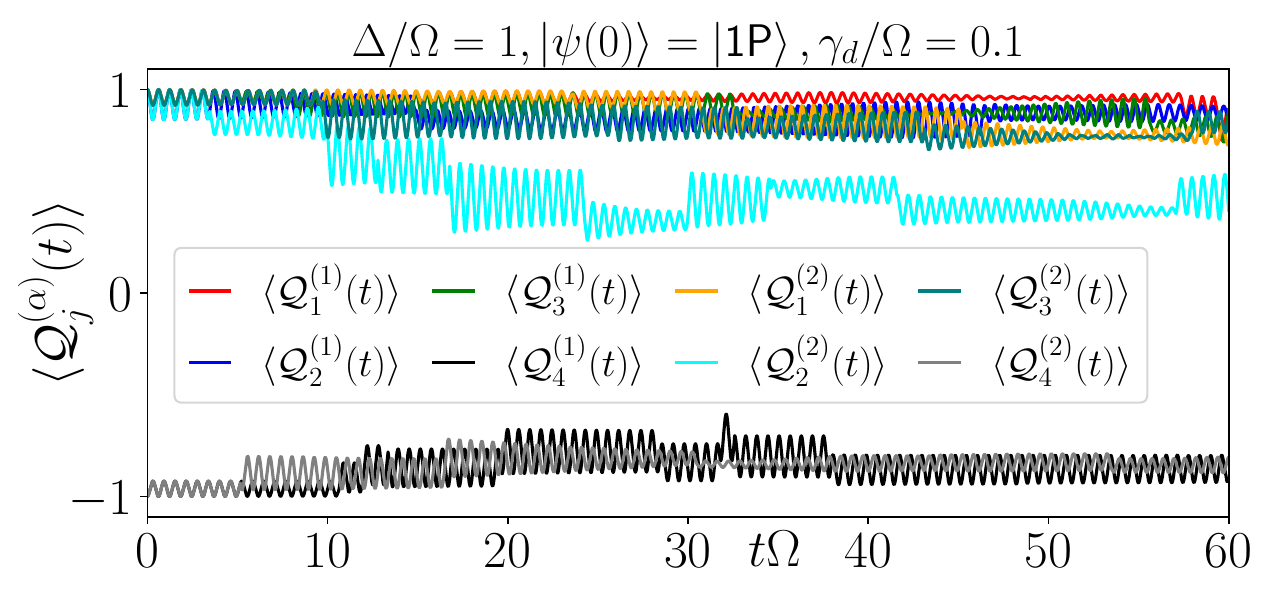}
		\caption[]{Time-evolution of quasi-conserved charges $\{ \hat{Q}_j^{\alpha} \}$ ($j=1,2,3,4$ stands for rung index in a 2-leg square ladder with $N=8$ atoms) with the initial state $\ket{\text{1P}}\equiv\ket{\substack{\circ\circ\circ\bullet \\ \circ\circ\circ\circ}}$ for $\Delta=5, \gamma_d=0.1$. Two individual quantum trajectories ($\alpha=1,2$) are shown. See text for details.}
		\label{fig:dephasing-QT-conservation-stability}
	\end{figure}
	
	As we illustrated earlier in Fig.~\ref{fig:app-emission-dephasing-conservation-law-stability}, the quasi-conserved charges are more robust against dephasing compared to spontaneous emissions. For this reason, it is interesting to look at individual quantum trajectories for the dephasing case. In Fig.~\ref{fig:dephasing-QT-conservation-stability} we show the time-evolution of all the approximately conserved charges $\{Q_j\}$, starting from the $\ket{\text{1P}}$ state, when a dephasing rate $\gamma_d=0.1$ is considered. This illustrates that even at the level of individual quantum trajectories, the approximate emergent conservation laws seem to be respected. Moreover, the sign of the quasi-conserved charges are unaffected for a very long duration, which further strengthens the classical bit storage capabilities illustrated in Appendix-\ref{app:classical_bit_storage}

	\section{Long-range interactions and validity of kinetic constraints\label{sec:long-range}}
	
	So far in this paper, we have been concerned with quantum dynamics of the kinetically constrained model \eqref{main:eq:hamiltonian_ladder}, with and without environmental loss channels. However, keeping in mind that this model is an idealized representation of physical Rydberg atom quantum simulator platforms, we shall now investigate the validity of the specific Rydberg blockade shown in Fig.~\ref{fig:model} in the context of out-of-equilibrium dynamics, by considering long-range vdW repulsive interactions between excited Rydberg atoms which are always present in the real experimental hardware. Our results presented below indicate that in practice, it could be challenging to implement the specific kinematical constraint illustrated in Fig.~\ref{fig:model} in a real platform for out-of-equilibrium quantum dynamics. We associate this difficulty to the fact that for an actual Rydberg atom quantum simulator configured in a 2-leg geometry, the strength of the second nearest-neighbor interactions --- i.e. the repulsion between the diagonally placed atoms when they are excited --- are not negligible compared to the same between immediately neighboring ones. These issues will be discussed extensively in this section and we shall see that the specific form of constraint taken is not straightforwardly realized in simulators with neutral atoms having $nS$-type Rydberg excited orbitals. Having said this, we do however observe persistent many-body oscillations in some cases (see Fig.~\ref{fig:PXPladd-vs-longrange} below), even in the presence of full long-range vdW interactions, although the mechanisms leading to these revivals remains to be understood. \\
	
	In a Rydberg atom quantum simulator setup \cite{Bernien2017,Bluvstein2022}, neutral alkali atoms are trapped and arranged in a desired lattice geometry (such as in Fig.~\ref{fig:model}) using an array of optical tweezers, and are driven by external Rydberg lasers with detunings $\Delta_{\vec{r}}$. This leads to effective Rabi oscillations between the electronic ground-state $\ket{G}$, and the Rydberg excited state $\ket{R}$ at a frequency set by the laser intensities. The electronic orbitals of the Rydberg excited states are substantially large and lead to long-range vdW interactions between excited pairs. Considering these effects, the full Hamiltonian of the system reads \cite{Browaeys2020}
	
	\begin{equation}
		\hat{H}_{\text{Ryd}}^{\text{Full}} = \sum_{\vec{r}} \left( \Omega \hat{\sigma}^x_{\vec{r}} - \Delta_{\vec{r}} \hat{\sigma}^z_{\vec{r}} \right) + \frac{V_0}{2} \sum_{\vec{r},\vec{r}'} \frac{\hat{n}_{\vec{r}}\hat{n}_{\vec{r}'}}{|\vec{r}-\vec{r}'|^6}
		\label{eq:hamiltonian_Rydberg_long_range}
	\end{equation}
	
	where $\vec{r}$ denotes the position of the trapped atoms, which can be thought of as sites of a lattice. Furthermore, we have assumed that the electronic excited states for the atom is $\ket{R}=\ket{nS}$ and thus Rydberg excited pairs experience isotropic long-range repulsive vdW interactions which falls off as $1/d^6$ where $d$ is the distance between them. The last double sum is taken over all possible pairs of atoms placed on sites $\vec{r},\vec{r}'$ of the ladder with $N=2L$ sites and $\hat{n}_{\vec{r}}=\left(\hat{\sigma}^z_{\vec{r}}+1\right)/2=\ket{R}_{\vec{r}}\prescript{}{\vec{r}}{\bra{R}}$ being the projection operator corresponding to the Rydberg excited state $\ket{R}$ at site $\vec{r}$.\newline
	
	In order to have an understanding of how accurately the idealized kinetically constrained models mimic the out-of-equilirbrium dynamics of the fully interacting long-range system, we present a direct comparison between them. For a quantitative comparison of the 1D atomic chain and the associated 1D PXP model see Appendix-\ref{sec:1Dchain-blockade-dynamics}.  
	
	\subsection{2-leg Rydberg Square Ladder\label{subsec:2leg-blockade}}
	
	In this section we shall investigate the validity of the specific Rydberg blockade constraint for the 2-leg square ladder system assumed throughout the main text (see Fig.~\ref{fig:model}). We shall consider the full long-range interacting system and its various approximations as in Appendix-\ref{sec:1Dchain-blockade-dynamics}. The fully interacting long-range system is now given by the 2-leg square ladder version of Eq.~\eqref{eq:hamiltonian_Rydberg_long_range} while the low-energy effective Hamiltonian in the limit $V_0 \gg \Delta_{j,a},\Omega$ now reads
	
	\begin{equation}
		\hat{H}^{\text{eff}}_{\text{Ryd}} = \sum_{j,a} \left(\Omega \hat{\tilde{\sigma}}^x_{j,a} -\Delta_{j,a} \hat{\sigma}^z_{j,a}\right) + \!\!\! \sum_{\vec{r},\vec{r}' > \text{NN}_{1}} \!\! \frac{V_0}{2} \frac{\hat{n}_{\vec{r}}\hat{n}_{\vec{r}'}}{|\vec{r}
			-\vec{r}'|^6}
		\label{eq:Rydberg-ladder-lowenergy-effective}
	\end{equation}
	
	It is crucial here to note that the second nearest neighbor distance between two atoms in this case is $\sqrt{2}$ (in lattice units). This implies that in this case, the conditions which the parameters $V_0,\Delta_{j,a},\Omega$ must satisfy so that the long-range interactions beyond the first neighbor could be safely neglected is $V_0 \gg \Delta_{j,a},\Omega \gg V_0/(\sqrt{2})^6$. However, this hard to satisfy, even in an order of magnitude sense. This can be seen, for example, by taking $\Omega,\Delta_{j,a}\sim1, V_0=10\Omega$ which is required to satisfy $V_0 \gg \Delta_{j,a},\Omega$, but makes it difficult to have $\Delta_{j,a},\Omega \gg V_0/(\sqrt{2})^6$. For this reason we expect that, the specific form of the constraint assumed throughout the main text, could be in practice difficult to achieve in actual experiments in a Rydberg atom quantum simulator platform, particularly in the context of out-of-equilibrium quench dynamics. We note however that when probing low-energy properties of the system, i.e. ground-state and few excited states, the failure to satisfy the condition $V_0 \gg \Delta_{j,a},\Omega \gg V_0/(\sqrt{2})^6$ is not as severe and has been studied theoretically \cite{Samajdar_PhysRevLett.124.103601,Ebadi2021,Sarkar_10.21468/SciPostPhys.14.1.004,Liao_PhysRevB.111.165154,Soto-Garcia_PhysRevResearch.7.013215,Yuzhou2025,Pierre_PhysRevB.106.155411} and experimentally \cite{Labuhn201_Long_range_Ising_model_expt,Schaub2012}. \newline
	
	\begin{figure}[!htpb]
		\centering
		\includegraphics[width=0.5\textwidth]{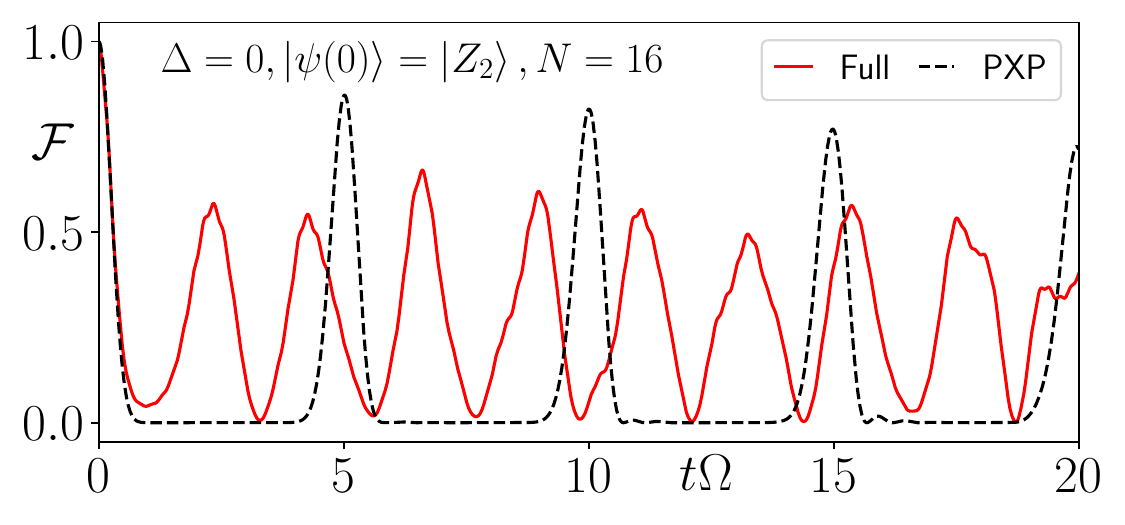}
		\includegraphics[width=0.5\textwidth]{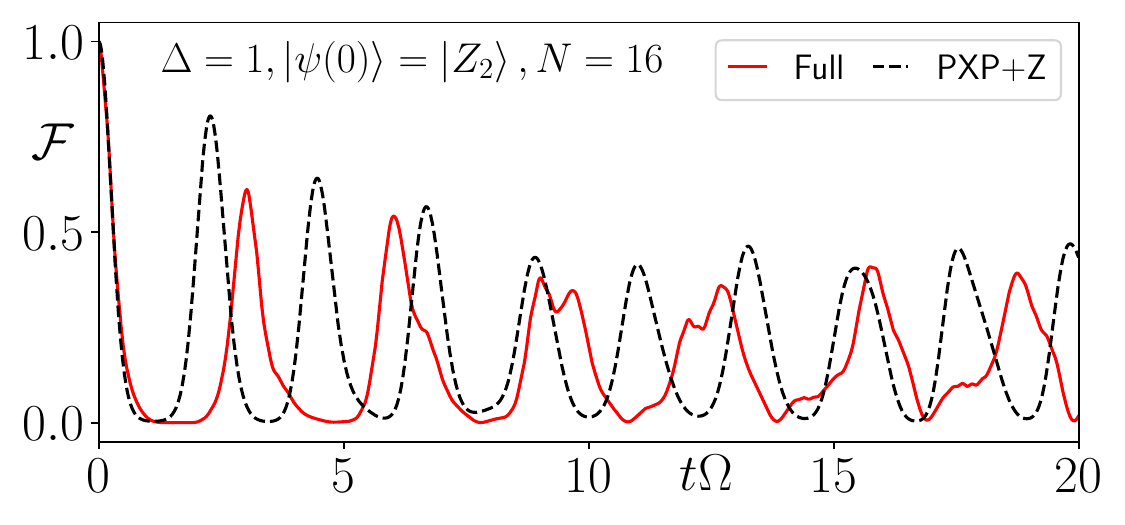}
		\includegraphics[width=0.5\textwidth]{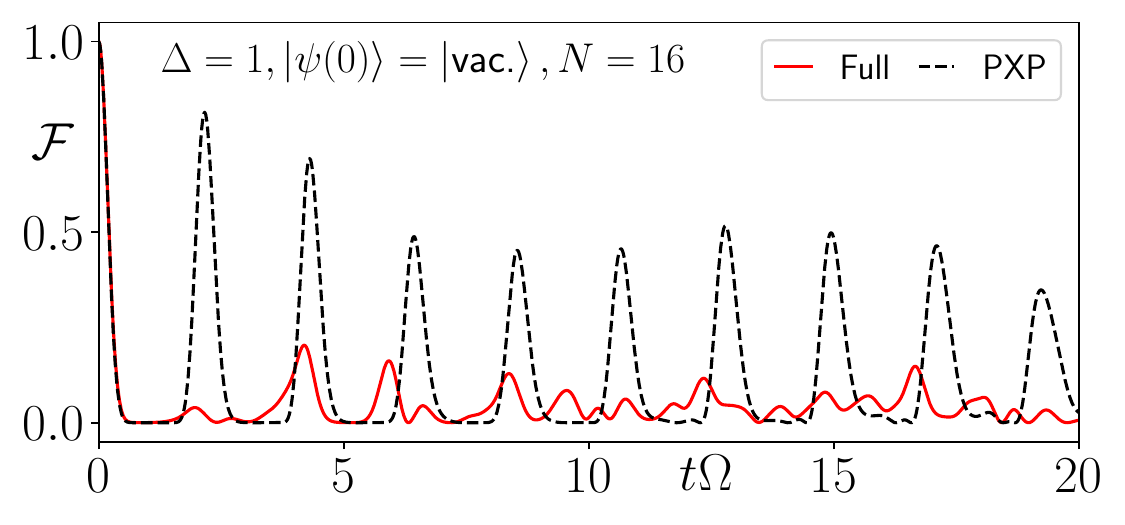}
		\caption{Comparison of the return probability $\mathcal{F}=|\langle \psi(0)|\psi(t)\rangle|^2$ by evolving an initial state under the fully interacting long-range Hamiltonian \eqref{eq:hamiltonian_Rydberg_long_range} (red solid line) taking $V_0=10\Omega$ and the idealized kinetically constrained system 2D PXP+Z model \eqref{main:eq:hamiltonian_ladder} (black dashed lines) for $N=16$ atoms (with periodic boundary condition). The specific initial conditions are as follows: (top panel) $\ket{\psi(0)}=\ket{\mathbb{Z}_2},\Delta=0$ (middle panel) $\ket{\psi(0)}=\ket{\mathbb{Z}_2},\Delta=1$ and (bottom panel) $\ket{\psi(0)}=\ket{\text{vac.}},\Delta=1$.}
		\label{fig:PXPladd-vs-longrange}
	\end{figure}

	In Fig.~\ref{fig:PXPladd-vs-longrange} we compare the evolution of the return probability starting from the N\'eel state, under the fully interacting long-range Hamiltonian Eq.~\eqref{eq:hamiltonian_Rydberg_long_range}, and the kinetically constrained Hamiltonian Eq.~\eqref{main:eq:hamiltonian_ladder}. This figure shows that even the qualitative nature of quantum dynamics of these two Hamiltonians are quite different from each other, which suggests that the kinetic constraint illustrated in Fig.~\ref{fig:model}, is not a faithful representation of the actual long-range interacting system that would be realized in an experimental setup with $nS$-type Rydberg excited state. From bottom panel of Fig.~\ref{fig:PXPladd-vs-longrange} it is clear that the $\ket{\text{vac.}}$ state does not show any anomalous revivals when the full long-range interacting system is considered. However, as Fig.~\ref{fig:PXPladd-vs-longrange} top and middle panel demonstrates, the $\ket{\mathbb{Z}_2}$ state shows a substantial degree of anomalous oscillation lasting several cycles, which cannot be explained from the PXP+Z approximation. 
	
	These oscillations can be instead qualitatively described by augmenting the PXP+Z model with second nearest neighbor vdW interactions as in Eq.~\eqref{eq:hamiltonian_Rydberg_ladder_lowenergy_NN2}, 
	
	\begin{equation}
		\hat{H}^{\text{eff,NN}_2}_{\text{Ryd}} = \sum_{j,a} \left(\Omega \hat{\tilde{\sigma}}^x_{j,a} - \Delta_{j,a} \hat{\sigma}^z_{j,a}\right) +\sum_{\vec{r},\vec{r}' \in \text{NN}_{2}}  \frac{V_0}{2}  \frac{\hat{n}_{\vec{r}}\hat{n}_{\vec{r}'}}{|\vec{r}
			-\vec{r}'|^6}
		\label{eq:hamiltonian_Rydberg_ladder_lowenergy_NN2}
	\end{equation}

\begin{figure}[!htpb]
		\centering
		\includegraphics[width=0.35\textwidth]{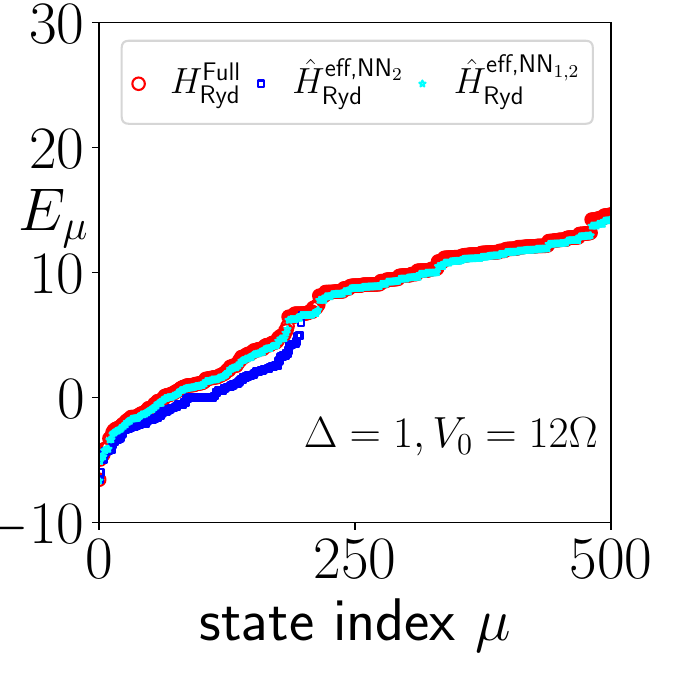}
		\caption{Comparison of the eigenvalues of the three Hamiltonians $\hat{H}^\text{Full}_\text{Ryd.}$, $\hat{H}^{\text{eff,NN}_2}_{\text{Ryd}}$ and $\hat{H}^{\text{eff,NN}_{1,2}}_{\text{Ryd}}$ for $\Delta=1,V_0=12\Omega$. Only the low-energy part of the spectrum has been displayed.}
		\label{fig:spectrum_comparison_H1H2H3}
	\end{figure}

\begin{figure}[!htpb]
		\centering
		\rotatebox{0}{\includegraphics[width=0.5\textwidth]{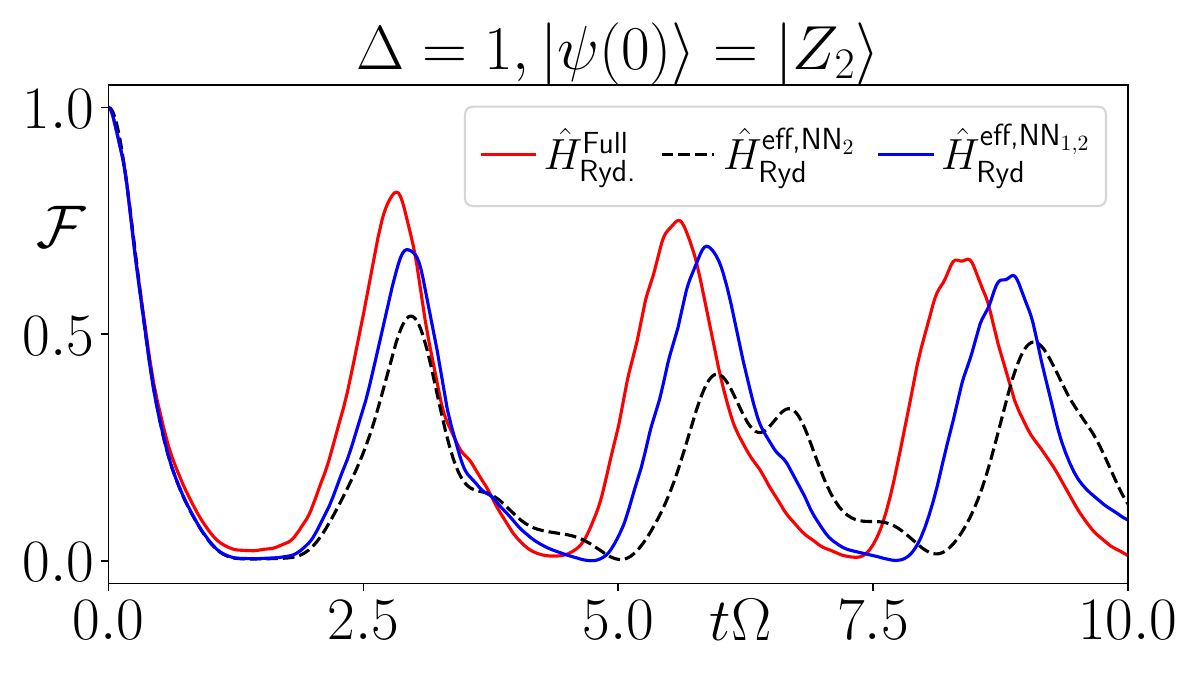}}
		\caption{Comparison of the return probability $\mathcal{F}=|\langle \psi(0)|\psi(t)\rangle|^2$ computed by evolving the initial state $\ket{\psi(0)}=\ket{\mathbb{Z}_2}$ state via three different Hamiltonians $\hat{H}^\text{Full}_\text{Ryd.}$, $\hat{H}^{\text{eff,NN}_2}_{\text{Ryd}}$ and $\hat{H}^{\text{eff,NN}_{1,2}}_{\text{Ryd}}$ for a system of $N=12$ atoms with $\Delta=1,V_0=12\Omega$ (see text for details).}
		\label{fig:oscillations-PXP-Z-NN2}
	\end{figure}

	This claim is supported by the fact that considering a short-range interacting system governed by Eq.~\eqref{eq:hamiltonian_Rydberg_ladder_lowenergy_NN2}, which is obtained from Eq.~\eqref{eq:Rydberg-ladder-lowenergy-effective} by keeping interactions only up to the second-nearest neighbors, seems to capture both the spectral and dynamical features of the fully interacting long-range Hamiltonian Eq.~\eqref{eq:hamiltonian_Rydberg_long_range} (see Fig.~\ref{fig:spectrum_comparison_H1H2H3}, Fig.~\ref{fig:oscillations-PXP-Z-NN2}). Furthermore, we can also compare our results against Eq.~\eqref{eq:hamiltonian_Rydberg_ladder_lowenergy_NN12} which truncates interactions in Eq.~\eqref{eq:hamiltonian_Rydberg_long_range} up to second nearest neighbors. In this case, we also do not work in the kinetically constrained Hilbert space, meaning the spin-flip terms are not dressed by additional projectors, and the Hamiltonian reads, \\
	
	\begin{equation}
		\hat{H}^{\text{eff,NN}_{1,2}}_{\text{Ryd}} = \sum_{j,a} \left(\Omega \hat{\sigma}^x_{j,a} - \Delta_{j,a} \hat{\sigma}^z_{j,a}\right) + \!\!\!\! \sum_{\vec{r},\vec{r}' \in \text{NN}_{1,2}} \frac{V_0}{2}  \frac{\hat{n}_{\vec{r}}\hat{n}_{\vec{r}'}}{|\vec{r}
			-\vec{r}'|^6}
		\label{eq:hamiltonian_Rydberg_ladder_lowenergy_NN12}
	\end{equation}    
	
	In Fig.~\ref{fig:oscillations-PXP-Z-NN2} the return probability $\mathcal{F}(t)=|\langle\psi(0)|\psi(t)\rangle|^2$ starting from the $\ket{\mathbb{Z}_2}$ state is shown as a function of time by evolving the state under Hamiltonian given by Eqs.~\eqref{eq:hamiltonian_Rydberg_long_range},~\eqref{eq:hamiltonian_Rydberg_ladder_lowenergy_NN2} and ~\eqref{eq:hamiltonian_Rydberg_ladder_lowenergy_NN12} for a system with $N=12$ atoms at $\Delta=1$. 
	
	Thus for reasons mentioned above the second nearest-neighbor vdW interactions are essential in understanding the qualitative features of quantum dynamics depicted by the full long-range interacting system.  \newline        
	
	\begin{figure}[!htpb]
		\centering
		\rotatebox{0}{\includegraphics[width=0.5\textwidth]{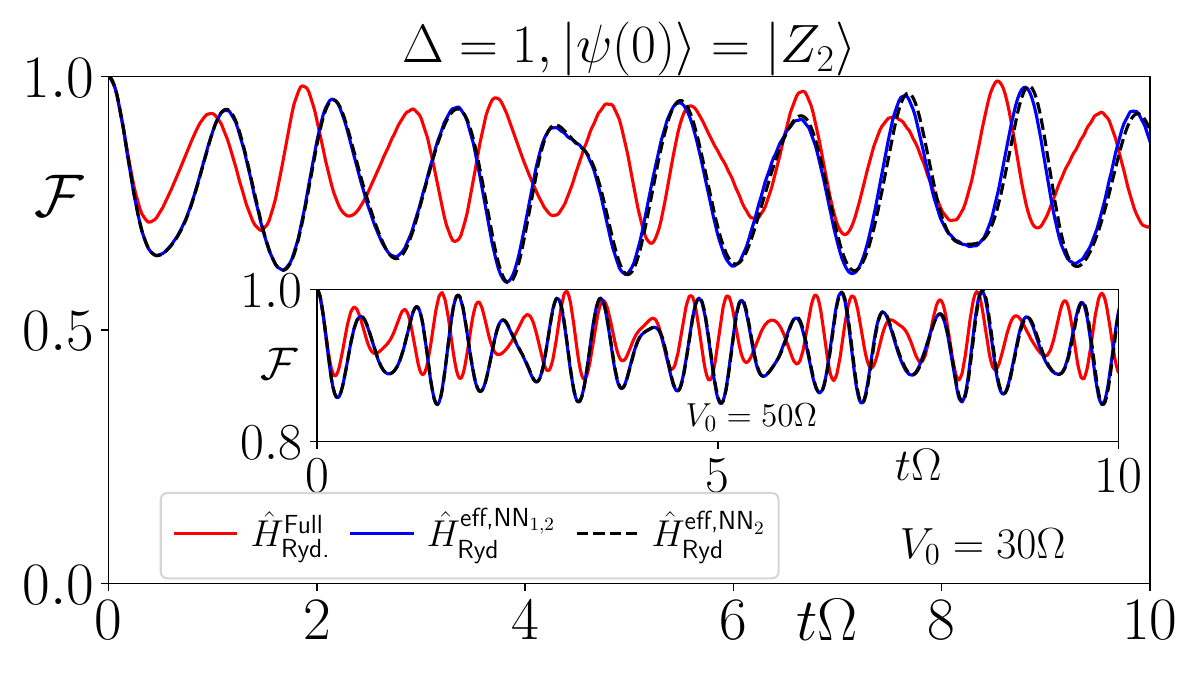}}
		\caption{Comparison of the return probability $\mathcal{F}=|\langle \psi(0)|\psi(t)\rangle|^2$ computed by evolving the initial state $\ket{\psi(0)}=\ket{\mathbb{Z}_2}$ state via three different Hamiltonians $\hat{H}^\text{Full}_\text{Ryd.}$, $\hat{H}^{\text{eff,NN}_2}_{\text{Ryd}}$ and $\hat{H}^{\text{eff,NN}_{1,2}}_{\text{Ryd}}$ for a system of $N=12$ atoms with $\Delta=1,V_0=30\Omega$ (main) and  $\Delta=1,V_0=50\Omega$ (inset) respectively (see text for details).}
		\label{fig:oscillations-PXP-Z-NN12}
	\end{figure}
	
	It is also worth noting that increasing the strength of the vdW repulsive interaction (or equivalently reducing the lattice constant), increases the configurational energy of the N\'eel state (i.e., $\langle \mathbb{Z}_2 | \hat{H}^{\text{Full}}_{\text{Ryd}}| \mathbb{Z}_2 \rangle$) and results in confinement of this state (see Fig.~\ref{fig:oscillations-PXP-Z-NN12}). From the comparison between the dynamics obtained from the three different Hamiltonians Eq.~\eqref{eq:hamiltonian_Rydberg_long_range}, Eq.~\eqref{eq:hamiltonian_Rydberg_ladder_lowenergy_NN12} and Eq.~\eqref{eq:hamiltonian_Rydberg_ladder_lowenergy_NN2}, we can conclude that the confinement is a consequence of next nearest neighbor interaction, since the dynamics of the N\'eel state obtained by evolving the state via Hamiltonians Eq.~\eqref{eq:hamiltonian_Rydberg_ladder_lowenergy_NN2}, Eq.~\eqref{eq:hamiltonian_Rydberg_ladder_lowenergy_NN12} agree qualitatively with that obtained from the fully interacting long-range system governed by Eq.~\eqref{eq:hamiltonian_Rydberg_long_range}.

	\subsection{Two Dimensional Square Lattice\label{subsec:2D_pxp_staggered_detuning}}
	
	The ideal two dimensional (2D) PXP model has been shown to host QMBS \cite{Lin_PhysRevB.101.220304}. However, the difficulty in realization of the kinetic constraints in out-of-equilibrium dynamics for a 2-leg ladder geometry with $nS$-type Rydberg excited neutral atoms (see Sec.~\ref{subsec:2leg-blockade}), brings into question the existence of persistent many-body revivals in a physical two-dimensional Rydberg atom simulator in which vdW interactions are always present. The Hamiltonian for the 2D PXP+Z model (with site-dependent detunings $\Delta_{j,a}=(-1)^j\Delta$) of dimension $L_x \times L_y$ is given by
	
	\begin{equation}
		\hat{H}^{\text{2D}}_{\text{PXP+Z}} =  \sum_{j=1}^{L_x}  \sum_{a=1}^{L_y} \left( \Omega \hat{\tilde{\sigma}}^x_{j,a} -\Delta (-1)^j \hat{\sigma}^z_{j,a} \right)
		\label{eq:2DPXP+Z}
	\end{equation}
	
	where $\hat{\tilde{\sigma}}^x_{j,a}\equiv\hat{P}^{\downarrow}_{j+1,a}\hat{P}^{\downarrow}_{j-1,a}\hat{P}^{\downarrow}_{j,a-1}\hat{P}^{\downarrow}_{j,a+1} \hat{\sigma}^x_{j,a}$ \newline
	
	\begin{figure}[!htpb]
		\centering
		\includegraphics[width=0.5\textwidth]{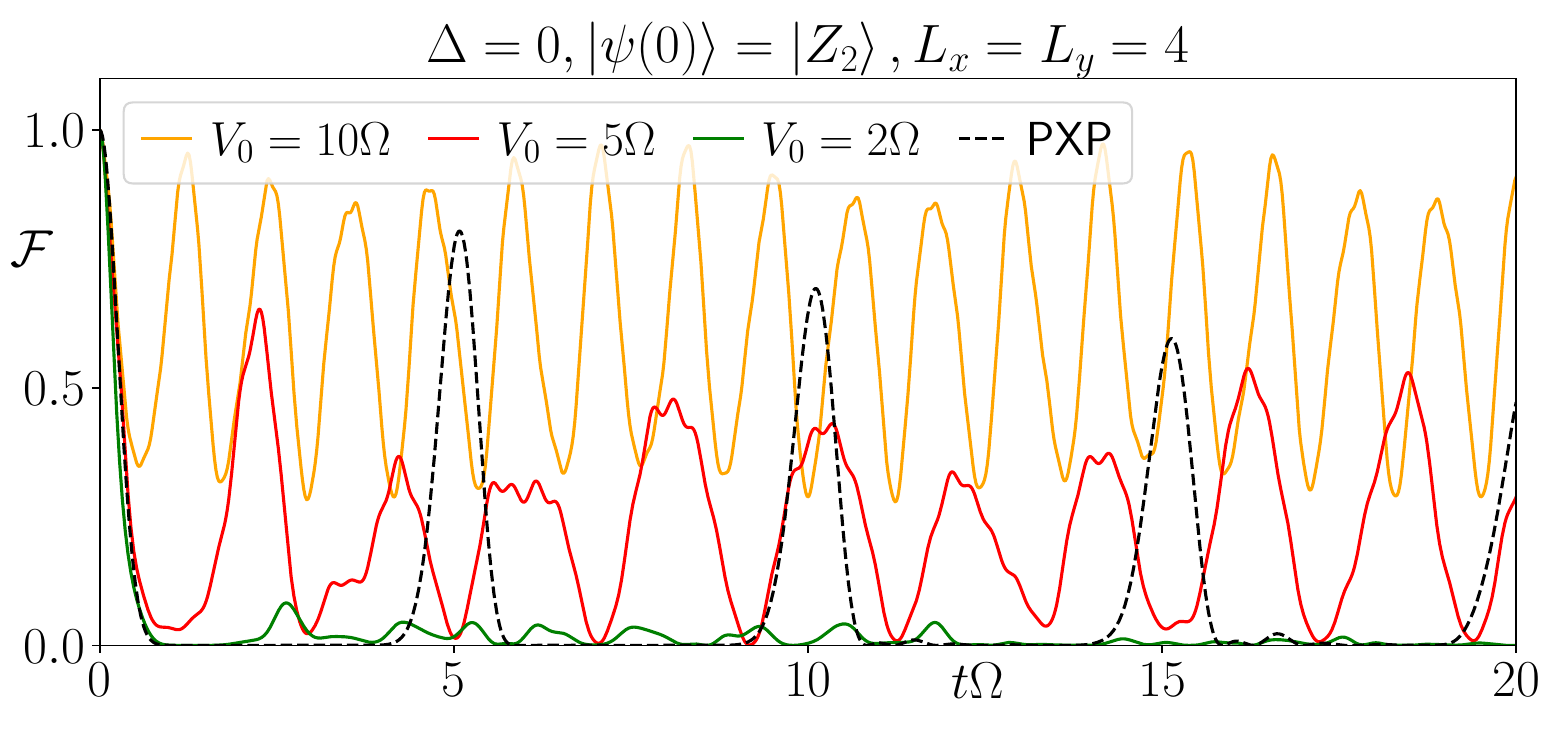}
		\caption{Comparison of the return probability $\mathcal{F}=|\langle \psi(0)|\psi(t) \rangle|^2$ as function of time $t$ starting from the N\'eel state  in the 2D geometry for a $4\times4$ square lattice, computed with (i) full long-range interacting system (solid curves) with different vdW interaction strengths $V_0=10\Omega$ (orange), $V_0=5\Omega$ (red), $V_0=2\Omega$ (green) and (ii) the idealized 2D PXP approximation (black dashed line) for $\Delta=0$(see text for details).}
		\label{fig:PXP2D-vs-longrange-Delta-0x000}
	\end{figure}
	
	Fig.~\ref{fig:PXP2D-vs-longrange-Delta-0x000} compares the return probability of the initial N\'eel state ($\ket{\mathbb{Z}_2}$) obtained via Hamiltonians Eq.~\eqref{eq:2DPXP+Z} with the corresponding results of the long-range interacting system for $V_0=10\Omega$, $V_0=5\Omega$. The parameter choices $V_0=10\Omega$ or $V_0=5\Omega$ satisfies the condition $V_0 \gg \Omega$ in a qualitative sense, however, the kinetically constrained model Hamiltonian does not capture the same dynamical behavior.
	
	We now turn on a non-zero staggered detuning $(\Delta\ne0)$ and ask if the anomalous revivals from the $\ket{\mathbb{Z}_2}$ and $\ket{\text{vac.}}$ state observed for the 2-leg square ladder geometry \cite{Pal_PhysRevB.111.L161101}, has any analogue in the 2D square lattice scenario. \newline
	
	\begin{figure}[!htpb]
		\centering
		\includegraphics[width=0.458\textwidth]{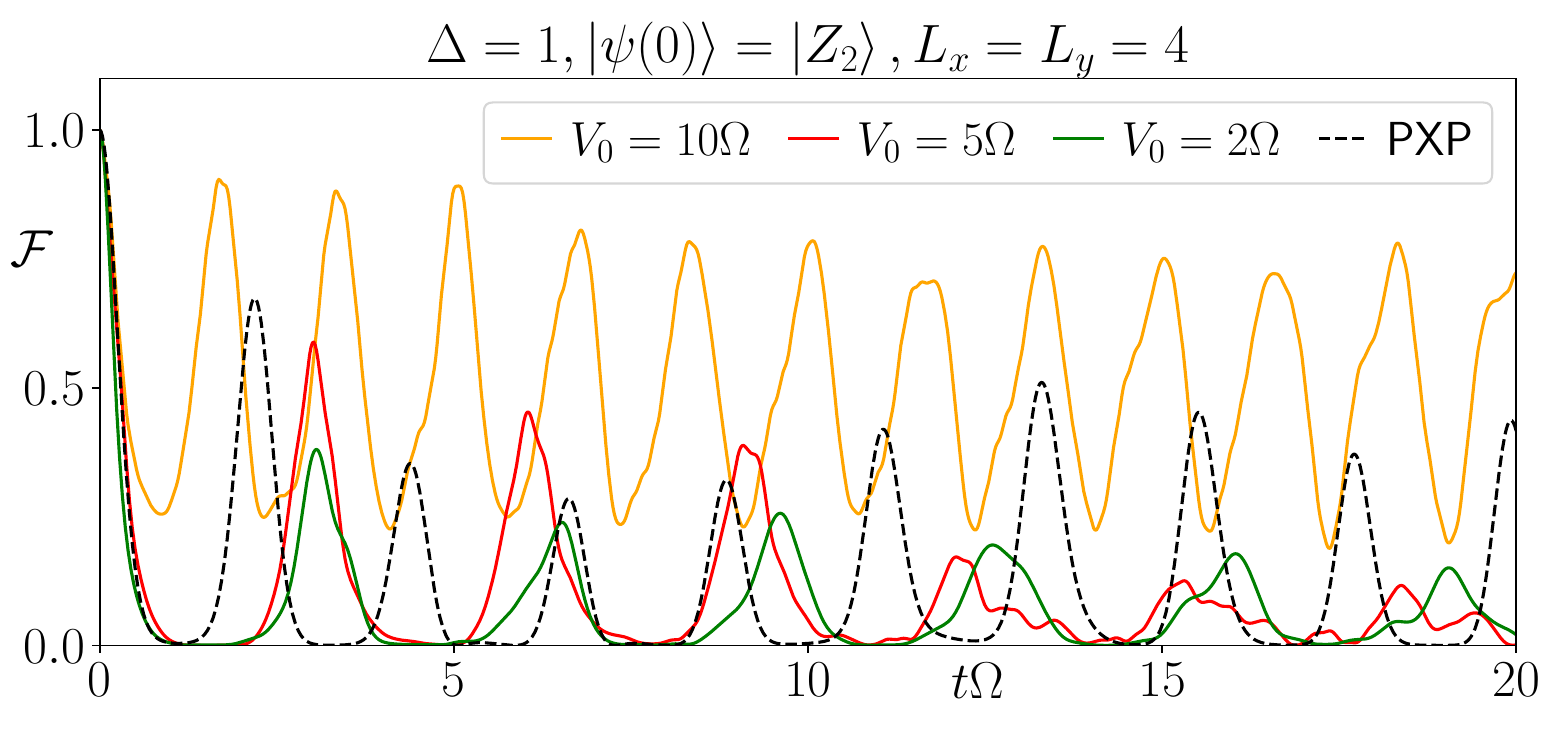}
		\includegraphics[width=0.458\textwidth]{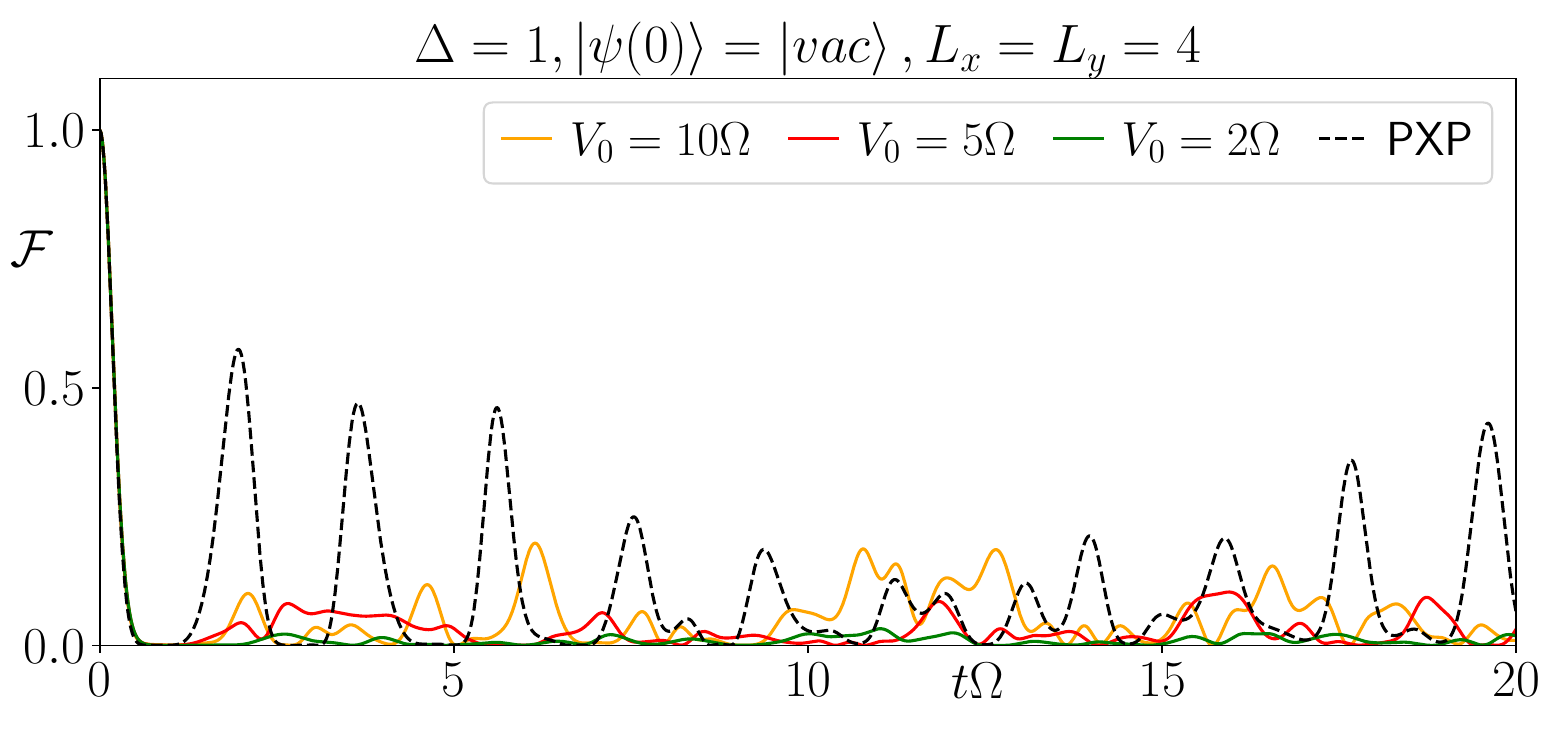}
		\caption{Comparison of the return probability $\mathcal{F}=|\langle \psi(0)|\psi(t) \rangle|^2$ as function of time $t$ starting from (top panel) the N\'eel state $\ket{\mathbb{Z}_2}$ and (bottom panel) the complete electronic ground-state $\ket{\text{vac.}}$ in the 2D geometry for a $4\times4$ square lattice, computed with (i) full long-range interacting system (solid curves) with different vdW interaction strengths $V_0=10\Omega$ (orange), $V_0=5\Omega$ (red), $V_0=2\Omega$ (green) and (ii) the idealized 2D PXP approximation given by Eq.~\eqref{eq:2DPXP+Z} (black dashed line) for $\Delta=1$ (see text for details).}
		\label{fig:PXP2D-vs-longrange-Delta-1x000}
	\end{figure}

	From Fig.~\ref{fig:PXP2D-vs-longrange-Delta-1x000} we can see that in 2D the PXP+Z approximation given by Eq.~\eqref{eq:2DPXP+Z} does not fully capture the features of the full long-range interacting system accurately for $V_0=2\Omega,5\Omega$ and $10\Omega$. The disagreement is more severe for the $\ket{\text{vac.}}$ initial state, as the PXP+Z predicts persistent oscillations lasting several cycles while the full long-range interacting system predicts this state does not have any noticeable revivals. This again leads us to conclude that for geometries in which the second nearest-neighbor distance is $\sqrt{2}$ (in lattice separation units), the idealized strong Rydberg blockaded models do not faithfully capture the dynamics in these systems.  \newline
	
	We end this section by noting that, realizing such geometric blockades, could in principle be made easier to achieve by leveraging anisotropic interactions as shown to exist for more complex Rydberg excited electronic states \cite{Vermersh_PhysRevA.91.023411_Zoller}, e.g. using complex electronic orbitals ($P,D$) as excited states, (bringing into play magic distances and anisotropic interactions \cite{Vermersh_PhysRevA.91.023411_Zoller}), dressed Rydberg states \cite{Kurdak_PhysRevLett.134.123404}, mixtures of atomic species \cite{Anand2024_dual_species_Rydberg_array}. However, such systems exhibit completely different kinds of effective spin interactions and the associated dynamics will exhibit vastly different phenomena. Nonetheless, as the model studied exhibits a wide range of out-of-equilibrium phenomena, it is worth exploring alternate approaches to engineer such constraints.
	
	\section{Conclusion and Outlook}
	
	In this paper we analyzed a model of two coupled 1D PXP chains with additional semi-staggered detuning profile. Using numerical exact diagonalization based techniques we found that for intermediate to strong values of staggering strength $\Delta$, the system has an extensive number of emergent approximate conserved charges leading to a number of anomalous features in quantum quench dynamics starting from simple initial states. Furthermore, by leveraging the underlying chirality operator we illustrated two different types of Floquet protocols which lead to dynamical signatures akin to discrete-time-crystalline order and exact periodic revivals. \\
	
	Since 1D PXP chains are naturally realized in the strong Rydberg blockade regime of Rydberg atom quantum simulator platforms, we investigated the effects of various forms of imperfections that can be present in an actual simulator. This includes dephasing of the atoms, spontaneous emission from the Rydberg excited atomic state and also protocol imperfections such as inexact many-body $\pi$-pulses. Our results indicate that the emergent conservation laws are stable against pure-dephasing but not against the consideration of the finite lifetime of the Rydberg-excited states. This conclusion was further supported by the robustness of the conservation laws for individual quantum trajectories. Moreover, we found that both the conservation laws and the quantum many-body scars are unstable in the presence of both dephasing and spontaneous emission from the Rydberg excited states.\\
	
	We then considered the effects of long-range vdW interactions between excited atoms. Our analysis suggested that the kinetic constrained imposed throughout this paper may be difficult to realize in an actual Rydberg atom simulator with only isotropic vdW repulsive interactions. However, since our analysis revealed that the kinetically constrained model contains several anomalous dynamical features, it would be worth searching for or engineering scenarios which mimics the exact constrained dynamics assumed. Although we only explored the connection of our model with Rydberg atom simulator platforms, it could also be interesting to explore other quantum simulator platforms to find a natural realization of the model considered. For example, recently a native, high-fidelity implementation of the 3-qubit $ZXZ$ gate has been realized in Ref.~\cite{xu2025paritycrossresonancemultiqubitgate} for superconducting transmon array platforms, which can be used to achieve digital simulation of constrained quantum dynamics, such as that of the PXP model, with fewer number of gates than shown in Ref.~\cite{Chen_PhysRevResearch.4.043027}.\\
	
	Thus, tuning the geometry \cite{Pal_PhysRevB.111.L161101}, dimensionality \cite{PhysRevX.10.021057_Zoller} and trying out different atomic species \cite{Anand2024_dual_species_Rydberg_array} by utilizing their hyperfine degrees of freedom, it may be possible to design and explore a broad class of interacting quantum many-body systems with novel phases and dynamical features. This may open up a gateway towards exploring minimal models of a rich variety of interacting quantum field theories and lattice gauge theories present ubiquitously in particle physics apart from those already observed in Refs.~\cite{Weimer2010,Realivistic_QFT_CIRAC_cold_atoms_PhysRevLett.105.190403,Lattice_gauge_theory_Dalmonte_PhysRevX.10.021041}.

	\section{Acknowledgments}
	The authors thank Arnab Sen, Immanuel Bloch, Vedika Khemani, Peter Zoller, Klaus M{\o}lmer and Shovan Dutta for stimulating discussions and insightful comments. The authors would also like to acknowledge valuable discussions in two consecutive programs held at ICTS-TIFR -- ``\textit{A Hundred Years of Quantum Mechanics}" (code: ICTS/qm100-2025/01) and ``\textit{Quantum Trajectories}" (code: ICTS/QuTr2025/01).

	\begin{appendix}
		
		\section{Symmetries of the Hamiltonian and Hilbert Space Dimensions}{\label{app-hsd}}
		
		In this section, we chart out the symmetries of the model \eqref{main:eq:hamiltonian_ladder}. By inspection we note that for $\Delta\ne0$, the Hamiltonian commutes with the operators $\hat{T}_x^2$ and $\hat{T}_y$. It is possible to construct a computational basis, in which the basis states are labeled by the eigenvalues of the symmetry operators $\hat{T}_x^2$ and $\hat{T}_y$ (see Ref.~\cite{Sandvik:2010lkj} for a review on construction of symmetry reduced Hilbert spaces). For reference, in Table \ref{tab:hsd_table}, we list the Hilbert space dimensions of some chosen sectors of these symmetry operators. The eigenvalues of $\hat{T_x^2}$ and $\hat{T_y}$ are denoted by $k_x$ and $k_{y}$ respectively in Table \ref{tab:hsd_table}.
		
		\begin{table}[!htpb]
			\centering
			\begin{tabular}{||c | c | c | c | c | c ||} 
				\hline
				$N$ & No Symmetry & $k_x=0$ & $k_{x,y}=0$ & $k_x=0,k_y=\pi$ & $k_{x,y}=\pi$ \\ [0.5ex] 
				\hline\hline
				4  & 7 & 7 & 5 & 2 & 0 \\
				\hline
				8 & 35 & 21 & 12 & 9 & 8 \\
				\hline
				12 & 199 & 71 & 36 & 35 & 32 \\
				\hline
				16 & 1,155 & 301 & 156 & 145 & 144 \\
				\hline
				20 & 6,727 & 1,351 & 676 & 675 & 672 \\ 
				\hline
				24 & 39,203 & 6,581 & 3,308 & 3,273 & 3,264 \\
				\hline 
				28 & 228,487 & 32,647 & 16,324 & 16,323 & 16,320 \\
				\hline 
				32 & 1,331,715 & 166,621 & 83,388 & 83,283 & 83,232 \\
				[1ex] 
				\hline
			\end{tabular}
			\caption{Hilbert space dimensions of \eqref{main:eq:hamiltonian_ladder} (with $\Delta\ne0$) for system size $N=4-32$ without any symmetry and for some specific symmetry sectors labeled by $k_x$ and $k_y$ (the eigenvalues of the symmetry operators $\hat{T}_x^2$ and $\hat{T}_y$ respectively).}
			\label{tab:hsd_table}
		\end{table}
		
		We note that the Hamiltonian \eqref{main:eq:hamiltonian_ladder} has additional lattice symmetries when $\Delta=0$. In this case the Hamiltonian \eqref{main:eq:hamiltonian_ladder} commutes with the operators $\hat{T}_x$, $\hat{T}_y$ and $\hat{P}_x$.
		
		\section{Entanglement Measures}
		
		\subsection{von Neumann entanglement entropy} \label{app:entanglement-entropy}
		
		The bi-partite von Neumann entanglement entropy \cite{Entanglement_RMP_Fazio_RevModPhys.80.517,Entanglement_general_Islam2015} for any (pure) quantum state $\ket{\psi}$, given the bi-partition $\mathcal{X},\mathcal{X}^c$ ($\mathcal{X} \cup \mathcal{X}^c = \mathcal{E}, \text{the entire system}$), is defined as,
		
		\begin{equation}
			S_{\text{ent}}^{\text{vN}}\left(\mathcal{X}\right) = -\text{Tr}_{\mathcal{X}}\left(\rho_{\mathcal{X}}\log\rho_{\mathcal{X}}\right)
			\label{eq:Sent_vN-defn}
		\end{equation}
		
		where $\rho_\mathcal{X}=\text{Tr}_{\mathcal{X}^c} \left(\ket{\psi}\bra{\psi}\right)$ is the reduced density matrix (RDM) of sub-system $\mathcal{X}$ with respect to its conjugate $\mathcal{X}^c$.\newline
		
		In the main text, we have presented results for two kinds of bi-partitions, namely (i) ``LR" (short for left-right) partition and (ii) ``UD" (short for up-down) partition. In the LR partition, $\mathcal{X}$ is either taken to be left half or the right half of the ladder. For setting a fixed convention, we take this bi-partition along the middle (in the sense of site indices) of the chain. For the UD partition, $\mathcal{X}$ is either the upper leg or the lower leg of the ladder.\newline
		
		For a kinetically constrained system like \eqref{main:eq:hamiltonian_ladder}, the Hilbert space does not have a direct product structure: for any bi-partition $\mathcal{X}$,$\mathcal{X}^c$ with $\mathcal{X} \cup \mathcal{X}^c = \mathcal{E}$, the full Hilbert space, say $\mathscr{H}_{\mathcal{E}}$, is not equivalent to the direct product of Hilbert spaces $\mathscr{H}_{\mathcal{X}}$ and $\mathscr{H}_{\mathcal{X}^c}$ i.e., $\mathscr{H}_{\mathcal{E}}$ $\neq$ $\mathscr{H}_{\mathcal{X}} \bigotimes \mathscr{H}_{\mathcal{X}^c}$ (both Hilbert spaces $\mathscr{H}_{\mathcal{X}}$ and $\mathscr{H}_{\mathcal{X}^c}$ are obtained by considering open boundary conditions).  In such a scenario, a quantum state $\ket{\psi}\in\mathscr{H}_{\mathcal{E}}$, can in principle be expressed in the product Hilbert space $\mathscr{H}_{\mathcal{X}} \bigotimes \mathscr{H}_{\mathcal{X}^c}$ as
		
		\begin{equation}
			\ket{\psi} = \sum_{\alpha=1}^{\mathcal{D}_{\mathcal{X}}} \sum_{\beta=1}^{\mathcal{D}_{\mathcal{X}^c}} \psi_{\alpha,\beta}^{\mathcal{X},\mathcal{X}^c}  \ket{\Phi^\mathcal{X}_\alpha} \otimes \ket{\Phi^{\mathcal{X}^c}_\beta}
			\label{eq:constrained_space_to_product_space}
		\end{equation}
		
		where $\{ \psi_{\alpha,\beta}^{\mathcal{X},\mathcal{X}^c}\}$ are complex numbers describing the state, $\{ \ket{\Phi^\mathcal{X}_\alpha}\},\{ \ket{\Phi^{\mathcal{X}^c}_\beta} \}$ are suitable basis for $\mathscr{H}_{\mathcal{X}}$ and $\mathscr{H}_{\mathcal{X}^c}$ respectively with Hilbert space dimensions $\mathcal{D}_\mathcal{X}$ and $\mathcal{D}_{\mathcal{X}^c}$. The Hilbert space constraint is taken care of by the additional condition that for indices $\alpha,\beta$, for which the state ket $\ket{\Phi^\mathcal{X}_\alpha} \otimes \ket{\Phi^{\mathcal{X}^c}_\beta} \notin \mathscr{H}_{\mathcal{E}}$, i.e. does not satisfy the Hilbert space constraint, we simply fix $\psi_{\alpha,\beta}^{\mathcal{X},\mathcal{X}^c}=0$. Once the state $\ket{\psi}$ has been written in the form of Eq.~\eqref{eq:constrained_space_to_product_space}, one can readily compute partial traces of $\ket{\psi}\bra{\psi}$ over the Hilbert space of $\mathcal{X}^c$ to obtain the sub-system RDM $\rho_{\mathcal{X}}$, required for computing the desired entanglement entropy via Eq.~\eqref{eq:Sent_vN-defn}. \newline
		
		\begin{figure}[!htpb]
			\centering
			\includegraphics[width=0.5\textwidth]{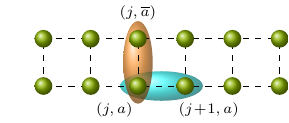}
			\caption{Schematic illustration of the mutual information computation for the horizontal (cyan) and vertical bonds (orange). Also see Eq.~\eqref{eq:mutual_info_defn} and text below it.}
            \label{app:fig:mutual_info_schematic}
		\end{figure}
		
		\subsection{Mutual Information} \label{app:mutual_info}
		
		In Sec.~\ref{subsec:emergent_integrability}, we also considered the evolution of mutual information between two sites. In general, the mutual information between any two subsystems, $\mathcal{X}$ and $\mathcal{Y}$ is defined as 
		
		\begin{equation}
			I(\mathcal{X},\mathcal{Y}) = S_{\text{ent}}^{\text{vN}}(\mathcal{X}) + S_{\text{ent}}^{\text{vN}}(\mathcal{Y}) - S_{\text{ent}}^{\text{vN}}(\mathcal{X}\cup\mathcal{Y})
			\label{eq:mutual_info_defn}
		\end{equation}
		
		This quantity $I(\mathcal{X},\mathcal{Y})$ for any state, for any bi-partition $\mathcal{X},\mathcal{Y}$, describes the total amount of correlation shared between $\mathcal{X}$ and $\mathcal{Y}$, which includes both classical and quantum correlations \cite{Mutual_information_PhysRevA.72.032317,Entanglement_vs_correlations_mutual_information_CIRAC_PhysRevLett.92.027901}. From a quantum information point of view, the mutual information encodes the amount of information about the sub-system $\mathcal{X}$ that can be obtained from the sub-system $\mathcal{Y}$. In this case, it is not necessary that $\mathcal{X}\cup\mathcal{Y}=\mathcal{E}$. For the specific results displayed in Sec.~\ref{subsec:emergent_integrability}, i.e. the mutual information of the horizontal bonds we take $\mathcal{X}=\{\left(j,a\right)\}$, $\mathcal{Y}=\{\left(j+1,a\right)\}$ and for vertical bonds we take $\mathcal{X}=\{\left(j,a\right)\}$, $\mathcal{Y}=\{\left(j,\overline{a}\right)\}$ in Eq.~\eqref{eq:mutual_info_defn} with $j=1,a=1$ (see Fig.~\ref{app:fig:mutual_info_schematic}).\newline
		
		\section{Effective Hamiltonian for $\Delta\gg1$\label{app:Heff_SW}}
		
		At $\Delta=\infty$, the spectrum of \eqref{main:eq:hamiltonian_ladder} is known exactly: all computational basis states (i.e. Fock states in the $\hat{\sigma}^z$ basis) are eigenstates of the Hamiltonian. To quantitatively study the effect of this integrable point near $\Delta\sim 1$, we construct a Schrieffer-Wolff (SW) rotation of the basis which shall make the effective Hamiltonian block diagonal in the eigenbasis of the diagonal part of the Hamiltonian $\hat{H}_z$ up to first order in $1/\Delta$. To this end we consider the generator of the SW rotation $i\hat{S}$, which satisfies $[i\hat{S},\hat{H}_z] =-\hat{H}_x$ and is of the following form
		
		\begin{equation}
			i\hat{S} = \frac{1}{2i\Delta} \sum_{j=1}^L \sum_{a=1}^2 
			(-1)^j \hat{\tilde\sigma}^y_{j,a}
			\label{eq:SWgenerator}
		\end{equation}

		The effective low-energy Hamiltonian, in the appropriate rotated basis reads
		
		\begin{equation}
			\begin{split}
				\hat{H}_{\text{eff}} & = \hat{P}_{Z_{\pi}} \left( e^{i\hat{S}}\hat{H} 
				e^{-i\hat{S}} \right) \hat{P}_{Z_{\pi}} \\
				& = \hat{H}_\text{eff}^{\left(0\right)} + \hat{H}_\text{eff}^{\left(1\right)} + \hat{H}_\text{eff}^{\left(2\right)}+\hat{H}_\text{eff}^{\left(3\right)} + ... 
			\end{split}
		\end{equation}
		
		Where
		
		\begin{equation}
			\begin{split}
				\hat{H}_\text{eff}^{\left(0\right)} & = \hat{P}_{Z_{\pi}}\hat{H}_z\hat{P}_{Z_{\pi}}, \hat{H}_\text{eff}^{\left(1\right)} = 0 \\
				\hat{H}_\text{eff}^{\left(2\right)} & = \hat{P}_{Z_{\pi}}\left(\frac{1}{2}\left[i\hat{S},\hat{H}_x\right]\right)\hat{P}_{Z_{\pi}} \\
				\hat{H}_\text{eff}^{\left(3\right)} & = \hat{P}_{Z_{\pi}}\left(\frac{1}{3}\left[i\hat{S},\left[i\hat{S},\hat{H}_x\right]\right]\right)\hat{P}_{Z_{\pi}} = 0\\
				\hat{H}_\text{eff}^{\left(4\right)} & = \hat{P}_{Z_{\pi}}\left(\frac{1}{8}\left[i\hat{S}\left[i\hat{S},\left[i\hat{S},\hat{H}_x\right]\right]\right]\right)\hat{P}_{Z_{\pi}}
			\end{split}
			\label{app:eq:Heff_formulas}
		\end{equation}
		
		In the above perturbation theory, all odd-order terms vanish and only even-order terms have a non-trivial contribution. By direct computation of the commutators, one arrives at the expression for $\hat{H}^{\left(2\right)}_\text{eff}$ (Eq.~\eqref{eq:Heff_SW}) \\
		
		At second order, the form of the effective Hamiltonian allows us to understand dominant physical processes taking place in the quench dynamics of the system. At this order, a Fock state is either completely frozen, or it undergoes Rabi oscillations induced via the $XX+YY$ terms. This effective Hamiltonian also suggests the consideration of a number of ``\textit{emergent}" constants of motion for a system with $2L$ degrees of freedom. It is worth mentioning that, for a deeper understanding, it should be explored whether one can explicitly cast the effective Hamiltonian up to second order in a Bethe integrable form. The second-order effective Hamiltonian also gives rise to slow dynamics due to existence of the following extensive number of conservation laws (see Eqs.~\eqref{eq:main:conservation_laws_Zpi},~\eqref{eq:main:conservation_laws_Qj}). Considering the conservation of the total energy, we have $L+2$ number of constants of motion, which suggests that the effective Hamiltonian (up to second order) is integrable, and it is possible to write down explicitly the eigenvalues and eigenvectors of $\hat{H}^{[2]}_{\text{eff}} = \hat{H}^{(0)}_{\text{eff}}+\hat{H}^{(2)}_{\text{eff}}$ in each charge sector labeled by the quantum numbers $\{z_{\pi},q_1,q_2,...,q_L\}$. In the following section, we present an outline for the process of enumeration of all the eigenvectors and the eigenvalues of $\hat{H}^{[2]}_{\text{eff}}$.
		
		\subsection{Full Spectrum of $\hat{H}^{[2]}_{\text{eff}}$ \label{subsec:Heff2spectrum}}
		
		First, we note by inspection that the vacuum state $\ket{\text{vac.}}$ is annihilated by $\hat{H}^{[2]}_{\text{eff}}$, the effective Hamiltonian up to second order i.e. $\hat{H}^{[2]}_{\text{eff}}=\hat{H}^{(0)}_{\text{eff}}+\hat{H}^{(2)}_{\text{eff}}$. Now, we consider adding particles on top of the $\ket{\text{vac.}}$ state by forming singlets/triplets of the following form
		
		\begin{equation}
			\ket{j,\pm}_{\text{1P}} = \frac{1}{\sqrt{2}}\left( \ket{\substack{...\bullet_{j}...\\...\circ_{j}...}} \pm \ket{\substack{...\circ_{j}...\\...\bullet_{j}...}} \right)
		\end{equation}
		
		where for brevity of notation, the dots ``$...$" within the symbol $\ket{}$, imply absence of any Rydberg excitations/particles. We now introduce the following notation,
		
		\begin{eqnarray*}
			\mathbf{0}_j \equiv \ket{\substack{\circ_j\\\circ_j}}, S_j^{(1)} \equiv \ket{\substack{\bullet_j\\\circ_j}}, S_j^{(2)} \equiv \ket{\substack{\circ_j\\\bullet_j}} \\
			\mathcal{N}_{j,2m}^{(a)} \equiv \bigotimes_{p=j}^{j+2(m-1) \; \text{step} \;2}  S_{p}^{(a)} S_{p+1}^{(\bar{a})} ,\\
			\mathcal{N}_{j,2m+1}^{(a)} \equiv \mathcal{N}_{j,2m}^{(a)} S_{j+2m}^{(a)}, 
		\end{eqnarray*}
		
		with $a=1,2$ and $m \in \mathbb{Z}^+$ are positive integers. Recall that $\bar{a}=1$ when $a=2$ and vice-versa. Note that the action of $\hat{H}_{\text{eff}}^{(2)}$ on states made out of blocks of $\mathbf{0}$, $\mathcal{N}$ and $S$ are easy to follow. Both $\mathbf{0}$, $\mathcal{N}$ remain frozen under $\hat{H}_{\text{eff}}^{(2)}$, while $S_{j}^{(a)}$ flips into $S_{j}^{(\bar{a})}$ up to an overall factor. Using the above notation, we can express $\ket{j,\pm}_{\text{1P}}$ as,
		
		\begin{equation}
			\ket{j,\pm}_{\text{1P}} =  \frac{1}{\sqrt{2}} \left(\otimes_{k<j} \mathbf{0}_k  \right) (S_j^{(1)} \pm S_j^{(2)}) \left(\otimes_{k>j} \mathbf{0}_k \right)
		\end{equation}
		
		The states $\ket{j,\pm}_{\text{1P}},\forall j=1,2,...,L$ are exact eigenstates of $\hat{H}^{[2]}_{\text{eff}}$ with eigenvalue $\pm2\Delta-3/4\Delta$. Similarly, one can construct two particle states on top of the $\ket{\text{vac.}}$ state, with particles placed on neighboring sites (on alternate rungs) in the following way,
		
		\begin{eqnarray}
			\ket{j,1}_{\text{2P}} & = & \ket{\substack{...\bullet\circ_{j}...\\...\circ\bullet_{j}...}} = \left(\otimes_{k<j} \mathbf{0}_k  \right) \mathcal{N}_{j,2}^{(1)} \left(\otimes_{k>j} \mathbf{0}_k  \right) \nonumber \\
			\ket{j,2}_{\text{2P}} & = & \ket{\substack{...\circ\bullet_{j}...\\...\bullet\circ_{j}...}} = \left(\otimes_{k<j} \mathbf{0}_k  \right) \mathcal{N}_{j,2}^{(2)} \left(\otimes_{k>j} \mathbf{0}_k  \right)
		\end{eqnarray}
		
		which are zero modes of $\hat{H}^{[2]}_{\text{eff}}$, $\forall j=1,2,...,L$. This construction can be generalized to many contiguous sites as follows, 
		
		\begin{eqnarray}
			\ket{j,1}_{\text{3P}} & = & \ket{\substack{...\bullet\circ\bullet_{j}...\\...\circ\bullet\circ_{j}...}}  = \left(\otimes_{k<j} \mathbf{0}_k  \right) \mathcal{N}_{j,3}^{(1)} \left(\otimes_{k>j} \mathbf{0}_k  \right) \nonumber \\
			\ket{j,2}_{\text{3P}} & = & \ket{\substack{...\circ\bullet\circ_{j}...\\...\bullet\circ\bullet_{j}...}} = \left(\otimes_{k<j} \mathbf{0}_k  \right) \mathcal{N}_{j,3}^{(2)} \left(\otimes_{k>j} \mathbf{0}_k  \right)
		\end{eqnarray}
		
		Again the states, $\ket{j,1}_{\text{3P}}$ and $\ket{j,2}_{\text{3P}}$, $\forall j=1,2,...,L$ are eigenvectors of $\hat{H}^{[2]}_\text{eff}$ and the energy eigenvalue has to be determined by inspection. This procedure can be continued to yield more eigenvectors of $\hat{H}^{[2]}_\text{eff}$, until every other site in the ladder is filled, which leaves us with two eigenvectors $\ket{\mathbb{Z}_2}$ and $\ket{\overline{\mathbb{Z}}_2}$. Alternatively, one may also put non-contiguous excitations, which will again be exact eigenstates of the effective Hamiltonian. For example, two such $N_\uparrow$-particle states with even $N_\uparrow$ can be written as follows,
		
		\begin{equation}
			\begin{split}
				& \ket{\left(j_1,1\right),(j_2,\pm)}_{N_\uparrow\text{-P}}  =  \frac{1}{\sqrt{2}}\left(\ket{\substack{...\bullet_{j_1}\circ\bullet...\bullet_{j_2}...\\...\circ_{j_1}\bullet\circ...\circ_{j_2}...}} \pm \ket{\substack{...\bullet_{j_1}\circ\bullet...\circ_{j_2}...\\...\circ_{j_1}\bullet\circ...\bullet_{j_2}...}} \right) \\
				& = \left(\otimes_{k<j_1} \mathbf{0}_k  \right) \mathcal{N}_{j_1,2m+1}^{(1)} (\otimes_{k>j_1+2m}^{j_2-1} \mathbf{0}_k  ) (S_{j_2}^{(1)} \pm S_{j_2}^{(2)}) \left(\otimes_{k>j_2} \mathbf{0}_k  \right)
			\end{split}
		\end{equation}
		\begin{equation}
			\begin{split}
				& \ket{\left(j_1,2\right),(j_2,\pm)}_{N_\uparrow\text{-P}} = \frac{1}{\sqrt{2}} \left( \ket{\substack{...\circ_{j_1}\bullet\circ...\bullet_{j_2}...\\...\bullet_{j_1}\circ\bullet...\circ_{j_2}...}} \pm \ket{\substack{...\circ_{j_1}\bullet\circ...\circ_{j_2}...\\...\bullet_{j_1}\circ\bullet...\bullet_{j_2}...}}  \right)  \\ 
				& = \left(\otimes_{k<j_1} \mathbf{0}_k  \right) \mathcal{N}_{j_1,2m+1}^{(2)} (\otimes_{k>j_1+2m}^{j_2-1} \mathbf{0}_k  ) (S_{j_2}^{(1)} \pm S_{j_2}^{(2)}) \left(\otimes_{k>j_2} \mathbf{0}_k  \right)
			\end{split}
		\end{equation}

		with $m=(N_\uparrow-2)/2$, and the condition that the block of contiguous spins based to the right of site $j_1$ have at least one insertion of $\ket{\substack{\circ\\\circ}}$ between the excitations placed at site $j_2$. Additionally, further non-zero modes of $\hat{H}^{[2]}_{\text{eff}}$ can be constructed by placing ``Bell-pair" like states which are separated by at least one insertion of $\ket{\substack{\circ\\\circ}}$ in between. Following these simple rules one may enumerate all eigenvectors of $\hat{H}^{[2]}_{\text{eff}}$.

		\subsection{Dynamics under $\hat{H}_{\text{eff}}^{[2]}$}
		\label{app:mutual_information_dynamics}
		
		The effective Hamiltonian $\hat{H}_{\text{eff}}^{[2]}$ represents the approximate quantum dynamics in a basis rotated by the SW generator $i\hat{S}$. Thus we need to evolve a rotated initial state $\ket{\tilde{\psi}(0)}$ via $\hat{H}_{\text{eff}}^{[2]}$ and rotate it back to the original basis to obtain $\ket{\psi(t)}$. When the SW rotation is perturbative, we can approximate the quantum state $\ket{\psi(t)}$ in the following manner (up to normalization).
		
		\begin{equation}
			\ket{\psi(t)} \approxeq (1-i\hat{S}) e^{-i\hat{H}_{\text{eff}}^{[2]}t} (1+i\hat{S}) \ket{\psi(0)}
		\end{equation}
		
\begin{figure}
			\centering
			\includegraphics[width=0.485\textwidth]{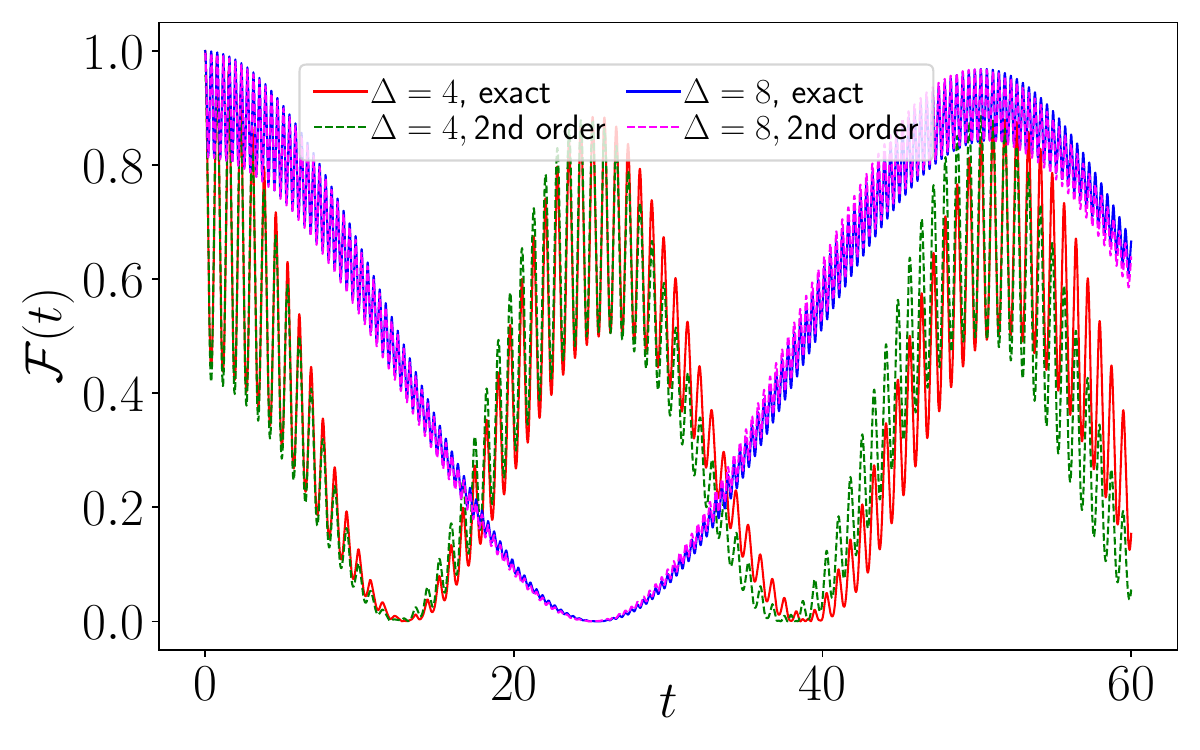}
			\caption{Return probability starting from the $\ket{\text{1P}}$ under the full Hamiltonian $\hat{H}$ and the second order effective Hamiltonian $\hat{H}^{[2]}_\text{eff}$ for $L=8$. As can be seen from this figure, $\mathcal{F}(t)$ contains a low-frequency component whose amplitude is independent of $\Delta$ and a high-frequency component whose amplitude is inversely proportional to $\Delta^2$ as suggested by Eq.~\eqref{app:eq:fid_1P_Heff2}}
			\label{app:fig:Heff2_1P_fid}
		\end{figure}

		We note here that since the rotated initial state $(1+i\hat{S}) \ket{\psi(0)}$ has a small but finite overlap between different $\hat{Z}_\pi$ sectors, we do not take into account the projectors $\hat{P}_{Z_\pi}$ when considering the effective Hamiltonian for real-time dynamics. This gives rise to the correct qualitative picture by capturing the high-frequency corrections to the dynamics due to hybridization of different $\hat{Z}_\pi$ sectors (see Fig.~\ref{app:fig:Heff2_1P_fid}).\\
		
		Considering the above approximate dynamics for a two-leg ladder with $L=2$ gives an understanding of the low- and high-frequency oscillations present in Fig.~\ref{fig:mutual_info_dynamics}. To this end, we consider the initial state in the original basis, $\ket{\psi(0)}=\ket{\text{1P}}=\ket{\substack{\bullet\circ\\\circ\circ}}$ for which we have the following
		
		\begin{equation}
			(1+i\hat{S}) \ket{\psi(0)} = \sum_{\mu=1}^3 c_\mu \ket{\phi_\mu}
		\end{equation}
		
		with $\ket{\phi_1}=\ket{\substack{\circ\circ\\\circ\circ}}$, $\ket{\phi_2}=\ket{\substack{\bullet\circ\\\circ\bullet}}$, $\ket{\phi_3}=\ket{\substack{\bullet\circ\\\circ\circ}}$, and real coefficients $c_1=c_2=-1/2\Delta$, $c_3=1$. Let us further introduce the states $\ket{\psi_\pm}=\frac{1}{\sqrt{2}}\left( \ket{\substack{\bullet\circ\\\circ\circ}} \pm \ket{\substack{\circ\circ\\\bullet\circ}} \right)$ for which $\hat{H}_{\text{eff}}^{[2]}\ket{\psi_\pm}=\mathcal{E}_\pm\ket{\psi_\pm}$, with $\mathcal{E}_\pm=2\Delta\pm1/2\Delta$. We also note that the states $\ket{\phi_1}$ and $\ket{\phi_2}$ are annihilated by $\hat{H}_{\text{eff}}^{[2]}$. This implies that the approximate quantum state $\ket{\tilde\psi(t)}$ in the rotated basis reads,
		
		\begin{equation}
			\ket{\tilde\psi(t)} \approxeq c_1\ket{\phi_1} + c_2\ket{\phi_2} + \frac{c_3}{\sqrt{2}} \left(e^{-i\mathcal{E}_+t}\ket{\psi_+}+e^{-i\mathcal{E}_-t}\ket{\psi_-}\right)
		\end{equation}
		
		Rotating back to the original basis, we find that the overlap with the initial state $\mathcal{F}(t)=|\langle \psi(0) | \psi(t) \rangle|^2$ is of the following form,
		
		\begin{equation}
			\begin{split}
				\mathcal{F}(t) & = |A+Be^{-i\mathcal{E}_+t}+Be^{-i\mathcal{E}_-t}|^2 \\
				& = A^2+2B^2 (1+\cos(\mathcal{E}_+-\mathcal{E}_-)t) \\
				& + 4AB\cos(\frac{\mathcal{E}_++\mathcal{E}_-}{2}t)\cos(\frac{\mathcal{E}_+-\mathcal{E}_-}{2}t)
			\end{split}
			\label{app:eq:fid_1P_Heff2}
		\end{equation} 
		
		where $A=-(c_2+c_1)/2\Delta$, $B=c_3/2$. Now since, $\mathcal{E}_++\mathcal{E}_- = 4\Delta$, $\mathcal{E}_+-\mathcal{E}_- = 1/\Delta$, $c_{1,2} \propto 1/\Delta$, and $A \propto 1/\Delta^2$, we see that the dynamics from the $\ket{1\text{P}}$ state under $\hat{H}^{[2]}_\text{eff}$ for $\Delta \gg 1$, contains a low-frequency term (second term in Eq.~\eqref{app:eq:fid_1P_Heff2}) whose amplitude is independent of $\Delta$ and a high-frequency term (third term in Eq.~\eqref{app:eq:fid_1P_Heff2}) whose amplitude decreases with increasing $\Delta$. This is consistent with our numerical results as illustrated in Fig.~\ref{app:fig:Heff2_1P_fid}. This analysis also explains qualitatively the presence of both low- and high-frequency oscillations in the mutual information dynamics (Fig.~\ref{fig:mutual_info_dynamics}).
		
		\subsection{Integrability breaking effects from $H_{\text{eff}}^{[4]}$} \label{app:Heff4_integrability_breaking}
		
		\begin{figure*}[!tpb]
			\rotatebox{0}{\includegraphics[width=0.985\textwidth]{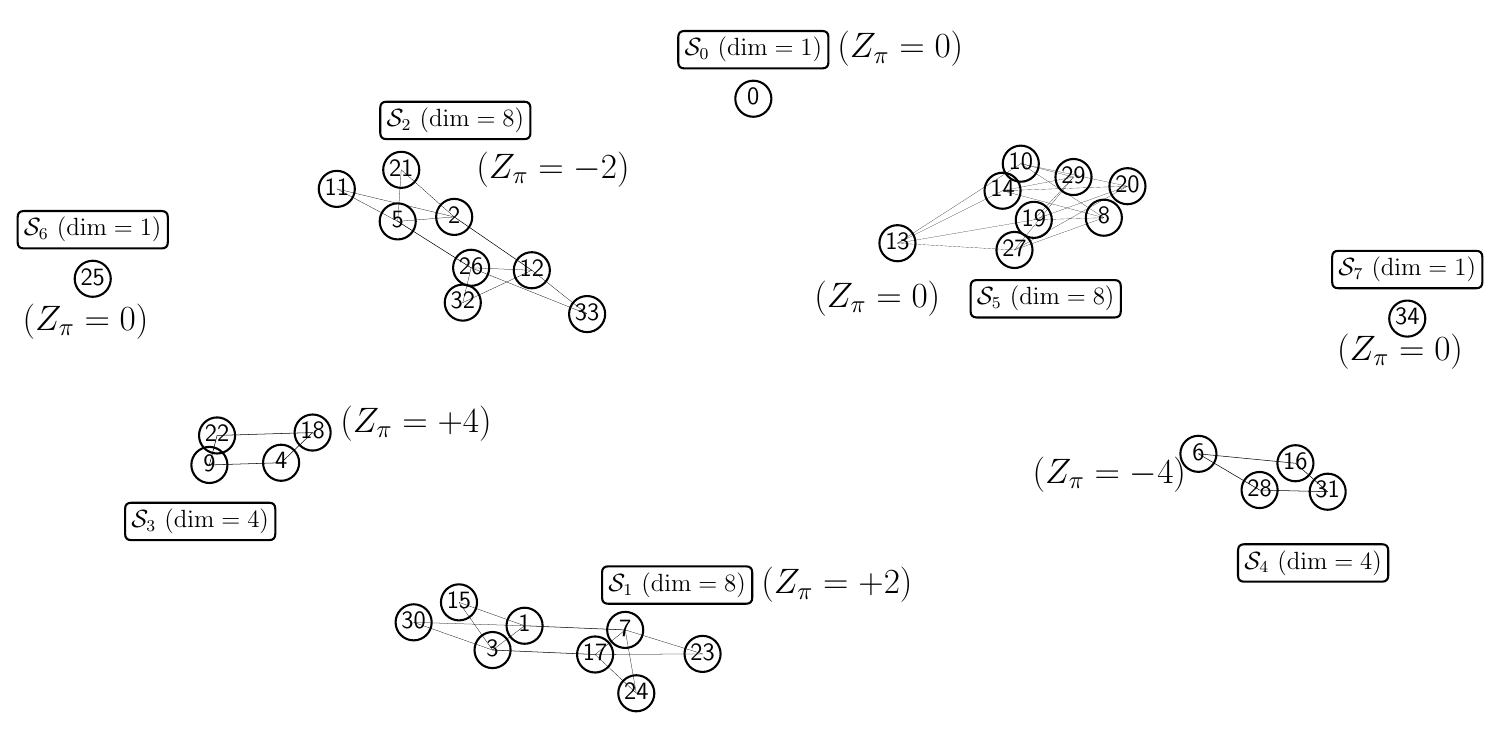}}
			\caption{Connectivity diagram of the basis states with $N=8$ sites under $\hat{H}_\text{eff}^{[4]}$. The basis states are labelled as 0-34 which correspond to the Fock states given in Table-\ref{app:table:Hilbert_Space_nodes_to_Fock_states_list}.}
			\label{app:fig:Heff4_conn_N_08}
		\end{figure*}
		
		The fourth-order corrections to the effective Hamiltonian is given by Eq.~\eqref{app:eq:Heff_formulas}. Since we explicitly know the generator $i\hat{S}$, and thus its matrix elements in the computational basis (the constrained Fock basis with no nearest neighbor excitations), we can readily construct $\hat{H}_\text{eff}^{[4]}$ in this basis and study its properties. In what follows, we illustrate the connectivity of the Fock states under $\hat{H}_\text{eff}^{[4]}$=$\hat{H}_\text{eff}^{(0)}$+$\hat{H}_\text{eff}^{(2)}$+$\hat{H}_\text{eff}^{(4)}$. First, we note that, similar to $\hat{H}_\text{eff}^{(2)}$, $\hat{H}_\text{eff}^{(4)}$ also annihilates certain Fock states, such as the $\ket{\mathbb{Z}_2}$ and $\ket{\overline{\mathbb{Z}}_2}$. However, unlike $\hat{H}_\text{eff}^{(2)}$, $\hat{H}_\text{eff}^{(4)}$ contains spin-flip processes which connects different particle number states within a fixed $\hat{Z}_\pi$ sector. This is illustrated in Fig.~\ref{app:fig:Heff4_conn_N_08} where we show the connectivity diagram of the entire Hilbert space for $N=8$ sites under $\hat{H}_\text{eff}^{[4]}$. For this system size, there are 35 states which satisfy the nearest neighbour strong Rydberg blockade constraint. These states are enumerated in Table-\ref{app:table:Hilbert_Space_nodes_to_Fock_states_list}. From the connectivity it is clear that even when we consider fourth order terms in the effective Hamiltonian (see Eq.~\eqref{app:eq:Heff_formulas}), Fock states belonging to the same $Z_\pi$ sector may not be connected to each other. Notably, the $Z_\pi=0$ sector has a total of 11, states 3 of which, namely, $\ket{\substack{\circ \circ \circ \circ\\\circ \circ \circ \circ}}$, $\ket{\substack{\circ \bullet \circ \bullet\\\bullet \circ \bullet \circ}}$ and $\ket{\substack{\bullet \circ \bullet \circ\\\circ \bullet \circ \bullet}}$ are isolated. This in part explains the higher stability of the $\ket{\text{vac.}}$ and $\ket{\mathbb{Z}_2}$ states when compared to the other Fock states in Fig.~\ref{fig:slow_dynamics} of the main text. Next, focusing on the sector $\mathcal{S}_1 (Z_\pi=2)$ we see that along with terms coming from $\hat{H}_\text{eff}^{[2]}$ such as $\ket{\substack{\bullet \circ \circ \circ\\\circ \circ \circ \circ}}$ $\leftrightarrow$ $\ket{\substack{\circ \circ \circ \circ\\\bullet \circ \circ \circ}}$, there are additional long-range spin-flip terms like $\ket{\substack{\bullet \circ \circ \circ\\\circ \circ \circ \circ}}$ $\leftrightarrow$ $\ket{\substack{\circ \circ \bullet \circ\\\circ \circ \circ \circ}}$ and also Rydberg excitation number non-conserving terms such as $\ket{\substack{\bullet \circ \circ \circ\\\circ \circ \circ \circ}}$ $\leftrightarrow$ $\ket{\substack{\bullet \circ \bullet \circ\\\circ \bullet \circ \circ}}$ arising from $\hat{H}_{\text{eff}}^{(4)}$, although they are still much weaker compared to the second order processes.
		
		\begin{table}
			\begin{tabular}{c l | c l | c l}
				\hline
				Node & Fock state & Node & Fock state & Node & Fock state \\
				\hline
				0 & $\ket{\substack{\circ \circ \circ \circ\\\circ \circ \circ \circ}}$ &
				1 & $\ket{\substack{\bullet \circ \circ \circ\\\circ \circ \circ \circ}}$ &
				2 & $\ket{\substack{\circ \bullet \circ \circ\\\circ \circ \circ \circ}}$ \\
				
				3 & $\ket{\substack{\circ \circ \bullet \circ\\\circ \circ \circ \circ}}$ &
				4 & $\ket{\substack{\bullet \circ \bullet \circ\\\circ \circ \circ \circ}}$ &
				5 & $\ket{\substack{\circ \circ \circ \bullet\\\circ \circ \circ \circ}}$ \\
				
				6 & $\ket{\substack{\circ \bullet \circ \bullet\\\circ \circ \circ \circ}}$ &
				7 & $\ket{\substack{\circ \circ \circ \circ\\\bullet \circ \circ \circ}}$ &
				8 & $\ket{\substack{\circ \bullet \circ \circ\\\bullet \circ \circ \circ}}$ \\
				
				9 & $\ket{\substack{\circ \circ \bullet \circ\\\bullet \circ \circ \circ}}$ &
				10 & $\ket{\substack{\circ \circ \circ \bullet\\\bullet \circ \circ \circ}}$ &
				11 & $\ket{\substack{\circ \bullet \circ \bullet\\\bullet \circ \circ \circ}}$ \\
				
				12 & $\ket{\substack{\circ \circ \circ \circ\\\circ \bullet \circ \circ}}$ &
				13 & $\ket{\substack{\bullet \circ \circ \circ\\\circ \bullet \circ \circ}}$ &
				14 & $\ket{\substack{\circ \circ \bullet \circ\\\circ \bullet \circ \circ}}$ \\
				
				15 & $\ket{\substack{\bullet \circ \bullet \circ\\\circ \bullet \circ \circ}}$ &
				16 & $\ket{\substack{\circ \circ \circ \bullet\\\circ \bullet \circ \circ}}$ &
				17 & $\ket{\substack{\circ \circ \circ \circ\\\circ \circ \bullet \circ}}$ \\
				
				18 & $\ket{\substack{\bullet \circ \circ \circ\\\circ \circ \bullet \circ}}$ &
				19 & $\ket{\substack{\circ \bullet \circ \circ\\\circ \circ \bullet \circ}}$ &
				20 & $\ket{\substack{\circ \circ \circ \bullet\\\circ \circ \bullet \circ}}$ \\
				
				21 & $\ket{\substack{\circ \bullet \circ \bullet\\\circ \circ \bullet \circ}}$ &
				22 & $\ket{\substack{\circ \circ \circ \circ\\\bullet \circ \bullet \circ}}$ &
				23 & $\ket{\substack{\circ \bullet \circ \circ\\\bullet \circ \bullet \circ}}$ \\
				
				24 & $\ket{\substack{\circ \circ \circ \bullet\\\bullet \circ \bullet \circ}}$ &
				25 & $\ket{\substack{\circ \bullet \circ \bullet\\\bullet \circ \bullet \circ}}$ &
				26 & $\ket{\substack{\circ \circ \circ \circ\\\circ \circ \circ \bullet}}$ \\
				
				27 & $\ket{\substack{\bullet \circ \circ \circ\\\circ \circ \circ \bullet}}$ &
				28 & $\ket{\substack{\circ \bullet \circ \circ\\\circ \circ \circ \bullet}}$ &
				29 & $\ket{\substack{\circ \circ \bullet \circ\\\circ \circ \circ \bullet}}$ \\
				
				30 & $\ket{\substack{\bullet \circ \bullet \circ\\\circ \circ \circ \bullet}}$ &
				31 & $\ket{\substack{\circ \circ \circ \circ\\\circ \bullet \circ \bullet}}$ &
				32 & $\ket{\substack{\bullet \circ \circ \circ\\\circ \bullet \circ \bullet}}$ \\
				
				33 & $\ket{\substack{\circ \circ \bullet \circ\\\circ \bullet \circ \bullet}}$ &
				34 & $\ket{\substack{\bullet \circ \bullet \circ\\\circ \bullet \circ \bullet}}$ &
				& \\
				\hline
			\end{tabular}
			\caption{Mapping of the node index to the allowed Fock states in a $N=8$ site system used in Fig.~\ref{app:fig:Heff4_conn_N_08} to show the Hilbert space connectivity under $\hat{H}_\text{eff}^{[4]}$.}
			\label{app:table:Hilbert_Space_nodes_to_Fock_states_list}
		\end{table}
		
		\subsection{Storing classical bits of information \label{app:classical_bit_storage}}
		
		Utilizing the slow dynamics induced by the extensive number of approximate conservation laws, it is possible to store classical information, encoded via a class of appropriately chosen initial states which can be labeled uniquely in terms of the eigenvalues of ``$L$"-quasi conserved charges in a square ladder with $N=2L$ Rydberg atoms. This stored information can be retrieved at later times by performing simple projective readout measurements. To see how this can be achieved, consider a string of $L$-classical bits $\left(c_1,c_2,...,c_{L}\right)$ where $c_i\in[-1,+1]$ $\forall i=1,2,...,L$, $i$ being the rung index. Now, this classical string of information can be encoded as initial values of the approximately conserved local charges $q_j$'s at all rungs as $q_{j}=c_j$. As an example, in a ladder with $N=2L=12$ atoms, one can encode the classical information ``bit-string" $\left(+1,+1,+1,-1,+1,-1,+1\right)$ in the initial quantum state $\ket{\psi(0)}=\ket{\substack{\circ\circ\bullet\circ\circ\circ\\\circ\circ\circ\circ\bullet\circ}}$. This choice is not unique and the states $\ket{\substack{\circ\circ\bullet\circ\bullet\circ\\\circ\circ\circ\circ\circ\circ}}$,$\ket{\substack{\circ\circ\circ\circ\circ\circ\\\circ\circ\bullet\circ\bullet\circ}}$,$\ket{\substack{\circ\circ\circ\circ\bullet\circ\\\circ\circ\bullet\circ\circ\circ}}$ are completely equivalent. In the course of full quantum dynamics via the Hamiltonian \eqref{main:eq:hamiltonian_ladder} for $\Delta \gg 1$, the expectation values of all $\hat{Q}_j$'s stay almost frozen, and more importantly do not change their sign, for times as large as $t\sim10^4$, which is much beyond the time to which coherence can be maintained in experimental platforms. By a simple readout procedure, one can determine the sign of $\langle \hat{Q}_{j}\rangle_t$ at any time $t$ during the experiment. By looking at this string of information, one can recover the encoded classical string $\left(c_1,...,c_{L}\right)$ as
		
		\begin{equation}
			c_i = \frac{\langle \hat{Q}_{i} \rangle_t }{|\langle \hat{Q}_{i} \rangle_t|} \quad \forall i=1,2,...,L
		\end{equation}
		
		From the structure of $H_{\text{eff}}^{(2)}$ (Eq.~\eqref{eq:Heff_SW}), it is easy to interpret the form of the initial Fock state required to encode the bit-string $\left(c_1,...,c_{L}\right)$. The study of quantum trajectories presented in the main text also suggests that the numbers $(c_1,....,c_L)$ are stable against decoherence mechanisms (see Fig.~\ref{fig:dephasing-QT-conservation-stability}).
		
		\subsection{Ensemble of relaxation: Gibbs vs Generalized Gibbs \label{app:GGE}}
		
		Under unitary quantum evolution via $\hat{H}^{[2]}_{\text{eff}}$, the system preserves the following operators: $\hat{H}^{[2]}_{\text{eff}}$, $\hat{Z_\pi}$ and $\hat{Q}_j$ $\forall j=1,2,...,L$. This implies, that if the system evolves purely via $\hat{H}^{[2]}_{\text{eff}}$, at late times it should relax to the generalized Gibbs ensemble (GGE) described by appropriate Lagrange multipliers associated with these conservations laws (exact or approximate), 
		
		\begin{eqnarray}
			\hspace{-5mm}\hat{\rho}_{\text{GGE}} = \frac{1}{Z_{\text{GGE}}} \exp\!\!\left(\!-\beta\hat{H}^{[2]}_{\text{eff}}-\lambda_\pi\hat{Z}_\pi-\sum_{j=1}^L \lambda_j \hat{Q}_j\!\right)
		\end{eqnarray}
		
		where $Z_{\text{GGE}}$ is a normalization factor which ensures that $\text{Tr}\left(\hat{\rho}_{\text{GGE}}\right)=1$. On the other hand, as the additional extensive number of conservation laws are only approximate, the system (which evolves via the Hamiltonian \eqref{main:eq:hamiltonian_ladder}) should relax (at very late times set by the fourth order effective Hamiltonian) to the Gibbs ensemble 
		
		\begin{equation}
			\hat{\rho}_{\text{GE}} = \frac{1}{Z_{\text{GE}}} \exp\left(-\beta\hat{H}\right)
		\end{equation}
		
		where $Z_{\text{GE}}$ is a normalization constant such that $\text{Tr}\left(\hat{\rho}_{\text{GE}}\right)=1$. For both cases, the relevant Lagrange multipliers i.e. $\{\beta,\lambda_{\pi}, \lambda_{1,2,...,L}\}$ for GGE and $\beta$ for GE, are determined by requiring the equality of the initial values of the conserved charges to the expectation values of conserved operators with respect to this ensemble. In practice, we have done this by minimizing the related cost functions $f_{\text{GGE}}(\{\beta,\lambda_{\pi}, \lambda_{1,2,...,L}\})$ and $f_{\text{GE}}(\beta)$ using numerical solvers \cite{2020SciPy-NMeth}
		
		\begin{equation}
			f_{\text{GE}}(\beta) = \left[ \text{Tr}\left(\hat{\rho}_{\text{GE}}\hat{H}\right)-\text{Tr}\left(\hat{\rho}_{0}\hat{H}\right)\right]^2
		\end{equation}
		
		\begin{equation}
			f_{\text{GGE}}(\{\beta,\lambda_{\pi}, \vec{\lambda}\}) = \sum_{\eta} \left[ \text{Tr}\left(\hat{\rho}_{\text{GGE}}\hat{K}_\eta\right)-\text{Tr}\left(\hat{\rho}_{0}\hat{K}_\eta\right)\right]^2
		\end{equation}
		
		Where $\{\hat{K}_\eta\}$ collectively stands for all the conserved quantities of $\hat{H}^{[2]}_{\text{eff}}$, $\hat{\rho}_0$ is the initial density operator, and the dependence of $\hat{\rho}_{\text{GE/GGE}}$ on the Lagrange multipliers is implicitly assumed.
		
		\subsection{Computation of infinite-time average and ETH values \label{app:sec:ETH_infty_time}}
		
		For an operator $\hat{O}$, the ETH value at inverse temperature $\beta$, and the infinite-time average $\overline{\hat{O}}$ with the initial state $\ket{\psi(0)}$ evolving under the Hamiltonian $\hat{H}$ are defined as follows
		
		\begin{subequations}
			\begin{align}
				\langle \hat{O} \rangle_\beta & = \text{Tr}\left(\hat{O}e^{-\beta\hat{H}}\right)/\text{Tr}\left(e^{-\beta\hat{H}}\right) \\
				\overline{\hat{O}} & = \lim_{T\rightarrow\infty} \frac{1}{T} \; \int_0^T \langle \psi(t) | \hat{O} | \psi(t) \rangle \; dt
			\end{align}
		\end{subequations}    
		
		When all eigenstates of $\hat{H}$, say $\{ \ket{E_\mu} \}$ are known, $\langle \hat{O} \rangle_\beta$ can by computed by evaluating the following expression,
		
		\begin{equation}
			\langle \hat{O} \rangle_\beta = \frac{\sum_{\mu} \langle E_\mu | \hat{O} | E_\mu \rangle e^{-\beta E_\mu}}{\sum_\mu e^{-\beta E_\mu}}
		\end{equation}
		
		We note that the initial state dependence of the ETH values comes in via the inverse temperature $\beta$ since we require $\langle \hat{H} \rangle_\beta = \langle \psi(0) | \hat{H} | \psi(0) \rangle $. Furhtermore we can express $\overline{\hat{O}}$ using the eigenstates of $\hat{H}$ as follows,
		
		\begin{equation}
			\overline{\hat{O}} =  \sum_{\mu,\nu} c_\mu^\ast O_{\mu\nu} c_\nu \left( \lim_{T\rightarrow\infty} \frac{1}{T}\int_0^T e^{i(E_\mu-E_\nu)t} \; dt \right)
		\end{equation}
		
		where we have defined $c_\mu = \langle E_\mu | \psi(0) \rangle$, $O_{\mu\nu} = \langle E_\mu | \hat{O} | E_\nu \rangle$. The integral reduces to zero, unless $E_\mu=E_\nu$. This happens trivially when $\mu=\nu$ and gives rise to the so-called diagonal ensemble average contribution to the infinite-time average. Additionally, in presence of exact degeneracies there are further contributions to the double sum above. For the particular Hamiltonian under consideration $\hat{H}$ (Eq.~\eqref{main:eq:hamiltonian_ladder}) the only \textit{exact} degeneracies (other than lattice symmetries) occur at zero energy due to the spectral reflection symmetry \cite{Pal_PhysRevB.111.L161101,Schecter_PhysRevB.98.035139}. Considering these two cases, we have the following expression which can be readily evaluated when all eigenstates of $\hat{H}$ are known, for example in an exact diagonalization setup,
		
		\begin{equation}
			\overline{\hat{O}}=\sum_\mu |c_\mu|^2 O_{\mu\mu} + \sum_{\mu,\nu \in \mathscr{H}_0} c_\mu^\ast c_\nu O_{\mu\nu}
		\end{equation}
		
		In the above equation $\mathscr{H}_0$ denotes the zero-mode sector of $\hat{H}$. We note that the emergent conservation laws of the effective Hamiltonian $\hat{H}_{\text{eff}}^{[2]}$ do not play any role in the computation of $\overline{\hat{O}}$.
		
		\section{Additional numerical results for protocol-II \label{app:protocol-II-additional-results}}
		
		In this section we present some additional results concerning the stability of the exact revivals discussed in Sec.-\ref{sec:floquet-flat-bands} when the many-body $\pi$-pulses are inexact. In particular we found that for the parameter regime explored in the main text, the $\ket{\text{vac.}}$ state was found to be more stable when compared to the $\ket{\mathbb{Z}_2}$ and the $\ket{\text{AR}}$ state. Considering other regimes of the parameter space illustrate that this is not a generic feature and there are some regimes where, in fact, for example, the $\ket{\mathbb{Z}_2}$ state becomes more stable compared to the $\ket{\text{vac.}}$ state. To see this, we shall consider the following measure of instability, $\mathcal{W}(\tau,\epsilon,\Delta) = 1-|\langle \psi(0)|\hat{U}_{F}^{\text{II}}(\tau,\epsilon,\Delta)|\psi(0)\rangle|^2$. Expanding $\hat{U}_{F}^{\text{II}}(\tau,\epsilon,\Delta)$ up to second order in $\epsilon$ we find,
		
		\begin{equation}
			\begin{split}
				\hat{U}_{F}^{\text{II}}(\tau,\epsilon,\Delta) & \approxeq (1+i\pi\epsilon\hat{N}-\frac{\pi^2\epsilon^2}{2}\hat{N}^2)e^{-i\hat{H}_{-\Delta_0}\tau/2}\\
				& (1+i\pi\epsilon\hat{N}-\frac{\pi^2\epsilon^2}{2}\hat{N}^2)e^{-i\hat{H}_{+\Delta_0}\tau/2}
			\end{split}
		\end{equation}
		
		Now, using the identity $\hat{C} e^{-i\hat{H}_{-\Delta_0}\tau/2} \hat{C} = e^{+i\hat{H}_{+\Delta_0}\tau/2}$, we find that the first order term in $\mathcal{W}$ is zero and the second order contribution to $\mathcal{W}(\tau,\epsilon,\Delta)$ reads,
		
		\begin{equation}
			\begin{split}
				\mathcal{W}^{[2]} & = -\pi^2\epsilon^2\langle \psi(0)| ( \hat{N} + \hat{U}_+^\dagger \hat{N} \hat{U}_+ ) | \psi(0) \rangle^2  \\
				& +\pi^2\epsilon^2\langle \psi(0)| ( \hat{N}^2 + \hat{U}_+^\dagger \hat{N}^2 \hat{U}_+ ) | \psi(0) \rangle  \\
				& + 2\pi^2\epsilon^2\text{Re}\left(\langle \psi(0) |\hat{N}\hat{U}_+^\dagger \hat{N} \hat{U}_+|\psi(0) \rangle\right)
			\end{split}
			\label{app:eq:floquet_instability}
		\end{equation}     
		
		In the above equation we have defined $\hat{U}_+=e^{-i\hat{H}_{+\Delta_0}\tau/2}$. We note that, as for the $\ket{\text{vac.}}$ state $\hat{N} \ket{\text{vac.}}=0$, Eq.~\eqref{app:eq:floquet_instability} further simplifies to,
		
		\begin{equation}
			\begin{split}
				\mathcal{W}^{[2]}_{\text{vac.}} & = -\pi^2\epsilon^2\langle \psi(0)|  \hat{U}_+^\dagger \hat{N} \hat{U}_+  | \psi(0) \rangle^2\\
				& +\pi^2\epsilon^2 \langle \psi(0)|  \hat{U}_+^\dagger \hat{N}^2 \hat{U}_+  | \psi(0) \rangle
			\end{split}
		\end{equation}
		
\begin{figure}[!htpb]
			\centering
			\rotatebox{0}{\includegraphics[width=0.485\textwidth]{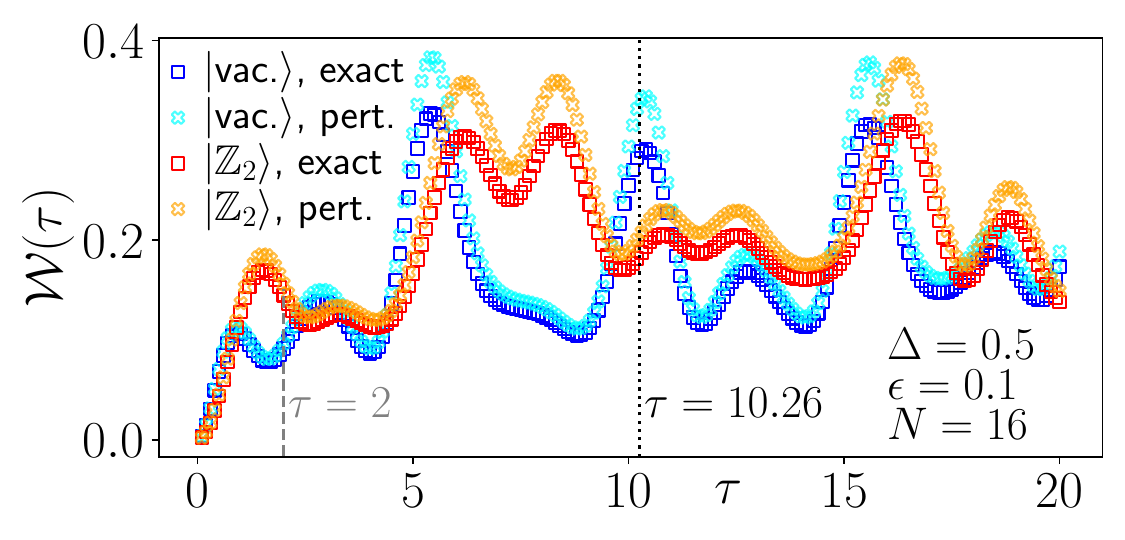}}
			\rotatebox{0}{\includegraphics[width=0.485\textwidth]{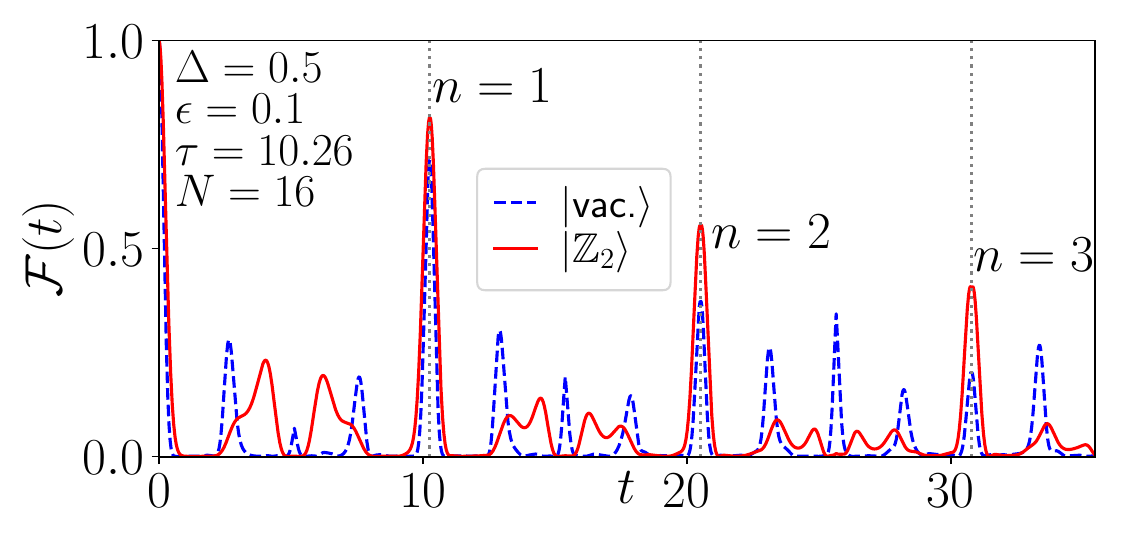}}
			\caption{\textit{Top:} Comparison of the instability $\mathcal{W}(\tau)$ for the initial states $\ket{\text{vac.}}$ and $\ket{\mathbb{Z}_2}$ with $N=16$ sites at $\Delta=0.5$ using exact time-evolution and Eq.~\eqref{app:eq:floquet_instability} (denoted as pert.) for $\epsilon=0.1$ for various values of $\tau$. \textit{Bottom:} Comparison of return probability $\mathcal{F}$ for $\ket{\mathbb{Z}_2}$ and $\ket{\text{vac.}}$ state at $\tau=10.26$ for first three cycles of the imperfect drive (other parameters are same as the top panel).}
			\label{app:fig:protocol-II-stability}
		\end{figure}

		In Fig.~\ref{app:fig:protocol-II-stability} we compare $\mathcal{W}$ (exact and perturbative) for $\ket{\mathbb{Z}_2}$ and $\ket{\text{vac.}}$ state for $\Delta=0.5$, $\epsilon=0.1$ (90\% accuracy in $\pi$-pulse) for values of $\tau \in [0,20]$ in a system with $N=16$ sites. The vertical gray dashed line corresponds to $\tau=2$ data shown in the main text. From this figure, it is clear that in some parameter regimes, for example at $\tau=10.26$ (black dotted vertical line), the $\ket{\mathbb{Z}_2}$ state can become more stable.
		
		\section{Generalizations to the 3-leg cylinder \label{app:3leg-cylinder}}
		
		The results obtained so far for the 2-leg staggered ladder can be generalized to lattices with more legs. To make this transparent, we now focus on the 3-leg generalization of the model \eqref{main:eq:hamiltonian_ladder} with a semi-staggered detuning pattern along the longer direction of the lattice and study the nature of quantum quench dynamics for a range of parameter values. As before, we consider periodic boundary conditions along both horizontal and vertical directions, and the Hamiltonian now reads 
		
		\begin{equation}
			\hat{H}_{\text{3-leg}} = \sum_{j=1}^{L_x}\sum_{a=1}^{L_y} \hat{\tilde\sigma}^x_{j,a} -\Delta \sum_{j=1}^{L_x} \sum_{a=1}^{L_y} (-1)^j \sigma^z_{j,a}
			\label{main:eq:hamiltonian_ladder_3leg}
		\end{equation}  
		
		where $L_y=3, L_x=L$ for a 3-leg cylinder of linear dimension $L$ and $\hat{\tilde{\sigma}}^\alpha_{j,a}\equiv\hat{P}^{\downarrow}_{j+1,a}\hat{P}^{\downarrow}_{j-1,a}\hat{P}^{\downarrow}_{j,a-1}\hat{P}^{\downarrow}_{j,a+1} \hat{\sigma}^\alpha_{j,a}$, with $\alpha=(x,y,z)$ 
		
		\begin{figure}[!t]
			\centering
			\includegraphics[width=0.45\textwidth]{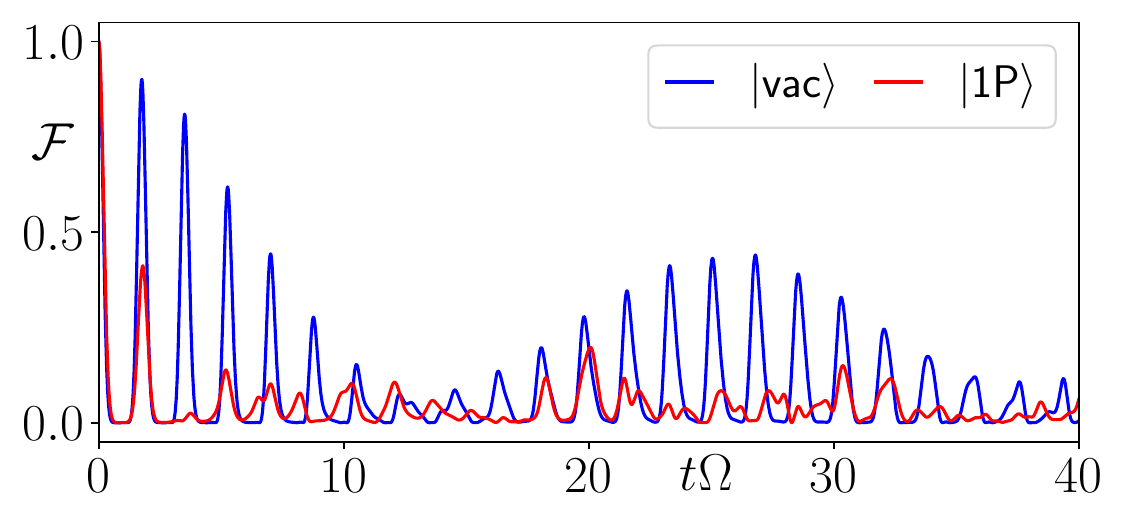}
			\caption{Return probability $\mathcal{F}(t) = |\langle \psi(0)|\psi(t) \rangle|^2$ starting from different initial states for $\Delta=1$ on a 3-leg cylinder (see Eq.~\eqref{main:eq:hamiltonian_ladder_3leg}) with periodic boundary conditions for a with linear size $L=6$ (total $N=L_y L_x= 18$ sites).}
			\label{fig:3leg-QMBS}
		\end{figure}       
		
		\subsection{Quantum many-body scars}
		
		We now consider quantum quench dynamics starting from simple initial product states (in the Fock/computational basis). Due to the periodic boundary condition, it is not possible to accommodate the N\'eel state $\ket{\mathbb{Z}_2}$ on a lattice with an odd number of legs without violating the Rydberg blockading constraint. For this reason, we do not consider the quantum quench dynamics at $\Delta=0$ in this case since, since for the 2-leg ladder, it was the N\'eel state that showed anomalous revivals for this parameter choice. We shall instead focus on the dynamics from other initial states that are allowed in a 3-leg cylinder geometry such as $\ket{\text{vac.}} \equiv \bigket{\substack{\circ\circ\circ\circ\circ\circ\\\circ\circ\circ\circ\circ\circ\\\circ\circ\circ\circ\circ\circ}}$ and the $\ket{\text{1P}} \equiv \bigket{\substack{\circ\circ\circ\circ\circ\circ\\\circ\circ\circ\bullet\circ\circ\\\circ\circ\circ\circ\circ\circ}}$ states at $\Delta=1$.\\

		For the $\ket{\text{vac.}}$ state, we have $\langle \text{vac.}|\hat{H}_{\text{3-leg}}|\text{vac.}\rangle=0$ which, combined with the fact that the spectrum of \eqref{main:eq:hamiltonian_ladder_3leg} has a spectral reflection symmetry, implies that the system should relax to an infinite temperature ($\beta_{\text{vac}}=0$) Gibbs ensemble. However, as Fig.~\ref{fig:3leg-QMBS} illustrates, the $\ket{\text{vac.}}$ state instead undergoes anomalous revivals at $\Delta=1$ which are qualitatively similar to the revivals in the 2-leg ladder scenario discussed in Ref.~\cite{Pal_PhysRevB.111.L161101}. In contrast, for the $\ket{\text{1P}}$ state, the initial energy density is $\langle \text{1P}|\hat{H}_{\text{3-leg}}|\text{1P}\rangle=-2\Delta$, which corresponds to a Gibbs ensemble of inverse temperature $\beta_{\text{1P}} \simeq -10^{-5}$, however this state does not show any anomalous persistent revivals in the return probability.\\

		\subsection{\texorpdfstring{Emergent Conservation Laws and Slow Dynamics $\Delta \gg 1$}{} }

		For $\Delta \gg 1$ the low-energy effective Hamiltonian can be obtained in the same way as in Appendix-\ref{app:Heff_SW} by considering a 3-leg version of the generator for SW rotations (see Eq.~\eqref{eq:SWgenerator}). As before, the rotation ensures there are no high-energy processes at first-order, and the second-order effective Hamiltonian now has the form \eqref{eq:HeffSW_3leg}. Similar to the 2-leg case, the effective low-energy Hamiltonian in this case \eqref{eq:HeffSW_3leg} only allows some specific spin-flip processes to take place if a ``blockading" condition is satisfied. This specific blockading constraint and the spin-flip process can be understood in the same way as for the 2-leg case by considering the terms involving $\hat{\sigma}^{x/y}$ in the second-order effective Hamiltonian given below (see Eq.~\ref{eq:HeffSW_3leg}). We have assumed periodic boundary conditions in both directions, which implies $a-2 \equiv a+1$ and $a+2 \equiv a-1$ for $a=1,2,3$ in a 3-leg cylinder.
		
		\begin{widetext}
			\begin{equation}
				\begin{split}
					\hat{H}^{[2]}_{\text{3-leg,eff}} = -\frac{1}{2\Delta} & \sum_{j,a} (-1)^j \hat{\tilde{\sigma}}^z_{j,a}-\frac{1}{4\Delta} \sum_{j,a} \sum_{j',a'} (-1)^j\left[ \hat{P}^{\downarrow}_{j-1,a-1} \hat{P}^{\downarrow}_{j+1,a-1} \hat{P}^{\downarrow}_{j,a-2} \hat{P}^{\downarrow}_{j-1,a} \hat{P}^{\downarrow}_{j,a+1}  \hat{P}^{\downarrow}_{j,a} \hat{P}^{\downarrow}_{j,a}\left(  \hat{\sigma}^y_{j,a} \hat{\sigma}^y_{j,a-1} + \hat{\sigma}^x_{j,a} \hat{\sigma}^x_{j,a-1} \right) \right] \\
					& + (-1)^j \left[ \hat{P}^{\downarrow}_{j-1,a+1} \hat{P}^{\downarrow}_{j+1,a+1} \hat{P}^{\downarrow}_{j-1,a} \hat{P}^{\downarrow}_{j+1,a} \hat{P}^{\downarrow}_{j,a-1} \hat{P}^{\downarrow}_{j,a+2} \hat{P}^{\downarrow}_{j,a} \left( \hat{\sigma}^y_{j,a} \hat{\sigma}^y_{j,a+1} +  \hat{\sigma}^x_{j,a} \hat{\sigma}^x_{j,a+1}\right)  \right]
				\end{split}
				\label{eq:HeffSW_3leg}
			\end{equation}
		\end{widetext}    
		
\begin{figure}[!htpb]
			\centering
			\includegraphics[width=0.5\textwidth]{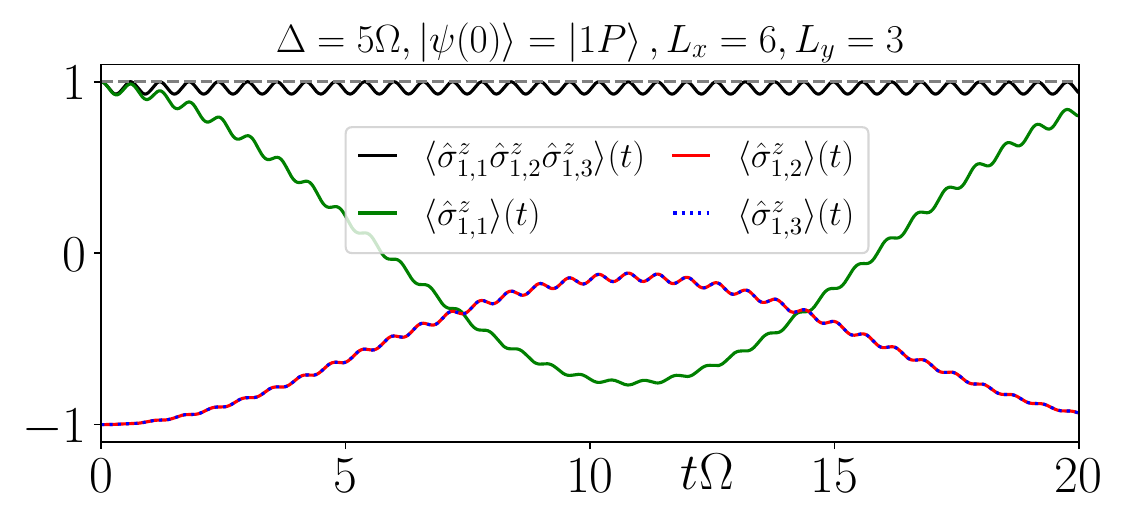}
			\caption[]{Illustration of 3-body conservation law $\langle\hat{\sigma}^z_{1,1}\hat{\sigma}^z_{1,2}\hat{\sigma}^z_{1,3}\rangle$ (black solid curve) in the 3-leg cylinder starting from a different one particle state, namely $\ket{\text{1P}} \equiv \bigket{\substack{\bullet\circ\circ\circ\circ\circ\\\circ\circ\circ\circ\circ\circ\\\circ\circ\circ\circ\circ\circ}}$. In this case, the product of $\hat{\sigma}^z$'s on each rung behaves as an emergent conserved quantity. For $\Delta \gg \Omega$, the 3-body conservation law would be exact and is shown by the horizontal dashed gray line for comparison.}
			\label{fig:3leg-conservation-law}
		\end{figure}

		Under this effective Hamiltonian, specific initial Fock states undergo simple evolution which can be described by considering only the ``blockaded-hopping" of the Rydberg excitations on the sites of the lattice allowed by $\hat{H}^{[2]}_{\text{3-leg,eff}}$, till times before the effects of the fourth-order terms of the perturbative Hamiltonian sets in (see bottom panel of Fig.~\ref{fig:3leg-conservation-law}). For the initial state $\ket{\text{1P}} \equiv \bigket{\substack{\bullet\circ\circ\circ\circ\circ\\\circ\circ\circ\circ\circ\circ\\\circ\circ\circ\circ\circ\circ}}$, evolving under the action of $\hat{H}^{[2]}_{\text{3-leg,eff}}$, at any instant of time $t$, the quantum state can be written in the form $\ket{\psi(t)}=\ell_1(t)\ket{(1,1)} + \ell_2(t)\ket{(1,2)}+\ell_3(t)\ket{(1,3)}$, with $|\ell_3(t)|=|\ell_2(t)|$ where $\ket{(j,a)}$ denotes the state one Rydberg excitation on site $(j,a)$ of the 3-leg cylinder which can be clearly seen in Fig.~\ref{fig:3leg-conservation-law}. 
		
		As a consequence of such a evolution the 3-body operator $\hat{\sigma}^z_{1,1}\hat{\sigma}^z_{1,2}\hat{\sigma}^z_{1,3}$ remains approximately conserved throughout the quantum evolution at $\Delta=5$. This is qualitatively similar to the 2-leg square ladder case, where product of 2-body spin operators remained approximately conserved. Thus, the results described in the main text for a $2$-leg square ladder can be easily extended for a $3$-leg cylinder with staggered detuning along the horizontal direction.

		\section{1D Atomic Chain vs. 1D PXP model \label{sec:1Dchain-blockade-dynamics}}
		
		The Rydberg blockading regime can be realized from Eq.~\eqref{eq:hamiltonian_Rydberg_long_range}, if the repulsive vdW interaction energy is much higher than other energy scales of the system. In such a scenario the system can be described by a simpler effective Hamiltonian obtained by perturbative arguments \cite{Bluvstein_doi:10.1126/science.abg2530}. In order to understand how accurately, the effective Hamiltonian in a constrained Hilbert space represents the dynamics of the full system as depicted by Eq.~\eqref{eq:hamiltonian_Rydberg_long_range}, we first focus on the paradigmatic PXP chain with site-dependent detunings $\Delta_j$ (henceforth referred to as the PXP+Z model) and the related fully interacting long-range model. The fully interacting long-range Hamiltonian for the 1D chain is the one dimensional analogue of Eq.~\eqref{eq:hamiltonian_Rydberg_long_range}. In the limit $V_0 \gg \Delta_j,\Omega$ the low-energy effective Hamiltonian reads \cite{Bluvstein_doi:10.1126/science.abg2530}
		
		\begin{equation}
			\hat{H}^{\text{eff}}_{\text{Ryd}} = \sum_{j=1}^L \left(\Omega\hat{\tilde\sigma}^x_{j} - \Delta_{j} \hat{\sigma}^z_{j}\right) + \frac{V_0}{2} \sum_{j,j' > \text{NN}_1} \frac{\hat{n}_{j}\hat{n}_{j'}}{|j-j'|^6} \label{eq:Rydberg-chain-lowenergy-effective}
		\end{equation}        
		
		Here the double sum in the equation above is over all pairs of sites which are not nearest neighbors of each other. If the parameters $V_0,\Omega,\Delta_j$ are such that $V_0 \gg \Delta_j,\Omega \gg V_0/2^6$, then the repulsive interaction term in Eq.~\eqref{eq:Rydberg-chain-lowenergy-effective} is extremely small and the evolution of the system is entirely captured by the PXP+Z model. Such a situation is in pratice achieved when $\Delta_j,\Omega\sim1$ and $V_0=10\Omega$, in which case the condition $V_0 \gg \Delta_j,\Omega \gg V_0/2^6$ is satisfied in an order of magnitude sense. However when $V_0/2^6$ is not negligible compared to $\Delta_j,\Omega$, the repulsive interaction terms play a significant role in the approximate evolution of the full system and needs to be considered. This is the case if, for example, the parameters are such that $V_0 \gg \Delta_j,\Omega \gg V_0/3^6$ and we can only effective ignore long-range interactions beyond two lattice separation units. In this case the Hamiltonian \eqref{eq:Rydberg-chain-lowenergy-effective} further simplifies to a short-range interacting Hamiltonian with the following form
		
		\begin{equation}
			\hat{H}^{\text{eff,NN}_2}_{\text{Ryd}} = \sum_{j=1}^L \left(\Omega\hat{\tilde\sigma}^x_{j} - \Delta_{j} \hat{\sigma}^z_{j}\right) + \sum_{j,j' \in \text{NN}_2} \frac{V_0\hat{n}_{j}\hat{n}_{j'}}{2|j-j'|^6}
			\label{eq:Rydberg-chain-lowenergy-shortrange}
		\end{equation}

		Here $j,j' \in \text{NN}_2$ implies that the double sum is restricted to $j,j'$ pairs separated by exactly two lattice sites.

		\begin{figure}[!htpb]
			\centering
			\rotatebox{0}{\includegraphics[width=0.5\textwidth]{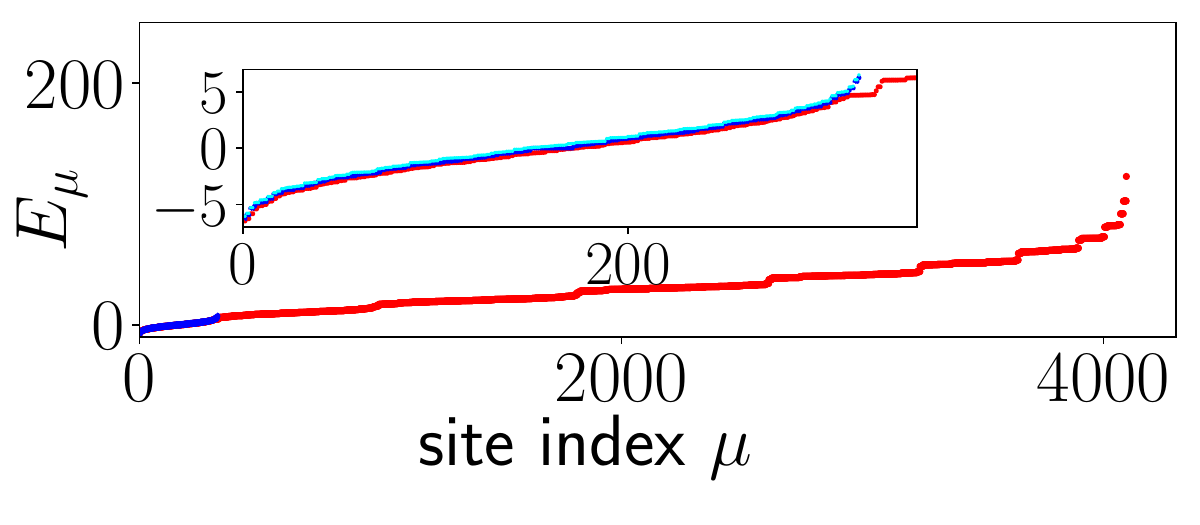}}
			\rotatebox{0}{\includegraphics[width=0.5\textwidth]{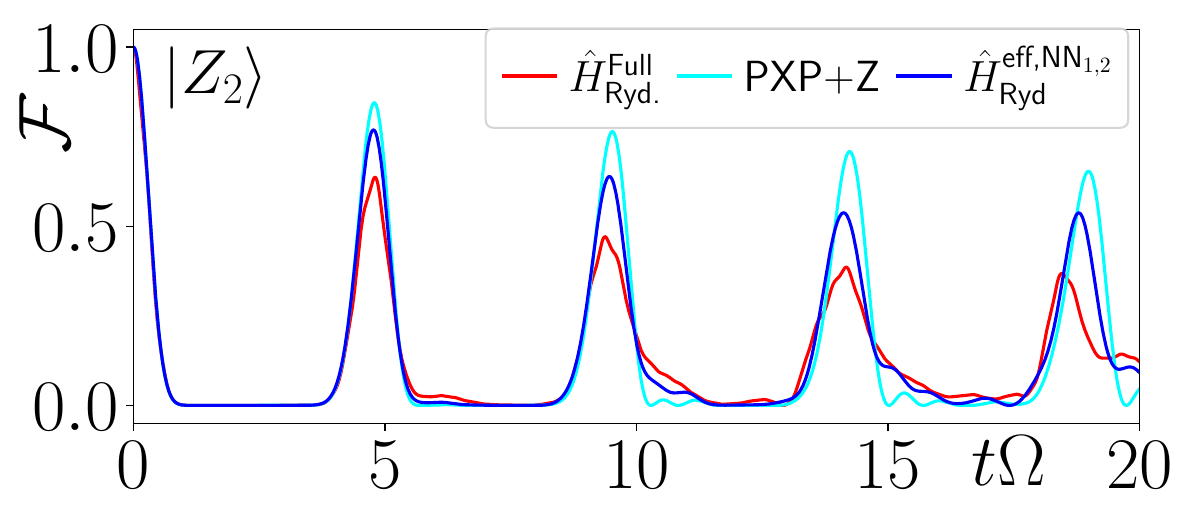}}
			\caption{1D atomic chain as captured via the Hamiltonians Eq.~\eqref{eq:hamiltonian_Rydberg_long_range},  Eq.~\eqref{eq:Rydberg-chain-lowenergy-shortrange} and the PXP+Z model for $N=12$ atoms and $V_0=10\Omega$: (top panel) many-body spectrum and (bottom panel) time-evolution of the return probability starting from the N\'eel state. See text for details.}
			\label{fig:1d-rydberg-chain-vs-effective-vs-PXP_Z}
		\end{figure}          
		
		In Fig.~\ref{fig:1d-rydberg-chain-vs-effective-vs-PXP_Z} we show a comparison of the Hamiltonians \eqref{eq:hamiltonian_Rydberg_long_range}, \eqref{eq:Rydberg-chain-lowenergy-shortrange}: the top panel shows that the low-energy part of the many-body spectrum of the full long-range interacting system \eqref{eq:hamiltonian_Rydberg_long_range} can be faithfully captured by the entire spectrum of the kinetically constrained Hamiltonian (PXP+Z model). As the inset of this panel demonstrates, there is no qualitative difference in the energy eigenvalues even when the second nearest neighbor interactions (Eq.~\eqref{eq:Rydberg-chain-lowenergy-shortrange}) are considered beyond the Rydberg blockade regime. The bottom panel of Fig.~\ref{fig:1d-rydberg-chain-vs-effective-vs-PXP_Z} shows the time-evolution of the return probability starting from the N\'eel state and we can conclude that, although the qualitative features remain the same, there are quantitative differences in the behaviors predicted by Eq.~\eqref{eq:hamiltonian_Rydberg_long_range}, Eq.~\eqref{eq:Rydberg-chain-lowenergy-shortrange} and the PXP+Z model. Comparison of both spectral as well as dynamical features of these three descriptions indicate that the PXP+Z model provides a good qualitative description of the full long-range interacting many-body system described by Eq.~\eqref{eq:hamiltonian_Rydberg_long_range}. \\

		\section{Scaling of the norm of rescaled, regularized Adiabatic Gauge Potential (AGP)\label{app:AGP_norm_scaling}}
		
		To understand the onset of thermalization and the finite size effects we have studied the behavior of adiabatic gauge potential (AGP) for different system sizes $N$ for various values of $\Delta$. It has been argued in earlier works \cite{Pandey_PhysRevX.10.041017}, that the scaling behavior of the norm of the regularized AGP is an extremely sensitive indicator of quantum chaos, and can detect its onset even in circumstances where widely used alternative probes such as the mean consecutive level spacing ratio \cite{Oganesyan_PhysRevB.75.155111} fails to do so. The scaling behavior of the regularized AGP can correctly predict the onset of quantum chaos even for very modest system sizes as this directly probes features of a many-body Hamiltonian at time scales which are exponentially large in system size. In this section, we consider this sensitive indicator of quantum chaos and illustrate that for all the parameter regimes discussed in the paper, the idealized kinetically constrained system \eqref{main:eq:hamiltonian_ladder}, is overall, quantum chaotic, although various forms of ergodicity breaking behavior persists for finite-size systems at finite times for specific initial states as discussed in Sec.~\ref{main-sec:quench_dynamics}.    
		
		The adiabatic gauge potential (AGP) is defined as the generator of adiabatic evolution of the eigenstates of the Hamiltonian, when some parameters of the Hamiltonian are varied. As in the rest of this paper, we are concerned with the behavior of the model \eqref{main:eq:hamiltonian_ladder}, when $\Omega=1$, and $\Delta$ is varied. We shall focus on the AGP $\hat{\mathcal{A}}_{\Delta}$ defined as
		
		\begin{equation}
			\hat{\mathcal{A}}_{\Delta} \ket{E_\mu(\Delta)} = i\partial_{\Delta}\ket{E_\mu(\Delta)}
		\end{equation}
		
		\begin{figure}[!t]
			\centering
			\includegraphics[width=0.4\textwidth]{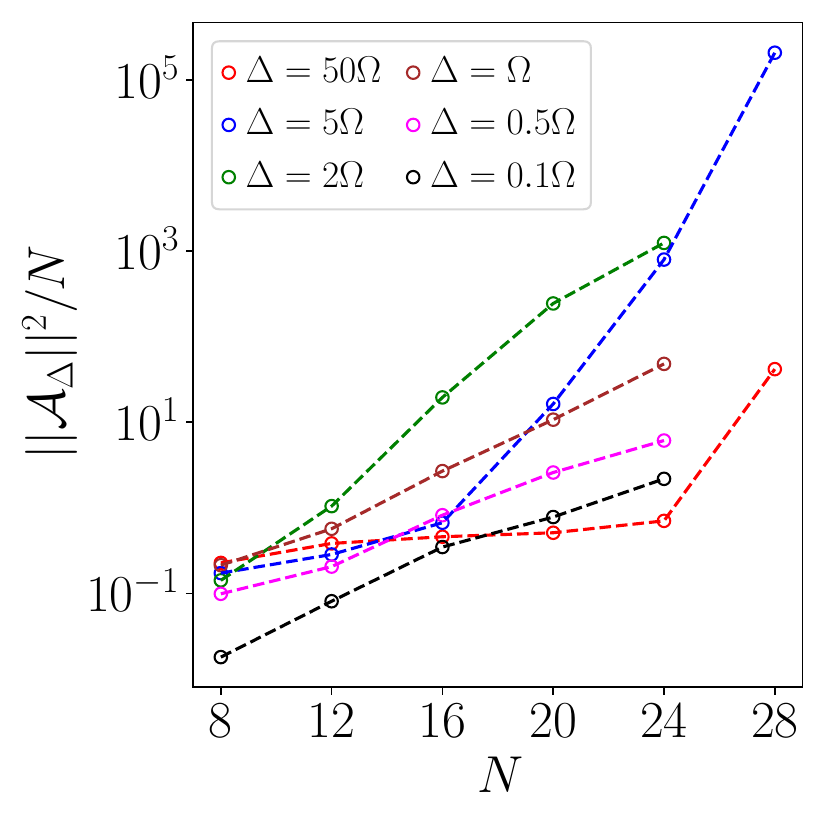}
			\caption{Behavior of the rescaled, regularized AGP norm for a range of values of $\Delta$ for system sizes $N=8,12,16,20,24,28$. In all cases $||\hat{\mathcal{A}}_\Delta||^2/N$ increases in an exponential fashion as larger system sizes are accessed, which indicates that the Hamiltonian \eqref{main:eq:hamiltonian_ladder} is as a whole, quantum chaotic, even if ergodicity breaking features of various forms exist for specific states and finite-size systems.}
			\label{fig:APGnorm-scaling}
		\end{figure}        
		
		Where $\hat{H}(\Delta)\ket{E_\mu(\Delta)} = E_\mu(\Delta)\ket{E_\mu(\Delta)}$. To understand whether the model given by Eq.~\eqref{main:eq:hamiltonian_ladder} is chaotic for finite values of $\Delta$ at times which are exponentially large in system size, we study the scaling properties of the regularized AGP norm \cite{Pandey_PhysRevX.10.041017}
		
		\begin{equation}
			|| \hat{\mathcal{A}}_\Delta ||^2 = \frac{1}{\mathcal{D}_H} \sum_{\nu} \sum_{\mu\ne\nu} \frac{\omega_{\mu\nu}^2}{(\omega_{\mu\nu}^2+\zeta^2)^2}|\langle E_\mu | \partial_\Delta \hat{H}| E_\nu \rangle|^2
			\label{eq:AGPnorm2-regularlized}
		\end{equation}
		
		In the above equation, the sum on $\mu,\nu$ runs over the Hilbert space (which may or may not be symmetry-reduced),  $\omega_{\mu\nu} \equiv E_\mu-E_\nu$ and $\zeta \propto N\log(N)$ is a cutoff introduced to regularize singularities arising due to degeneracies in the spectrum. Introducing the ultraviolet regulator $\zeta$, ensures that even if the symmetries of the Hamiltonian are not completely resolved, the regularized AGP norm gives a meaningful answer. In contrast almost all alternative measures of spectral statistics which are readily used as probes of the onset of quantum chaos (see Ref.~\cite{LucaDAlessio03052016}), the knowledge and resolution of all symmetries of the Hamiltonian is essential. We note that the spectral form factor is an important exception, but necessarily relies on the introduction of disorder in the system to extract a reliable signature (the dip-ramp-plateau structure \cite{LucaDAlessio03052016}).\\

		In Fig.~\ref{fig:APGnorm-scaling} we show the behavior of the rescaled norm of the regularized AGP i.e. $||\hat{\mathcal{A}}_\Delta||^2/N$, for a range of system sizes $N=8-28$. This figure illustrates that $||\hat{\mathcal{A}}_\Delta||^2/N$ grows with $N$ in an exponential fashion for all regimes of $\Delta$. This exponential growth becomes noticeable only for larger system sizes when we are in the regime of emergent approximate conservation laws ($\Delta \gg 1$) as seen by the data for $\Delta=5$ (blue open circles) and $\Delta=50$ (red open circles). In contrast for lower values of $\Delta$ such as $\Delta=2$ (green open circles), $\Delta=1$ (brown open circles), $\Delta=0.5$ (magenta open circles) and $\Delta=0.1$ (black open circles) the exponential growth seems noticeable even for small system sizes.

	\end{appendix}

	\bibliography{references}
	
\end{document}